\documentclass[12pt,journal]{IEEEtran}
\usepackage{amsmath}
\usepackage{amsfonts}
\usepackage{amssymb}
\usepackage{graphicx}
\usepackage{epstopdf}
\usepackage[cmintegrals]{newtxmath}
\usepackage{multirow}
\usepackage{cite}
\usepackage{graphicx}
\usepackage{csquotes}
\usepackage{makecell}
\usepackage[keeplastbox]{flushend}
\usepackage[table,xcdraw]{xcolor} 
\usepackage[acronym,nopostdot]{glossaries}
\usepackage{xurl}
\usepackage{balance}
\usepackage{tikz-qtree}
\usepackage[keeplastbox]{flushend} 
\usetikzlibrary{arrows.meta,shapes.geometric,positioning,shadows,trees,shadings,backgrounds,fit,automata}
\usepackage[edges]{forest}
\usepackage{pgf}
\usetikzlibrary{shapes.multipart} 
    \tikzstyle{every node}=[thin,anchor=west, minimum height=3.1em]
\usetikzlibrary{arrows,shapes,positioning,shadows,trees,scopes}
\newcommand{\RN}[1]{%
\textup{\uppercase\expandafter{\romannumeral#1}}
}
\forestset{
  direction switch/.style={
    for tree={edge+=thick, font=\sffamily},
    where level=0{
       edge from parent fork down,
      l sep'=1.2cm,
      s sep'=1cm, draw=black, rounded corners=6pt, text width=15cm,text centered,  fill=blue!20, level distance=3cm, minimum height=3.1em,
    }{
      grow'=0
    },
    where level=1{folder, grow'=0, draw, text width=3.5cm , text centered,  level distance=3cm,}{for children=forked edge},
    where level=2{folder, grow'=0, draw, fill=pink!30, text width=3.6cm , text centered,   l sep=8mm}{},
    where level=3{text width=2.9cm, draw,fill=yellow!15, text centered}{},
  }
}

\makeglossaries 
\newacronym{dsn}{DSN}{deep-space network}
\newacronym{dtn}{DTN}{delay/disruption tolerant networking}
\newacronym{bp}{BP}{bundle protocol}
\newacronym{bpsec}{BPSec}{bundle protocol security}
\newacronym{cgr}{CGR}{contact graph routing}
\newacronym{ietf}{IETF}{Internet Engineering Task Force}
\newacronym{irtf}{IRTF}{Internet Research Task Force}
\newacronym{ccsds}{CCSDS}{Consultative Committee for Space Data Systems}
\newacronym{ssi}{SSI}{solar system Internet}
\newacronym{darpa}{DARPA}{Defense Advanced Research Projects Agency}
\newacronym{dro}{DRO}{distant retrograde orbit}
\newacronym{elfo}{ELFO}{elliptical lunar frozen orbit}
\newacronym{edl}{EDL}{entry, descent, and landing}
\newacronym{edrs}{EDRS}{European data relay system}
\newacronym{esa}{ESA}{European Space Agency}
\newacronym{fcc}{FCC}{Federal Communications Commission}
\newacronym{geo}{GEO}{geostationary Earth orbit}
\newacronym{hydron}{HydRON}{High thRoughput Optical Network}
\newacronym{hydron ds}{HydRON DS}{HydRON demonstration system}
\newacronym[\glslongpluralkey={inter-satellite links}]{isl}{ISL}{inter-satellite link}
\newacronym{lct}{LCT}{laser communication terminal}
\newacronym[\glslongpluralkey={Laser Inter-Satellite Links}]{lisl}{LISL}{Laser Inter-Satellite Link}
\newacronym{leo}{LEO}{low Earth orbit}
\newacronym{lco}{LCO}{lunar circular orbit}
\newacronym{marco}{MarCO}{Mars Cube One}
\newacronym{nasa}{NASA}{National Aeronautics and Space Administration}
\newacronym{nrho}{NRHO}{near-rectilinear halo orbit}
\newacronym[\glslongpluralkey={optical inter-satellite links}]{oisl}{OISL}{optical inter-satellite link}
\newacronym{osc}{OSC}{Optical Satellite communications Consortium}
\newacronym{scylight}{ScyLight}{SeCure and Laser communication Technology}
\newacronym{swap}{SWaP}{size, weight, and power}
\newacronym{sda}{SDA}{Space Development Agency}
\newacronym[\glslongpluralkey={very high throughput satellites}]{vhts}{VHTS}{very high throughput satellite}
\newacronym{vleo}{VLEO}{very low Earth orbit}
\newacronym{meo}{MEO}{medium Earth orbit}
\newacronym[\glslongpluralkey={Radio Frequency Inter-Satellite Links}]{rfisl}{RF ISL}{Radio Frequency Inter-Satellite Link}
\newacronym[\glslongpluralkey={Free-Space Optical Inter-Satellite Links}]{fsoisl}{FSO ISL}{Free-Space Optical Inter-Satellite Link}
\newacronym{atp}{ATP}{acquisition, tracking, and pointing}
\newacronym{pat}{PAT}{pointing, acquisition, and tracking}
\newacronym{rf}{RF}{radio frequency}
\newacronym{fso}{FSO}{free-space optical}
\newacronym{ut}{UT}{User Terminal}
\newacronym[\glslongpluralkey={free-space optical satellite networks}]{fsosn}{FSOSN}{free-space optical satellite network}
\newacronym[\glslongpluralkey={orbital planes}]{op}{OP}{orbital plane}
\newacronym{imdd}{IMDD}{intensity modulation and direct detection}
\newacronym{ook}{OOK}{on-off keying}
\newacronym{hft}{HFT}{high-frequency trading}
\newacronym[\glslongpluralkey={mobile network operators}]{mno}{MNO}{mobile network operator}
\newacronym{fov}{FoV}{field of view}
\newacronym[\glslongpluralkey={user equipments}]{ue}{UE}{user equipment}
\newacronym{sdn}{SDN}{software defined network}
\newacronym{sdsn}{SDSN}{software defined satellite network}
\newacronym{nfv}{NFV}{network function virtualization}
\newacronym{vnf}{VNF}{virtual network function}
\newacronym{td}{TD}{topology discovery}
\newacronym{embb}{eMBB}{enhanced mobile broadband}
\newacronym{vsat}{VSATs}{very small aperture terminals}
\newacronym{ta}{TA}{tracking area}
\newacronym{harq}{HARQ}{hybrid automatic repeat request}
\newacronym{rlc}{RLC}{radio link control}
\newacronym{tti}{TTI}{transmission time interval} 
\newacronym{rtt}{RTT}{round trip time} 
\newacronym{arq}{ARQ}{automatic repeat request} 
\newacronym{acm}{ACM}{automatic coding and modulation}  
\newacronym{rar}{RAR}{random access response}
\newacronym{papr}{PAPR}{peak-to-average power ratio}
\newacronym{prach}{PRACH}{physical random access channel}
\newacronym{vap}{VAP}{virtual attachment point}
\newacronym{ptrs}{PT-RS}{phase tracking reference signal}
\newacronym{pls}{PLS}{physical layer security}
\newacronym{tdd}{TDD}{time division duplex}
\newacronym{fdd}{FDD}{frequency division duplex}
\newacronym{cpofdm}{CP-OFDM}{cyclic prefix – orthogonal frequency division multiplexing}
\newacronym{nocc}{NOCC}{network operation and control centre}
\newacronym{3gpp}{3GPP}{3^{rd} 	\textrm{generation partnership project}}
\newacronym{iot}{IoT}{Internet of things}
\newacronym{capex}{CAPEX}{capital expenditures}
\newacronym{opex}{OPEX}{operational expenditures}
\newacronym{qos}{QoS}{quality of service}
\newacronym{qoe}{QoE}{quality of experience}
\newacronym{mec}{MEC}{mobile edge computing}
\newacronym{si}{SI}{study item}
\newacronym{tsg}{TSG}{technical specification group}
\newacronym{nr}{NR}{new radio}
\newacronym{wi}{WI}{work item}
\newacronym{ran}{RAN}{radio access network}
\newacronym{ct}{CT}{core network and terminal}
\newacronym{sa}{SA}{service and system aspects}

\newacronym[\glslongpluralkey={Earth orbit satellites}]{eos}{EOS}{Earth orbit satellite}
\newacronym[\glslongpluralkey={Earth orbit satellite constellations}]{eosc}{EOSC}{Earth orbit satellite constellation}
\newacronym[\glslongpluralkey={laser inter-space node links}]{lisnl}{LISNL}{laser inter-space node link}
\newacronym[\glslongpluralkey={inter-space node links}]{isnl}{ISNL}{inter-space node link}
\newacronym[\glslongpluralkey={space nodes}]{sn}{SN}{space node}
\newacronym[\glslongpluralkey={radio frequency inter-space node links}]{rfisnl}{RFISNL}{radio frequency inter-space node link}
\newacronym[\glslongpluralkey={free-space optical inter-space node links}]{fsoisnl}{FSOISNL}{free-space optical inter-space node link}
\newacronym[\glslongpluralkey={optical inter-space node links}]{oisnl}{OISNL}{optical inter-space node link}
\newacronym{losc}{LOSC}{lunar orbit satellite constellation}
\newacronym[\glslongpluralkey={lunar orbit satellites}]{los}{LOS}{lunar orbit satellite}
\newacronym[\glslongpluralkey={Earth orbit satellite networks}]{eosn}{EOSN}{Earth orbit satellite network}
\newacronym[\glslongpluralkey={space network terminals}]{snt}{SNT}{space network terminal}
\newacronym[\glslongpluralkey={high-altitude platform systems}]{haps}{HAPS}{high-altitude platform system}
\newacronym[\glslongpluralkey={non-terrestrial networks}]{ntn}{NTN}{non-terrestrial network}
\newacronym[\glslongpluralkey={unmanned aerial vehicles}]{uav}{UAV}{unmanned aerial vehicle}

\ifCLASSINFOpdf
\else
\fi
\begin{document}
%
\title{Future Space Networks: \\Toward the Next Giant Leap for Humankind\\ \Large{Invited Paper}}
%
%
%
\author{Mohammed~Y.~Abdelsadek,~\IEEEmembership{Senior Member,~IEEE,}
        Aizaz~U.~Chaudhry,~\IEEEmembership{Senior Member,~IEEE,}
        Tasneem~Darwish,~\IEEEmembership{Senior Member,~IEEE,}
        Eylem~Erdogan,~\IEEEmembership{Senior Member,~IEEE,}
        Gunes~Karabulut-Kurt,~\IEEEmembership{Senior Member,~IEEE,}
        Pablo~G.~Madoery,~\IEEEmembership{Member,~IEEE,}
        Olfa~Ben~Yahia,~\textit{Member,~IEEE,}
        and~Halim~Yanikomeroglu,~\IEEEmembership{Fellow,~IEEE}
\thanks{M. Y. Abdelsadek, A. U. Chaudhry, T. Darwish, P. G. Madoery, and H. Yanikomeroglu are with the Systems and Computer Engineering Department, Carleton University, Ottawa, ON K1S 5B6, Canada. Emails: \{mohammedabdelsadek; auhchaud; tasneemdarwish; pablomadoery; halim\} @sce.carleton.ca. M. Y. Abdelsadek is also with (on leave) the Department of Electrical Engineering, Assiut University, Assiut 71516, Egypt.}
\thanks{E. Erdogan is with the Department of Electrical and Electronics
Engineering, Istanbul Medeniyet University, 34720 Istanbul, Turkey. Email: eylem.erdogan@medeniyet.edu.tr.}
\thanks{G. Karabulut-Kurt is with the Department of Electrical Engineering, Polytechnique Montréal, Montréal, QC H3T 1J4, Canada. She is also an Adjunct Research Professor in the Department of Systems and Computer Engineering, Carleton University, Ottawa, ON K1S 5B6, Canada. Email: gunes.kurt@polymtl.ca.}
\thanks{O. Ben Yahia is with the Department of Electrical Engineering, Polytechnique Montréal, Montréal, QC H3T 1J4, Canada. Email: olfa.b.yahia@gmail.com}
}
%
%

\markboth{}%
{}
%



\maketitle

{\color{black}
\begin{abstract}
Due to the unprecedented advances in satellite fabrication and deployment,  innovative communications and networking technologies, ambitious space projects and programs, and the resurgence of interest in satellite networks, there is a need to redefine space networks (SpaceNets) to incorporate all of these evolutions. This paper introduces a vision for future SpaceNets that considers advances in several related domains. First, we present a reference architecture that captures the various network entities and terminals in a holistic manner. Based on this, space, air, and ground use cases are studied. Then, the architectures and technologies that enable the envisaged SpaceNets are investigated. In so doing, we highlight the activities and projects of different standardization bodies, satellite operators, and national organizations towards the envisioned SpaceNets. Finally, the challenges, potential solutions, and open issues from communications and networking perspectives are discussed.
\end{abstract}
}

\glssetwidest{HAPS-SMBS}
\hspace{1cm}\printglossary[style=alttree,type=\acronymtype,title=\textbf{Abbreviations},nogroupskip, nonumberlist]

%
\IEEEpeerreviewmaketitle

%
\section{Introduction}
\label{sec:Introduction}
%
%
%
%

\IEEEPARstart{T}{he} curiosity of humans to know about the space outside Earth's atmosphere has started since ancient times by means of observatory astronomy using naked eyes. The development of different types of telescopes revolutionized astronomy by offering more accurate images of the cosmos. With the advances in astronautics, humans began deploying spacecraft in outer space,  which has enabled space communications, Earth observation, and deep space exploration. One major benefit of using such spacecraft is the \glspl{eosn}. The idea of using satellites to relay radio signals was first introduced in Clarke's seminal article, ``Extra-Terrestrial Relays," published in Wireless World magazine in 1945 \cite{clarke1966extra}. The first satellite, Sputnik 1,  was launched in 1957. Since then,  thousands of satellites have been launched into space for a wide range of applications. From a networking perspective, EOSNs used to be considered the exterior layer of terrestrial and aerial networks; today, they are mainly used to backhaul these networks when fiber optic is not a feasible solution (e.g., along oceans).  This is due to the nature of \glspl{eos}  being deployed at significantly higher altitudes compared to terrestrial and aerial access points. Besides, some space-related activities (e.g., deep-space exploration) opt for direct connectivity between terrestrial stations and space. That is, they operate independently of EOSNs and are handled by different entities. This has been the state of satellite networks and space missions for a long time.

However, recent advances in space-related activities and EOSNs are changing the way we view space networks (SpaceNets). For example, several nations have been involved in deep space exploration and have created \glspl{dsn}, such as USA, Russia, China, European Union, India, and Japan. The spacecraft of these DSNs directly connect to ground stations from space, which entails severe transmission losses due to the long distance to Earth. This also requires multiple stations on the ground with huge dish antennas. However, utilizing relaying satellites (e.g., orbiting Mars and Earth) would be more efficient and support higher data rate transmissions. This has led to a new paradigm called the interplanetary Internet, which is being investigated in a number of studies \cite{ccsds2014ssi, alhilal2019thesky}. Besides, researching permanent settlement of humans on other planets (e.g., Mars) is being carried out by several space agencies (e.g., the National Aeronautics and Space Administration (NASA), China National Space Administration (CNSA), European Space Agency (ESA), and Roscosmos) and private organizations such as SpaceX, Lockheed Martin, and Boeing. This is in addition to other ambitious projects, such as space tourism, asteroid mining, and space farming, that have started to come into the picture. On another front, there is significant progress in EOSNs. In 2016, the total number of active satellites orbiting the Earth was $1,459$ \cite{pratt2019satellite}. Yet by September 2021 the number of active satellites had risen to $4,550$ \cite{UCSActiveSats} (i.e., more than $210\%$ increase in less than 5 years). Furthermore, more than $100,000$ satellites are planned to be deployed by 2030. A closer look reveals that the main game-changers of the satellite industry are the unprecedented advances in microelectronics, innovative commercial off-the-shelf technology solutions, and rocket launch platforms, which have reduced the cost of manufacturing and deploying satellites.

These developments and others have triggered a redefinition of SpaceNets in a more holistic way. The new definition should incorporate different types of spacecraft (e.g., satellites, space stations, space hotels) and any spacecraft that can be used as network nodes, which we will refer to as \glspl{sn}.  These SNs will serve various \glspl{snt}, which include ground and air user terminals (UTs), UTs in a space hotel, a rover on Mars, a space probe, a crewed space capsule, a space telescope that studies distant stars and galaxies, etc. Moreover, EOSNs should be considered as the edge of SpaceNets that enable us to interface with other space nodes while providing a complete set of services for terrestrial and air applications. That is, the antennas of EOSNs should be pointing towards Earth and space as well.

Accordingly, future SpaceNet will be the network that interconnects all of these entities (i.e., SNs, EOSN, and SNTs) and connects them with other networks (e.g., the Internet, aerial networks, and cellular networks). This network of networks will enable a plethora of applications spanning terrestrial (e.g., integrated terrestrial-satellite access), aerial (e.g., integrated aerial-satellite communications), and space (e.g., solar system Internet, space tourism, and moon village) use cases that would improve human life on Earth and connect them to space. However, many technical challenges need to be addressed to realize such a network. For instance, due to the motion of SNs, the topology of SpaceNet is highly dynamic, which entails several issues for mobility management, inter-SN links, and routing of data packets. In this paper, we present a vision and definition for future SpaceNets that consider the aforementioned new developments and investigate what it will take to get us there. Besides, we present how this SpaceNet can change human life by presenting the potential use cases and applications. Moreover, the challenges from a communications and networking perspective and potential solutions are studied. 

\subsection{Paper Organization and Contributions}
\label{ssec:ContributionOrg}

\begin{figure*}[!htb]
	\centerline{\includegraphics[scale=1.01]{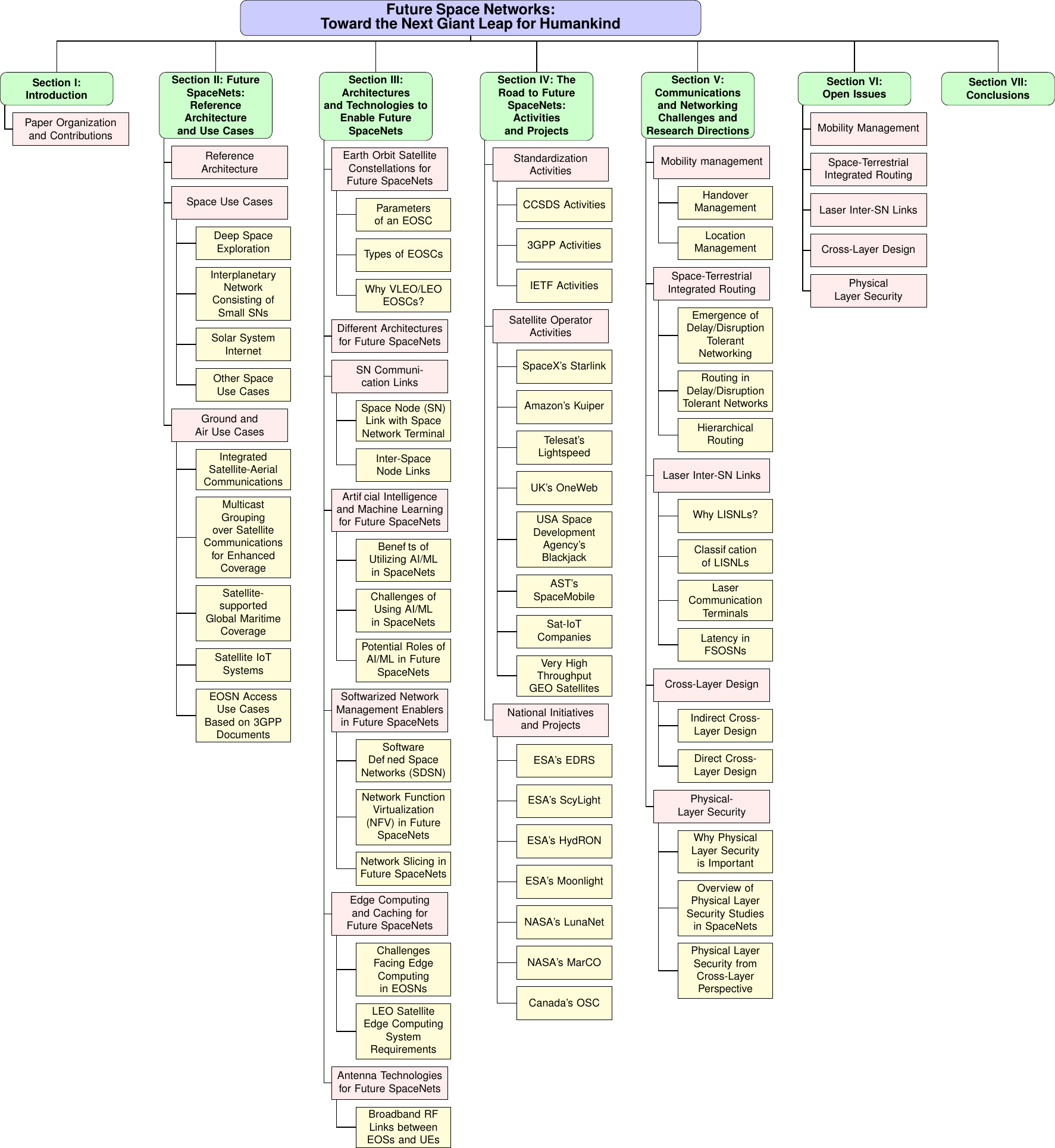}}
    \caption{Organization of the paper.}
    \label{fig:paperOrganization}
\end{figure*}

{\color{black} 
The structure of this paper is depicted in Fig. \ref{fig:paperOrganization}. In Section \ref{sec:UseCases}, we present a novel reference architecture for future SpaceNets. Based on this architecture, the potential space use cases and services are discussed, and we consider how future EOSNs can revolutionize networking, computing, sensing, navigation, and positioning applications on the ground and in the air and connect the unconnected of our world. In Section \ref{sec:Enablers}, we investigate the architectures and technologies that will enable the envisaged SpaceNets and use cases. Section \ref{sec:Projects} highlights the activities of standardization bodies, satellite operators, and national organizations as they pertain to the envisioned SpaceNets. Section \ref{sec:ResearchDirections} investigates the challenges, potential solutions, and research directions to realize such future SpaceNets from the communications and networking perspectives. Finally, Section \ref{sec:OpenIssues} presents the open issues in the literature in these directions, and Section \ref{sec:Conclusions} concludes the paper.

The topics included in this paper are precisely selected to provide a complete picture for the envisioned SpaceNets to cover the gaps in the literature as they pertain to different aspects of SpaceNets. Table \ref{tab:RelatedWorkMatrix}  summarizes the coverage level of each topic compared to the most related visionary and survey works from the literature. As the table reveals, the selected topics are not well-covered in the literature on satellite communications and networking.

In comparison with these existing survey works, the specific contributions of our paper are summarized hereafter. In Section \ref{ssec:basics}, we review general parameters of an \acrfull{eosc}, general types of EOSCs, different types of EOSCs based on altitude, and pros and cons of \acrfull{vleo}/\acrfull{leo} EOSCs. The types of EOSCs with respect to altitude are discussed by Kodheli \emph{et al}. \cite{kodheli2020satellite} and Li \emph{et al}. \cite{li2019physical}. In Section \ref{ssec:DifArch}, we outline the different components of the 3-D layered architecture for future SpaceNets. Only the work in \cite{kodheli2020satellite} has briefly presented a general description of the envisioned architecture for future networks. In \ref{sssec:UpDownLinks}, we outline the different channel impairments with mitigation techniques for the SN links with a space network terminal. Some of the satellite channel impairments including delay and Doppler have been provided in \cite{guidotti2020architectures}. However, non of the reference works have discussed the mitigation techniques. In Section \ref{sssec:ISLs}, we discuss the logic behind the use of \glspl{isnl}, types of ISNLs, the Iridium NEXT EOSC that employs ISNLs, and challenges that are faced in the creation of ISNLs, including cross-seam communication, pointing, acquisition, and tracking, and Doppler shift. Some related works on Doppler effect in LEO \acrshort{eos}-based system are provided by Kodheli \emph{et al}. \cite{kodheli2020satellite}. The Iridium system is mentioned by Li \emph{et al}. \cite{li2019physical}. Radhakrishnan \emph{et al}. \cite{radhakrishnan2016survey} have discussed the benefits of inter-EOS communications, Iridium as an example of a multiple EOS mission with inter-EOS communications, and a related work on Doppler shift in EOS formation flying.

In Section \ref{ssec:AI_ML}, we discuss AI/ML as a technology to enable future SpaceNets. AI/ML was briefly discussed by Rinaldi \textit{et al.} \cite{rinaldi2020non} under future and open issues. However, in this paper, we discuss the benefits that AI/ML can bring to SpaceNets, the challenges of using it, and the potential roles that it can play in SpaceNets. In section \ref{sec:Enablers}-E and \ref{sec:Enablers}-F, we discussed the current advances on softwarized network management and edge computing as two main enabler technologies for future satellite networks. Also, we highlighted some gaps and points that require further investigation.

In Section \ref{sssec:spacemobile}, we examine the antenna technology being developed by AST for realizing broadband \acrfull{rf} links between EOSs in space and \glspl{ue} on ground since AST is planning an EOSC named SpaceMobile consisting of LEO EOSs with a very large aperture phased array antenna to directly communicate with off-the-shelf mobile phones on ground. A. Guidotti \emph{et al}. have mentioned beam layout parameters for multi-beam LEO and \acrfull{geo} EOSs from 3GPP’s technical report 38.821 on solutions for New Radio to support non-terrestrial networks \cite{guidotti2020architectures}. A discussion on antennas suitable for very small CubeSat EOSs is provided by Burleigh \emph{et al}. \cite{burleigh2019connectivity}. Multi-beam phased array antenna operating in multiple frequency bands is discussed as a technology enabler on the ground user terminal side by Ravishankar \emph{et al}. \cite{ravishankar2021next}. Kodheli \emph{et al}. have mentioned the use of active antennas with direct radiating array architecture for LEO EOSs \cite{kodheli2020satellite}. A discussion on modular antenna arrays for \glspl{rfisnl} between CubeSats is given by Radhakrishnan \emph{et al}. \cite{radhakrishnan2016survey}.

In Sections \ref{sssec:ccsds} and \ref{sssec:ietf}, we summarize the protocols and standardization activities carried out by CCSDS and IETF, respectively. 
An additional contribution to the reference papers \cite{kodheli2020satellite, rinaldi2020non, ravishankar2021next, guidotti2020architectures, guidotti2019architectures, burleigh2019connectivity, li2019physical, radhakrishnan2016survey} is a description of the recommended standards, including those of the \acrfull{dtn} suite, and a description of the \acrfull{ssi} concept. In addition, unlike reference works, emphasis is placed on the working groups dealing with issues related to incorporating satellite communications.

In Section \ref{ssec:SatOps}, we discuss four largest upcoming commercial EOSCs, including Starlink, Kuiper, Lightspeed, and OneWeb. We review Blackjack and SpaceMobile EOSCs that are intended for military purposes and for broadband RF communications, respectively. We summarize upcoming CubeSat EOSCs meant for IoT applications and EOS projects on very high throughput EOSs. The upcoming EOSC initiatives from SpaceX, Amazon, OneWeb, and Telesat are mentioned by Kodheli \emph{et al}. \cite{kodheli2020satellite} but no discussion is provided. OneWeb and Starlink are briefly discussed by Burleigh \emph{et al}. \cite{burleigh2019connectivity}. In Section \ref{ssec:IandP}, we review significant near-Earth, Lunar, and Martian projects, including ESA's \acrfull{edrs}, \acrfull{scylight}, \acrfull{hydron}, and Moonlight, and NASA's LunaNet and \acrfull{marco}. We also discuss the activities under Canada’s \acrfull{osc} initiative. MarCO has been discussed by Burleigh \emph{et al}. \cite{burleigh2019connectivity} with respect to delay-tolerant networking.

In Section \ref{ssec:MobilityManagement}, we discuss the mobility management for future SpaceNets, which was briefly discussed as an open issue by Rinaldi \textit{et al.} \cite{rinaldi2020non} and the types of handover were discussed by Ravishankar \textit{et al.} \cite{ravishankar2021next}. However, in this paper, we discuss the two major categories of mobility management (i.e., handover management and location management) in detail. In this respect, after defining each type, we discuss the challenges associated with it in SpaceNets and highlight the potential solutions that were proposed in the literature to address those challenges. Moreover, we summarize the open issues associated with mobility management in SpaceNets in Section \ref{ssec:OpenIssuesMM}

Section \ref{ssec:space-terrestrial-routing} provides an overview of routing mechanisms that emerged in the Delay/Disruption Tolerant Networking paradigm and that are intended to integrate terrestrial and space networks. Ravishankar et al. \cite{ravishankar2021next} describe routing algorithms that optimize different metrics for different types of satellite constellations. However, they don't provide an in-depth survey of routing in the context of DTN, the techniques to use them for integrating terrestrial and space networks, and the related open issues.

In Section \ref{ssec:LaserISLs}, we discuss the emergence of \glspl{lisnl}, the classification of LISNLs, the capabilities of currently available \glspl{lct} for creating LISNLs, types of delays that exist in \glspl{fsosn}, and recent works that have endeavoured to address the issue of latency in FSOSNs. Some discussion on related work for optical feeder links is provided by Kodheli \emph{et al.} \cite{kodheli2020satellite}. Burleigh \emph{et al}. \cite{burleigh2019connectivity} have mentioned the use of LCTs for ISNLs with respect to systems like EDRS. Some demonstrations performed in recent years for optical uplink and downlink are also mentioned by Burleigh \emph{et al}. [10]. A related work is discussed by Li \emph{et al}. \cite{li2019physical} on the investigation of optical communication link between a GEO EOS and an \acrfull{uav}. RFISNLs are discussed in detail by Radhakrishnan \emph{et al}. \cite{radhakrishnan2016survey}.

In Section \ref{ssec:CrossLayerDesign}, we discuss the two major types of cross-layer design in SpaceNets (i.e., direct and indirect cross-layer design) and the open issues in Section \ref{ssec:OpenIssuesCrossLayerDesign}. This aspect was highlighted very briefly by Kodheli \textit{et al.} \cite{kodheli2020satellite}, Rinaldi \textit{et al.} \cite{rinaldi2020non}, Guidotti \textit{et al.} \cite{guidotti2020architectures}, and Li \textit{et al.} \cite{li2019physical}.

In Section \ref{sssec:PLS}, we highlight the importance of physical layer security for SpaceNets and summarized the existing literature. This perspective was only discussed by \cite{li2019physical}.

\begin{table*}[t]
\renewcommand{\arraystretch}{1.3}
\caption{Related works. (N: Not covered, L: Low coverage, M: Medium coverage, H: High coverage)}
\label{tab:RelatedWorkMatrix} 
\centering
\begin{tabular}{c|c|c|c|c|c|c|c|c|c|c|c|c|c|c|c|c|c}
 & \multicolumn{2}{c|}{\textbf{Section II}}  & \multicolumn{7}{c|}{\textbf{Section III}}  &  \multicolumn{3}{c|}{\textbf{Section IV}} & \multicolumn{5}{c}{\textbf{Section V}} \\
\hline
\parbox[t]{2mm}{\multirow{2}{*}{\rotatebox[origin=c]{90}{\textbf{Reference}}}} &   \parbox[t]{2mm}{\rotatebox[origin=c]{90}{Space   use cases}} & {\rotatebox[origin=c]{90}{Ground  and air use cases}}  & \parbox[t]{2mm}{\rotatebox[origin=c]{90}{EOS constellations}}  &
\parbox[t]{2mm}{\rotatebox[origin=c]{90}{Different architectures}}  &
\parbox[t]{2mm}{\rotatebox[origin=c]{90}{SN communication links}}  &
 \parbox[t]{2mm}{\rotatebox[origin=c]{90}{AI/ML }}                    & \parbox[t]{2mm}{\rotatebox[origin=c]{90}{Softwarized network management}} & \parbox[t]{2mm}{\rotatebox[origin=c]{90}{Edge computing and caching}} & \parbox[t]{2mm}{\rotatebox[origin=c]{90}{Antenna technologies}}  \parbox[t]{2mm} & \parbox[t]{2mm}{\rotatebox[origin=c]{90}{Standardization activities}}        & \parbox[t]{2mm}{\rotatebox[origin=c]{90}{Satellite   operator activities}} & \parbox[t]{2mm}{\rotatebox[origin=c]{90}{National initiatives and projects}} & \parbox[t]{2mm}{\rotatebox[origin=c]{90}{Mobility management}}    &
\parbox[t]{2mm}{\rotatebox[origin=c]{90}{ Space-terrestrial integrated routing }} &
\parbox[t]{2mm}{\rotatebox[origin=c]{90}{Laser inter-SN links}}& \parbox[t]{2mm}{\rotatebox[origin=c]{90}{Cross-layer design}} & \parbox[t]{2mm}{\rotatebox[origin=c]{90}{Physical-layer security}}    \\
\hline
{\color{black} Kodheli \textit{et al.} \cite{kodheli2020satellite}   }    &  \cellcolor[HTML]{FFCB2F}M & \cellcolor[HTML]{009901}H  & \cellcolor[HTML]{FFCB2F}M &\cellcolor[HTML]{FFCB2F}M & \cellcolor[HTML]{F56B00}L & \cellcolor[HTML]{F56B00}L &  \cellcolor[HTML]{009901}H   & \cellcolor[HTML]{F56B00}L & \cellcolor[HTML]{F56B00}L  & \cellcolor[HTML]{FFCB2F}M & \cellcolor[HTML]{F56B00}L         & \cellcolor[HTML]{CB0000}N           & \cellcolor[HTML]{CB0000}N &
\cellcolor[HTML]{FFCB2F}M & 
\cellcolor[HTML]{CB0000}N     & \cellcolor[HTML]{F56B00}L & \cellcolor[HTML]{CB0000}N  \\
\hline
{\color{black} Rinaldi \textit{et al.} \cite{rinaldi2020non}  }      &  \cellcolor[HTML]{CB0000}N & \cellcolor[HTML]{FFCB2F}M  & \cellcolor[HTML]{CB0000}N & \cellcolor[HTML]{009901}H & \cellcolor[HTML]{CB0000}N &\cellcolor[HTML]{CB0000}N & \cellcolor[HTML]{CB0000}N       & \cellcolor[HTML]{CB0000}N & \cellcolor[HTML]{CB0000}N & \cellcolor[HTML]{FFCB2F}M & \cellcolor[HTML]{CB0000}N         & \cellcolor[HTML]{CB0000}N           & \cellcolor[HTML]{FFCB2F}M &
\cellcolor[HTML]{F56B00}L &
\cellcolor[HTML]{CB0000}N     & \cellcolor[HTML]{F56B00}L & \cellcolor[HTML]{009901}H   \\
\hline 
{\color{black} Ravishankar \textit{et al.} \cite{ravishankar2021next} }   &  \cellcolor[HTML]{CB0000}N & \cellcolor[HTML]{F56B00}L  & \cellcolor[HTML]{CB0000}N &\cellcolor[HTML]{F56B00}L  & \cellcolor[HTML]{CB0000}N & \cellcolor[HTML]{CB0000}N & \cellcolor[HTML]{F56B00}L       & \cellcolor[HTML]{CB0000}N & \cellcolor[HTML]{F56B00}L & \cellcolor[HTML]{FFCB2F}M & \cellcolor[HTML]{CB0000}N         & \cellcolor[HTML]{CB0000}N           & \cellcolor[HTML]{F56B00}L & 
\cellcolor[HTML]{FFCB2F}M &
\cellcolor[HTML]{CB0000}N     & \cellcolor[HTML]{CB0000}N & \cellcolor[HTML]{CB0000}N  \\
\hline 
{\color{black} Guidotti \textit{et al.} \cite{guidotti2020architectures}  }      &  \cellcolor[HTML]{CB0000}N & \cellcolor[HTML]{FFCB2F}M  & \cellcolor[HTML]{CB0000}N &\cellcolor[HTML]{F56B00}L  & \cellcolor[HTML]{CB0000}N & \cellcolor[HTML]{CB0000}N & \cellcolor[HTML]{CB0000}N       & \cellcolor[HTML]{CB0000}N & \cellcolor[HTML]{F56B00}L & \cellcolor[HTML]{FFCB2F}M & \cellcolor[HTML]{CB0000}N         & \cellcolor[HTML]{CB0000}N           & \cellcolor[HTML]{CB0000}N & 
\cellcolor[HTML]{CB0000}N &
\cellcolor[HTML]{CB0000}N     & \cellcolor[HTML]{F56B00}L & \cellcolor[HTML]{CB0000}N  \\
\hline
{\color{black}Guidotti \textit{et al.} \cite{guidotti2019architectures}  }     &  \cellcolor[HTML]{CB0000}N  & \cellcolor[HTML]{FFCB2F}M  &  \cellcolor[HTML]{CB0000}N & \cellcolor[HTML]{F56B00}L  & \cellcolor[HTML]{CB0000}N  & \cellcolor[HTML]{CB0000}N & \cellcolor[HTML]{CB0000}N       & \cellcolor[HTML]{CB0000}N & \cellcolor[HTML]{CB0000}N & \cellcolor[HTML]{FFCB2F}M & \cellcolor[HTML]{CB0000}N         & \cellcolor[HTML]{CB0000}N           & \cellcolor[HTML]{CB0000}N & 
\cellcolor[HTML]{CB0000}N &
\cellcolor[HTML]{CB0000}N     & \cellcolor[HTML]{CB0000}N & \cellcolor[HTML]{CB0000}N  \\
\hline
{\color{black} Burleigh   \textit{et al.} \cite{burleigh2019connectivity} }     & \cellcolor[HTML]{FFCB2F}M & \cellcolor[HTML]{FFCB2F}M  &  \cellcolor[HTML]{CB0000}N & \cellcolor[HTML]{CB0000}N  & \cellcolor[HTML]{FFCB2F}M  & \cellcolor[HTML]{CB0000}N & \cellcolor[HTML]{F56B00}L       & \cellcolor[HTML]{F56B00}L & \cellcolor[HTML]{F56B00}L & \cellcolor[HTML]{F56B00}L & \cellcolor[HTML]{F56B00}L         & \cellcolor[HTML]{F56B00}L           & \cellcolor[HTML]{CB0000}N & 
\cellcolor[HTML]{F56B00}L &
\cellcolor[HTML]{F56B00}L     & \cellcolor[HTML]{CB0000}N & \cellcolor[HTML]{CB0000}N \\
\hline
{\color{black}Li   \textit{et al.} \cite{li2019physical} }        & \cellcolor[HTML]{CB0000}N & \cellcolor[HTML]{F56B00}L  &  \cellcolor[HTML]{FFCB2F}M & \cellcolor[HTML]{CB0000}N & \cellcolor[HTML]{F56B00}L  & \cellcolor[HTML]{CB0000}N & \cellcolor[HTML]{F56B00}L       & \cellcolor[HTML]{F56B00}L & \cellcolor[HTML]{CB0000}N  & \cellcolor[HTML]{F56B00}L & \cellcolor[HTML]{CB0000}N         & \cellcolor[HTML]{CB0000}N           & \cellcolor[HTML]{CB0000}N & 
\cellcolor[HTML]{CB0000}N &
\cellcolor[HTML]{CB0000}N     & \cellcolor[HTML]{F56B00}L & \cellcolor[HTML]{CB0000}N \\
\hline
{\color{black}Radhakrishnan \textit{et al.}  \cite{radhakrishnan2016survey} } &  \cellcolor[HTML]{F56B00}L & \cellcolor[HTML]{F56B00}L  &  \cellcolor[HTML]{CB0000}N &\cellcolor[HTML]{CB0000}N & \cellcolor[HTML]{FFCB2F}M & \cellcolor[HTML]{CB0000}N & \cellcolor[HTML]{CB0000}N       & \cellcolor[HTML]{CB0000}N & \cellcolor[HTML]{F56B00}L  & \cellcolor[HTML]{CB0000}N & \cellcolor[HTML]{CB0000}N         & \cellcolor[HTML]{CB0000}N           & \cellcolor[HTML]{CB0000}N & 
\cellcolor[HTML]{F56B00}L &
\cellcolor[HTML]{CB0000}N     & \cellcolor[HTML]{CB0000}N & \cellcolor[HTML]{CB0000}N  \\
\hline
{\color{black}This Paper}      &\cellcolor[HTML]{009901}H& \cellcolor[HTML]{009901}H & \cellcolor[HTML]{009901}H &\cellcolor[HTML]{009901}H& \cellcolor[HTML]{009901}H & \cellcolor[HTML]{009901}H  & \cellcolor[HTML]{009901}H & \cellcolor[HTML]{009901}H      & \cellcolor[HTML]{FFCB2F}M  & \cellcolor[HTML]{009901}H & \cellcolor[HTML]{009901}H & \cellcolor[HTML]{009901}H         & \cellcolor[HTML]{009901}H           & \cellcolor[HTML]{009901}H & 
\cellcolor[HTML]{009901}H &
\cellcolor[HTML]{009901}H     & \cellcolor[HTML]{FFCB2F}M 
\end{tabular}
\end{table*}
%

\section{Future SpaceNets: Reference Architecture and Use Cases}
\label{sec:UseCases}
%
\subsection{Reference Architecture}
\label{ssec:VisionRefArch}
Fig. \ref{fig:architecture} shows the reference architecture for our vision of future SpaceNets. As we can see, future SpaceNets will act as a network of networks. That is, it will interconnect various homogeneous networks that have common properties, such as cellular, aerial (e.g., \gls{haps} and unmanned aerial vehicle (UAV)), Internet, EOSN, near space, interplanetary, and deep-space networks. This sophisticated network will encompass different types of SNs, which include satellites (orbiting Earth or other planets), space stations, space hotels, and any other spacecraft that can be used as a network node. The interconnection of these SNs composes the SpaceNet. Therefore, each SN will be capable of directing the data to other network nodes as a router in addition to acting as an access point. That is, the SNs are equipped with advanced antenna systems and serve various kinds of networks and SNTs, including ground and air UTs (e.g., very small aperture terminals (VSATs), Internet-of-things (\acrshort{iot}) devices, handheld devices, airplanes, ships), UTs in space hotels, rovers on planets, space telescopes, planetary probes, crewed space capsules, among others. For example, a space hotel serves the SNTs within it as an access point and acts as a network node that can be used to route the data packets from a farther satellite. 

\begin{figure*}[t]
\centering
\includegraphics[scale=0.5]{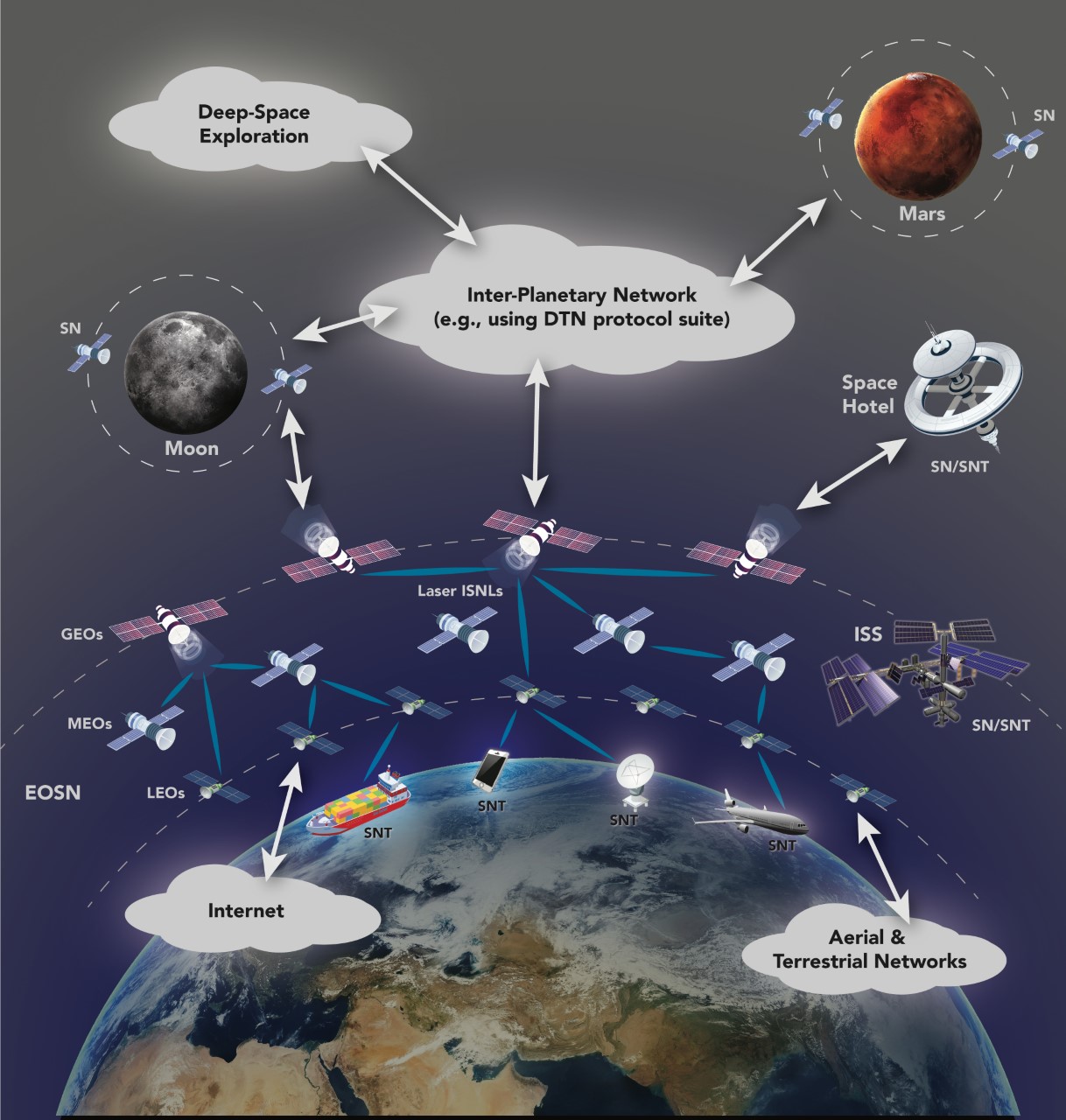}
\caption{Reference architecture for future SpaceNets. EOSN: Earth Orbit Satellite Network, SN: Space Node, SNT: Space Network Terminal, ISNL: Inter SN Link, DTN: Delay Tolerant Network, ISS: International Space Station.}
\label{fig:architecture}
\end{figure*}

This network of networks will provide many services to networks and terminals spanning a plethora of terrestrial, aerial, and space applications. In so doing, future SpaceNets will deploy advanced technologies to handle these services. First, the majority of the SNs will be interconnected via high-speed, low-latency, and high-reliability inter-SN links (ISNLs), utilizing technologies such as \gls{fso} communications. This will enable the reliable exchange of vast amounts of data among these SNs.  In addition, SpaceNets will be equipped with advanced and enormous computing resources that will be deployed on the SNs and ground stations to be able to process huge amounts of data. Moreover, future SpaceNets will be autonomous and require minimum intervention from humans. Therefore, sophisticated AI algorithms will be deployed in almost all levels of network management and operation. Furthermore, due to the lifetime of SNs and the high cost of deployment, software-based configuration, operation, networking, and management will be prevalent.

The most important and capable part of this SpaceNet is the EOSN, which acts as the edge of SpaceNets that provides services for Earth residents and objects on the ground and in the air. Besides, EOSNs enable us to interface with other SNs (and the SNTs connected to those SNs) in space in an efficient and secure manner. Towards that end, EOSNs should have advanced features. These networks should have an ultra-dense deployment of different kinds of EOSs, at different orbits (i.e., \gls{leo}, \gls{meo}, high Earth orbit (HEO), and \gls{geo}), with various capabilities (i.e., from low-cost CubeSats to complicated big GEO satellites), with advanced antenna systems directed towards the Earth and space, and operated by several entities. Table \ref{tab:EnvisionedEOSN} summarizes the main features of the envisioned EOSN in future SpaceNets.


\begin{table*}[t]
\renewcommand{\arraystretch}{1.1}
\centering
\caption{Features of the EOSN part of the Envisioned SpaceNet}
\label{tab:EnvisionedEOSN}
\begin{tabular}{| p{3cm} | p{6cm} | p{6cm} |} 
\hline
\textbf{Aspect of Technology} & \textbf{Conventional EOSN}                                                                                                                                                                                        & \textbf{Envisioned EOSN}                                                                            \\ 
\hline
EOSs type                            & Mainly GEO satellites                                                                                                                                                                                                     & Mainly LEO satellites, such as SmallSats, CubeSats, and NanoSats  \\ 
\hline
Inter-EOS connectivity    & No radio frequency (RF)- or FSO-based inter-EOS connectivity between GEO satellites; FSO-based communication between European Data. Relay System GEO satellites and client LEO satellites & RF- and FSO-based communication between SmallSats                 \\ 
\hline
Inter-EOS link data rate  & Up to 1.8 Gbps for FSO links between European Data. Relay System GEO satellites and client LEO satellites                                                                         & Up to 10 Gbps for FSO links between SmallSats                                                               \\ 
\hline
EOS Altitude                        & $\sim$ 36,000 km for GEO satellites                                                                                                                                                                            & Less than 2,000 km for LEO satellites                                                                       \\ 
\hline
Path loss                       & Very high                                                                                                                                                                                                                 & Low                                                                                                         \\ 
\hline
Latency                         & Very high                                                                                                                                                                                                                 & Low                                                                                                         \\ 
\hline
EOS Size                            & Very large                                                                                                                                                                                                                & Small                                                                                                       \\ 
\hline
Cost per EOS              & Very high                                                                                                                                                                                                                 & Low                                                                                                         \\ 
\hline
EOS deployment cost                 & Very high                                                                                                                                                                                                                 & Low                                                                                                         \\
\hline
Computing capabilities                 & Low                                                                                                                                                                                                                 & High                                                                                                         \\
\hline
Antennas direction                 & Towards Earth only & Towards Earth and space \\
\hline
\end{tabular}
\end{table*}

%
\subsection{Space Use Cases}
This section describes the most interesting space use-cases of satellite communications.
\subsubsection{Deep Space
Exploration} 
Deep space exploration started in $1977$ when two twin spacecrafts, Voyager $1$ and $2$ were launched. In $2012$, Voyager $1$ reached the outer edge of the solar system and entered into an interstellar space, whereas Voyager $2$ entered the interstellar region in $2018$. Both spacecraft are still active and sending information from outer space \cite{bryn2013science}. After Voyager $1$ and $2$, the deep space networks (DSNs) have relied on the NASA's giant sites that were built in Goldstone (California), Madrid (Spain), and Canberra (Australia). Each are separated by approximately 120 degrees to provide $24$/$7$ coverage. Furthermore, current DSN technology focuses primarily on scientific exploration and the collection of scientific data that is transmitted back to Earth. In this context, DSNs mainly consist of landers, probes, orbiters, and rovers \cite{deColaDeepSpace}. By contrast, next-generation space exploration will involve manned deep space exploration, lunar mining, solar system Internet, space tourism, Mars-Earth service links, and exploration of the outer parts of our galaxy. The new vision of space activities can be expressed by the term `Space 4.0', in which many different organizations aspire to establish a sustained human presence elsewhere in our solar system. This goal can be accomplished by setting reliable, ubiquitous, and enhanced capacity communication links and outposts between Earth and other planets in our system, and in next-generation DSN technology, satellites will play a vital role.  


\subsubsection{Interplanetary Network Consisting of Small SNs} 
An interplanetary network (IPN) can be established by deploying hundreds and thousands of small satellites equipped with propulsion, telecommunication systems, and sensor payloads in the solar system \cite{VelazcoIPN}. In this vision, an interplanetary superhighway (Fig. \ref{fig:IPSH}) could be used to deploy small satellites. An interplanetary superhighway refers to a network of gravitationally determined paths through the solar system that a spaceship could follow with a relatively minimal energy expenditure. In this envisioned IPN model, the interplanetary superhighway could be used to carry a number of small satellites to the moon, Mars, Jupiter, and other planets where gravitational forces between the planet and the sun provide centripetal force for spacecraft to orbit. CubeSats which have great potential in transmission with low cost can provide excellent scientific results.  In the IPN model given above, CubeSats can be deployed with an optical communicator, attitude determination system, and solar panels so that they can operate autonomously. The application areas of the envisioned IPN model can be summarized in the four following points: 
\begin{itemize}
    \item High-speed communication can be established between Earth and outer planets.
    \item A massive number of CubeSats deployed in lunar orbit can detect exoplanets where life can exist.
    \item Charged particles streaming off the sun can be detected, and analyzed.
    \item Delay-free communication infrastructure can be organized for the mining purposes in space \cite{LunarMining}.
\end{itemize}

\subsubsection{ Solar System Internet}

The requirement for robust SpaceNets will become unavoidable as human space travel increases. 
Over the next 100 years, the Interplanetary Networking Special Interest Group (IPNSIG) ambition is to extend networking to space, moving away from a point-to-point and bent-pipe communication design toward a store-and-forward system in which multiple nodes and networks communicate in an automated way. As described in \cite{ccsds2014ssi}, this concept is referred to as the \acrfull{ssi} and is conceived as a network of regional internets that implement a generalized suite of protocols in order to achieve end-to-end communication through multiple heterogeneous regions in a variable-delay and frequently disrupted environment. This suite of protocols is known as \acrfull{dtn} and, as explained in Section \ref{sec:Projects}, its standardization is being carried out by different bodies, such as the \acrfull{ietf} and the \acrfull{ccsds}. In addition to technical specifications, the IPNSIG Strategy Working Group (SWG) has issued a roadmap \cite{ssi_estrategy} that
proposes short-, mid-, and long-term scenarios for the evolution of the SSI, as well as key properties required for its connectivity architecture to be decentralized, sustainable, open, and neutral. Basically, the SWG's vision is that the SSI should evolve in the same way that the terrestrial Internet has evolved. That is, from a government-funded network to a commercial network funded by different stakeholders.





\subsubsection{Other Space Use Cases}

\paragraph{Asteroid mining} Asteroid mining can be defined as the extraction of resources from asteroids and other small planets, including near-Earth objects. So far, two asteroid sample-return missions—Hayabusa and Hayabusa2—have been completed by the Japanese aerospace exploration agency (JAXA). With the advance of `Space 4.0', asteroid mining will gain industry attention again, and SpaceNets can be used to provide ubiquitous connectivity for spacecrafts in asteroid mining facilities. 
    
\paragraph{Space farming} Space farming refers to the growth of crops in space or on other planets. Space farming is expected to gain importance as food delivery to space stations and other long-duration missions will likely be cumbersome and expensive. It will pick up with the advance of next generation space satellites when humans begin spending more time in space.
    
\paragraph{Space tourism} Human space flight for touristic purposes is known as space tourism. Orbital, suborbital, and lunar space tourism are examples of several examples of space tourism. In $2007$, space tourism was believed to be one of the first commercial spaceflight businesses to develop. In space tourism, uninterrupted communication can be vital; this can be established with the help of EOSN.

\paragraph{Moon village} The Moon Village Association (MVA) is a new non-governmental organization located in Vienna. Its mission is to establish a permanent informal form for the development of the Moon Village by interested parties, including governments, industry, academia, and the general public. The MVA encourages public and commercial partnerships for existing or prospective global lunar exploration projects.

\begin{figure*}[htbp]
	\centerline{\includegraphics[scale=0.75]{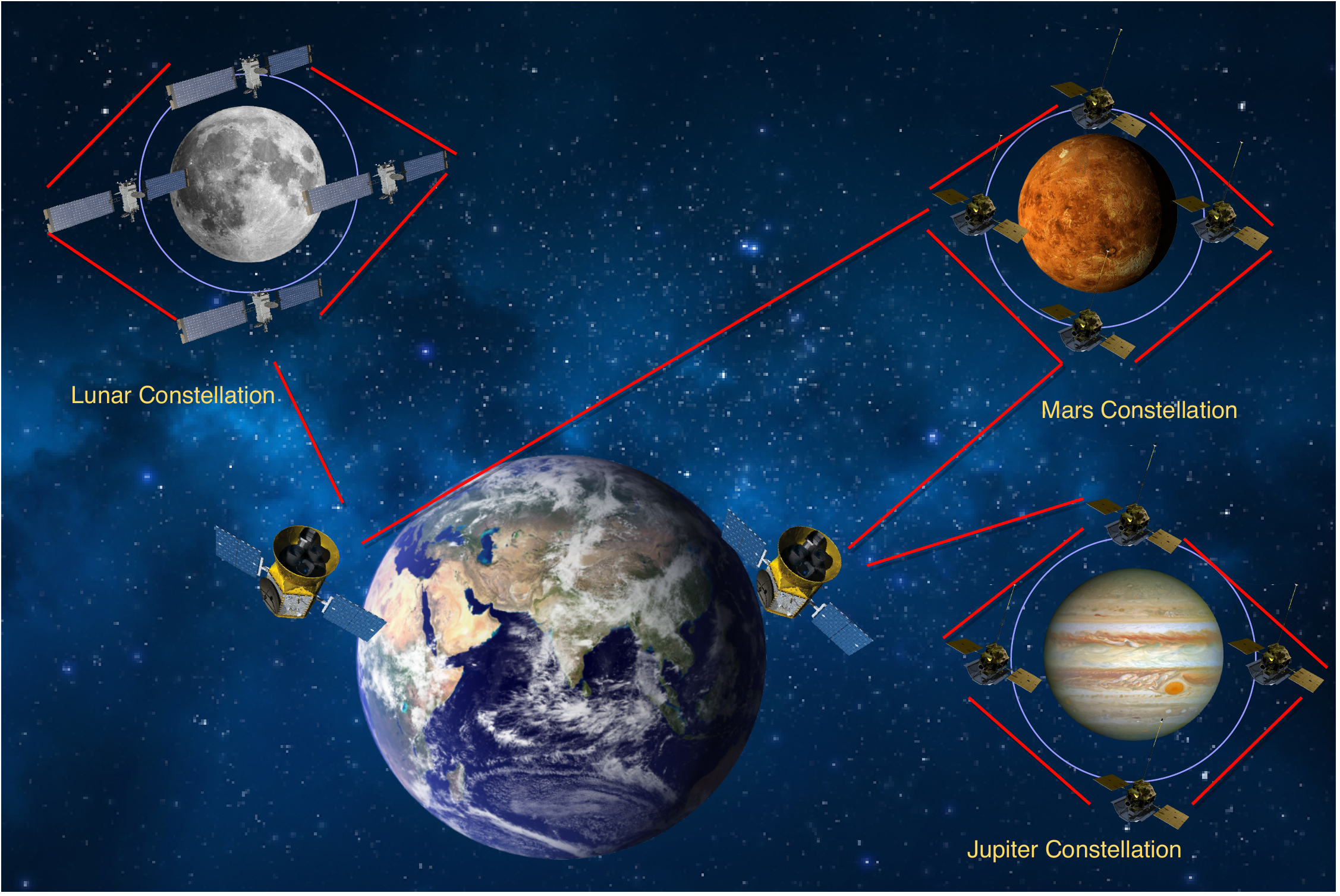}}
	\caption{Inter-planetary network consisting of satellites adapted from \cite{VelazcoIPN}.}
	\label{fig:IPSH}
\end{figure*}

}

%
\subsection{Ground and Air Use Cases}
According to the Internet World Stats \cite{InternetStats}, the Internet penetration rate in July 2022 is estimated to be $69\%$ of the world population. That is, more than 2.4 billion people are still not connected. Therefore, the main driver of several organizations is to provide connectivity for remote and rural areas (i.e., connecting the unconnected). However, a plethora of other applications and use cases can be provided by the EOSN utilizing the increasing investments in this domain in a more efficient manner. This can be achieved by exploiting recent technologies in fields such as networking, computing, caching, edge computing, sensing, and AI. These developments can open the door for many other applications that can be developed and supported by such ubiquitous networks. In what follows, we discuss use cases of SpaceNets on the ground and air. In doing so, we use the term satellite to refer to an EOS.

\subsubsection{Integrated Satellite-Aerial Communications}
The number of airplanes and UAVs is growing day by day as they can be used for flying base stations that can provide perfect line-of-sight connectivity to ground users. UAVs can maneuver and adjust their altitudes in harsh weather conditions. Their use cases extend from traffic monitoring, search and rescue missions, and remote sensing to covering large political events, music events or anywhere else that ubiquitous connectivity is needed \cite{erdogan2020cognitive}. Recent studies have shown that semi-autonomous drones will be used in aerial cargo delivery systems in densely populated metropolitan areas \cite{KarabulutDronDelivery}. In addition, the integration of UAVs with SpaceNets has attracted considerable interest in the literature more recently. More specifically, the authors of \cite{vondra2018integration} investigated the integration of satellite and UAV communications for heterogeneous flying vehicles, particularly for areas like seas or oceans where ground station coverage can not be established. The authors also analyzed the limits of satellite connectivity in terms of maximum capacity and maximum range, which showed that the capacity depends on the number of satellites that are available for UAV integration. Like \cite{vondra2018integration}, the authors in \cite{li2020enabling} noted the difficulty of deploying communications infrastructure that would cover oceans, as the $71\%$  of the Earth's surface is water-covered. As a solution for maritime ubiquitous communication networks, \cite{li2020enabling} proposed to integrate UAVs with satellites and shore-based terrestrial base stations. The paper showed that the characteristics of vessels and the UAV agility can meet on-demand coverage for maritime communication. Along similar lines, \cite{li2018investigation} investigated the UAV-satellite optical communication systems, which involved constructing a theoretical model of a UAV-to-satellite optical communication and analyzing the Doppler effect, pointing error, and the atmospheric turbulence induced fading on the communication performance. In a somewhat different direction, \cite{jia2020leo} and \cite{ma2021uav} proposed the UAV-EOSN integration to collect and process Internet-of-Remote-Things (IoRT) data. More precisely, in \cite{jia2020leo}, a UAV collected the data from the IoRT sensors, and an LEO satellite was used to provide lower transmission delay for delay sensitive IoRT data. In \cite{ma2021uav}, the authors investigated a resource allocation problem for a two-hop uplink UAV-LEO integrated data collection for IoRT networks, where numerous UAVs gathered data from Internet-of-Things (IoT) devices and transmit the IoT data to LEO satellites. In \cite{nguyen2021uav}, authors used an UAV to forward signals from a satellite to ground users, where hardware impairments were studied on the performance of integrated UAV-LEO satellites network.

Similar to the UAV-EOS integrated architectures, HAPSs can be integrated with SpaceNets \cite{erdoganCooperation}. A HAPS system, which is defined by the International Telecommunications Union (ITU) as an aerial node that is inherently quasi-stationary, can stay in the stratosphere at an altitude of about $20$ to $50$ km. Compared to UAVs, HAPS systems have larger footprints, more computational power, and they provide better line-of-sight links \cite{kurt2021vision}. Therefore, HAPS-assisted satellites can support terrestrial networks, provide enhanced connectivity, rapidly solve capacity improvement problems, and they introduce agility to address problems of high and variable traffic demands. In this regard, the integration of HAPS systems with EOSNs has attracted considerable interest in the literature, particularly in \cite{7341013,6941426,9174937,5010541,1391449,9446153,9129403,9177315} and the references therein. In \cite{7341013}, the authors considered an integrated HAPS-satellite system model, where energy consumption was modeled, and the adaptive algorithm for the energy-efficient path was obtained for energy-efficient transmission. Likewise, in \cite{6941426}, the authors considered an LEO satellite with a HAPS system, where both harvested solar energy to power their operations. In this concept, a new algorithm was proposed to minimize energy usage while delivering a given amount of information from the satellite. \cite{9174937} focused on the cooperation of LEO satellites and HAPS systems for massive access and data backhaul for remote area users. The authors proposed an algorithm to guarantee the revenue of satellites as well as the number of served users with faster execution time. In \cite{5010541}, the authors established optical links between HAPS systems and mobile satellites with a limited number of optical transmitters. They found the optimum solution for matching HAPS systems and satellites so that the utilization of HAPS systems was maximized. \cite{1391449} studied resource allocation and traffic management issues for the integration of  HAPS systems with EOSs, whereas \cite{9446153,9129403,9177315} obtained outage and error probabilities for various EOS-HAPS system architectures.

\subsubsection{Multicast Grouping over Satellite Communications for Enhanced Coverage}
A multicast group can be defined as a group of hosts that intend to receive packets at a specific multicast group address according to the international telecommunication union (ITU) recommendations \cite{ ITUMulticast}. In this way, bandwidth intensive real-time data can be delivered more economically and efficiently than traditional unicast transmission. In multicast communication, only one data stream is delivered simultaneously to multiple users, and it is replicated as many times as necessary so that the registered users can access that information. Due to its aforementioned capabilities, multicast transmission or multicast grouping can be an important enabler in SpaceNets for enhanced broadcast and coverage \cite{zhu2018cooperative}. In this regard, \cite{MulticastShi} proposed a new multicast algorithm for SpaceNets to decrease delay and minimize energy consumption, whereas \cite{MulticastAraniti} proposed using multicast resource allocation schemes and application layer
joint coding to boost the performance of video streaming in satellite systems. Similarly, in \cite{MulticastAraniti2}, the authors studied a low-computational multicast algorithm to optimize the data rates and to maximize the capacity for satellite systems, and \cite{YinMulticast} focused on the rate-splitting multiple access problems to achieve max-min fairness among multicast groups, where the proposed strategy managed to decrease inter-beam interference for multibeam satellites. In \cite{zhu2018cooperative}, the authors investigated a downlink multigroup multicast transmission where base stations and an LEO satellite work in a cooperative manner to provide multicast services for the terrestrial groups while reusing the entire bandwidth. In \cite{9806158}, the authors integrated a HAPS system and an LEO satellite for multicast communication for a group of users and obtained the outage probability for two practical HAPS-LEO satellite integrated schemes.  \cite{MulticastYuzhou} proposed a satellite-maritime network model where satellites used multicast transmission to provide ubiquitous broadband coverage for the sea area. In their proposed model, the satellites used multicast services for groups of users and a relay re-transmitted the signal to the other users. 

\subsubsection{Satellite-Supported Global Maritime Coverage}
The global maritime industry is expected to grow at a compound annual growth rate of around 6\% between $2021$ and $2026$. The recent evolution in wireless technologies, including $5$G, and space communications can help this development as the most important challenges in the maritime industry lie in the field of surveillance and seamless wide-area maritime information coverage \cite{SoldiMaritime1}. Until now, maritime surveillance, which is used in search and rescue missions, sea traffic, fishery, migration monitoring, and security issues, employs MF/HF/VHF-band communications and satellite systems \cite{MagicNet}. In maritime communications, an automatic identification system (AIS) is used to identify ships regardless of their size, and a navigational telex (NAVTEX) system provides navigation warnings, weather information, and other data to the vessels \cite{NAVTEX}. Furthermore, the most recent international maritime satellite communication system (INMARSAT) consists of four geosynchronous Earth orbit (GEO) satellites to provide high-speed communications and global information coverage \cite{INMARSAT2}. Even though there is a wide range of appealing technologies for maritime surveillance, terrestrial base stations deployed on the shores have limited coverage, whereas satellite networks can provide high coverage with limited capacity. For example, INMARSAT can provide $50$ Mbit/s data rates and wider coverage with the aid of Ka-band satellite communications; however, users face the problem of high cost and latency \cite{INMARSAT}. Moreover, GEO satellite enabled maritime communications can be affected by weather conditions due to long distances and higher frequencies. By taking advantage of both networks, an integrated satellite-terrestrial network can be a promising solution for maritime communications and surveillance. Along these lines, \cite{Borelli2009} proposed an intelligent middleware for high-speed maritime mesh networks with satellite communications, whereas \cite{Kessab2016} provided statistical tools to estimate the number of gateway stations in a cluster and obtain the propagation delays considering GEO, medium Earth orbit (MEO) and LEO satellite systems. In addition, in \cite{Dhivvya2017}, authors implemented an OceanNet backhaul Link selection (OBLS) algorithm in a test-bed and they found that the proposed algorithm can improve the packet delivery ratio up to $ 70\%$. In \cite{LiMaritime20}, the authors considered enhancing coverage by coordinating UAVs with satellites and terrestrial users. Similarly,  \cite{Xia2019} utilized an electromagnetic power flux density mask to decrease the self-interference in maritime IoT systems.

\subsubsection{ Satellite IoT Systems} 
The IoT can be described as a network of interrelated devices, machines, sensors, and other things that exchange data over the Internet \cite{CentenaroSatelliteIOT}. Today, there are $35$ billion connected IoT devices, and this number is expected to be $83$ billion by $2024$ \cite{SatIoT}. 
The IoT has great potential to bring new opportunities to current wireless networks by integrating and exchanging information from different devices without the need for human control. In an attempt to expand the benefits of IoT systems in rural areas where it would otherwise not be feasible due to economic or technical reasons, integrated satellite-IoT systems have been proposed for global coverage solutions \cite{CentenaroSatelliteIOT}. In these remote areas, IoT systems can be used to collect and process data from many different industries including logistics, transportation, farming, maritime, and many others. For this reason, satellite-IoT systems have become a popular field both in industry and academia \cite{kodheli2020satellite}. Related to this, two paradigms have been developed: the IoRT and the Internet of everything everywhere (IoEE). In IoRT, satellite communication provides a cost-effective solution for the interconnection of sensors and actuators in IoT systems \cite{SanctisSatIoT}. IoEE, by contrast, tackles the problem of connecting terrestrial IoT network segments to IoT systems via satellite broadband. In IoEE, \cite{PalattellaIoT} studied the market opportunities by using satellite broadband to connect IoT network segments to the satellite network to provide cost-effective solutions. In terms of IoRT, \cite{SanctisSatIoT} discussed MAC protocols for satellite routed sensor networks, efficient
IPv6 support, heterogeneous network inter-operability, and quality-of-service (\acrshort{qos}) issues by considering integrated satellite IoT systems. Furthermore, \cite{GavrilaSatIoRT} proposed a software defined radio based satellite gateway for IoRT systems to decrease costs and increase flexibility.

\subsubsection{EOSN Access Use Cases Based on 3GPP Documents \cite{3gpp38811,3gpp22822}}
\label{sec:UseCases:A:5}
As satellite networks are going to be an integral part of 5G and beyond communication networks, several EOSN access use cases were proposed in 3GPP documents \cite{3gpp38811,3gpp22822}. In \cite{darwish2022leo}, a detailed discussion of the 3GPP use cases was provided. The following is a summary of 3GPP EOSN access use cases:
\begin{itemize}
    \item To support roaming between terrestrial and EOSNs. 
    \item Broadcast and multicast with an EOS overlay. 
    \item Edge network content delivery.
\item Broadcast to mobile UE directly.
\item Support public safety by providing continuity of service for  emergency responders. 
\item Supporting Internet of Things (IoT) with an EOSN. 

\item Temporary use of an EOS component during disasters.
\item Supports network resilience to prevent complete network connection outage on critical network links that require high availability. 
\item Trunking to interconnect various 5G local access network islands. 
\item Optimal routing or steering over an EOSN (multi connectivity). 
\item EOS trans-border service continuity. 
\item Global EOS overlay to reduce delays on long distance communications. 
\item Indirect connection through a 5G EOS access network. 
\item 5G fixed backhaul between NR and the 5G core. 
\item 5G moving platform backhauling. 
\item 5G to premises for customers in unfavourable geographical areas with old terrestrial network infrastructure. 
\item To support off-shore wind farms. 
\end{itemize}
%
%
\section{Architectures and Technologies to Enable Future SpaceNets}
\label{sec:Enablers}
Having discussed the reference architecture and use cases of future SpaceNets, in this section we focus on the different architectures and technologies that will enable the envisioned network. We start by discussing the general parameters of an \acrfull{eosc} and the general types of EOSCs in Section \ref{ssec:basics}. Then, in Section \ref{ssec:DifArch}, we describe the main components of the different architectures envisaged for future SpaceNets. In Section \ref{ssec:CommunicationLinks}, the SN communication links are discussed, including uplink/downlink and inter-SN links. The use of artificial intelligence (AI) as an enabler technology is presented in Section \ref{ssec:AI_ML} and the softwarized network management techniques (e.g., software defined networks, network function virtualization, and network slicing) are discussed in Section \ref{ssec:NetworkManagement}. Finally, we investigate computing and caching services and antenna technologies in Sections \ref{ssec:cachingMEC} and \ref{ssec:AntennaTechnologies}, respectively.
%
\subsection{Earth Orbit Satellite Constellations for Future SpaceNets}
\label{ssec:basics}

\par In this subsection, we list and explain the general parameters of an \acrshort{eosc} and general types of EOSCs for future SpaceNets. We also review different types of EOSCs based on altitude and discuss pros and cons of \acrfull{vleo} and \acrfull{leo} EOSCs.

\subsubsection{Parameters of an EOSC}
\label{sssec:parameters}

\par EOSCs are generally designed so that \glspl{eos} have similar orbits, altitude, and inclination to maintain the EOSC without excessive station keeping, to ensure an equal amount of fuel usage at EOSs, and to achieve similar EOS lifetime. Circular orbits are generally used for EOSs in an EOSC since EOSs in such orbits have a constant altitude resulting in constant ground coverage, constant path loss, and constant signal strength to communicate with users on Earth \cite{Ma2013design}.

\par A uniform EOSC with EOSs in circular orbits is generally defined by the following four main parameters:

\begin{itemize}
	\item Inclination in degrees (or $i$);
	\item Altitude in km (or $h$);
	\item Total number of EOSs in the EOSC (or $T$); and
	\item Total number of \glspl{op} in the EOSC (or $P$).
\end{itemize}

\par The cost of an EOSC and its coverage area on the surface of the Earth depend on the number of EOSs in the EOSC. The number of orbital planes and the number of EOSs per orbital plane are the two factors that determine the EOSC’s coverage. The number of EOSs in each orbital plane is equal in a uniform EOSC and can be calculated using $T$/$P$. Furthermore, in a uniform EOSC, the spacing between the orbital planes is the same and is calculated using 360º/$P$, while the spacing between EOSs within an orbital plane is also the same and is calculated using 360º/($T$/$P$) \cite{chaudhry2020fso}.

\par
\begin{figure*}[htbp]
	\centerline{\includegraphics[scale=0.72]{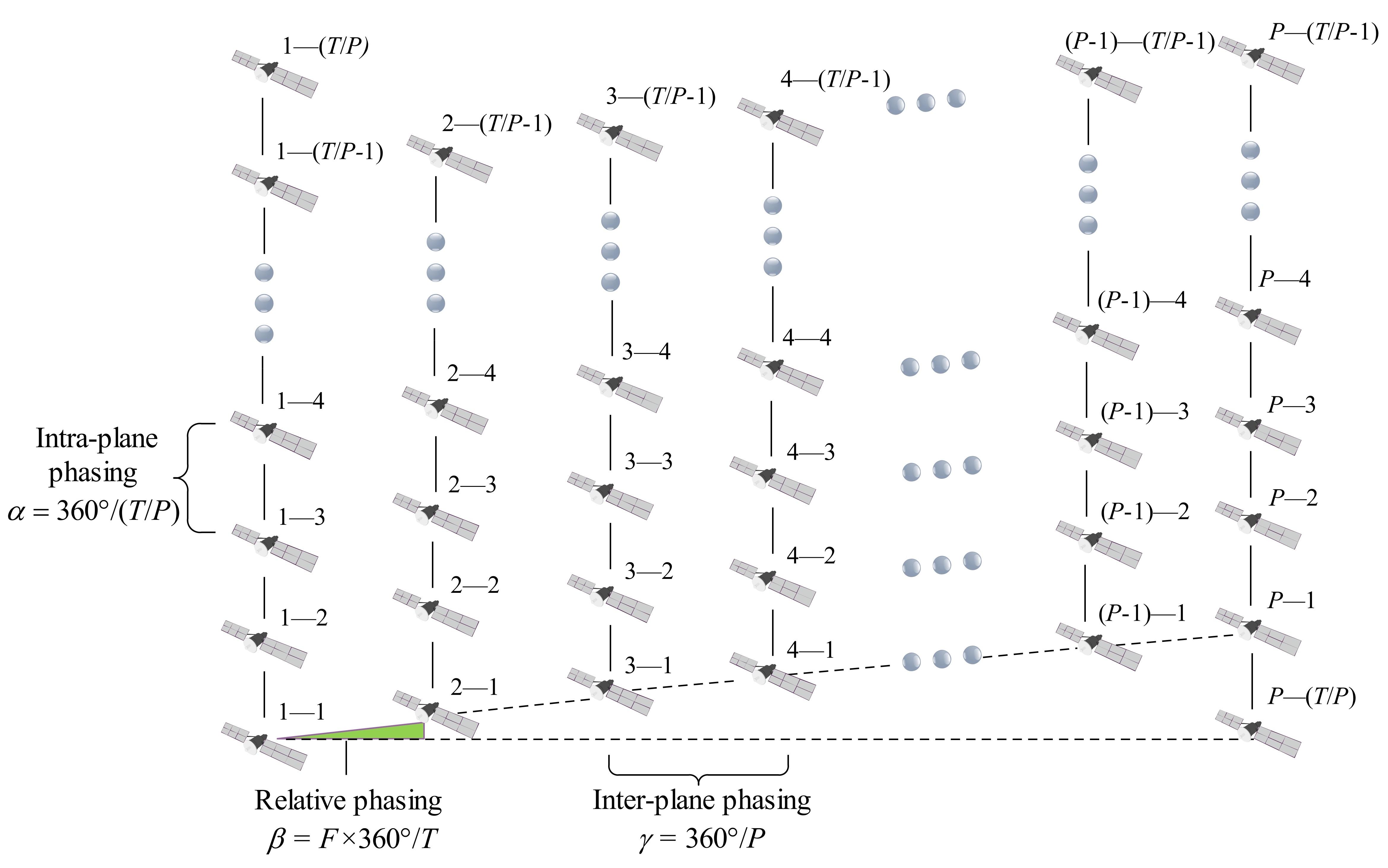}}
	\caption{Different parameters of an EOSC adapted from \cite{liang2021phasing}.}
	\label{fig:paramSatCon}
\end{figure*}

 Besides the four main parameters mentioned above, the phasing parameter (usually denoted by $F$) is also considered to be an important parameter, and its possible values range from 0 to $P$-1. This parameter is used to calculate the relative phasing $\beta$ between EOSs in adjacent orbital planes and affects the relative position of EOSs within the EOSC. While designing an EOSC to ensure intra-EOSC EOS collision avoidance, a value of $F$ needs to be chosen from among all possible values of $F$ that provides the maximum value of the minimum distance between EOSs. For complex EOSCs consisting of hundreds of EOSs, like the upcoming Starlink, Lightspeed, and Kuiper EOSCs, careful analysis and selection of $F$ are needed to avoid collisions between EOSs within the EOSC to ensure their safety, as EOSs are the most vital part of an EOSC\cite{liang2021phasing}. The different parameters of an EOSC, including intra-plane phasing $\alpha$ (i.e., angular spacing between EOSs in the same orbital plane), relative phasing $\beta$ (i.e., the relative phase difference between EOSs in adjacent orbital planes), and inter-plane phasing $\gamma$ (i.e., the angular spacing between adjacent orbital planes), are shown in Fig. \ref{fig:paramSatCon}. The number of EOSs in an orbital plane ranges from 1 to $T$/$P$; for example, the EOS 3---2 in this figure represents the second EOS on the third orbital plane. 

\subsubsection{Types of EOSCs}
\label{sssec:types}

\par A Walker EOSC pattern is generally used for EOSCs that are uniform. The EOSs in such EOSCs have the same inclination, altitude, and zero eccentricity, which means that the EOSs have circular orbits. Based on the inclination of EOSs in a Walker-type EOSC, there are generally two types of Walker EOSCs \cite{walker1971orbit},\cite{walker1984constellations}: Walker Star and Walker Delta.

\par In Walker Star EOSCs, the inclination of EOSs is around 90º with respect to the Equator, and such EOSCs are also known as polar or near-polar EOSCs. In these EOSCs, a ground user at the Equator will see overhead EOSs at regular intervals that are either orbiting from north to south or from south to north. Half of the EOSs on one side of the globe are moving from south to north in ascending orbital planes while the other half on the other side of the globe are moving in descending orbital planes from north to south. The “Star” designation is due to the fact that if a Walker Star EOSC is viewed from space, from above a Polar region, its orbital planes cross at Polar regions forming a star pattern \cite{wood2001constellations}. Due to the nature of their EOSC pattern, Walker Star EOSCs generally provide global coverage or coverage for the entire surface of the Earth including the Poles. 

\par The inclination of EOSs in Walker Delta EOSCs is generally less than 90º with respect to the Equator, and they are also referred to as inclined EOSCs. Unlike Walker Star EOSCs, where the EOSs in ascending and descending orbital planes remain separate, Walker Delta EOSCs have EOSs that cross in ascending and descending orbital planes, which means their coverage continuously overlaps. The “Delta” name comes from the fact that for such an EOSC with three orbital planes, a triangle pattern is formed when viewed from space above a Polar region. Due to the inclined nature of these EOSCs, they generally provide coverage below certain latitudes and do not cover the Poles. 

\par To describe an EOSC, Walker used the following notation: $i$:$T$/$P$/$F$ where $i$ is the inclination, $T$ is the total number of EOSs in the EOSC, $P$ is the number of orbital planes in the EOSC, and $F$ is the phasing parameter. The Iridium NEXT EOSC \cite{iridium} is an example of a Walker Star EOSC that is currently in operation. Its Walker notation is 86.4º:66/6/2, its altitude (or $h$) is 780 km, and it has 11 EOSs in each orbital plane. Similarly, the EOSC for the latest version of Starlink’s Phase I \cite{SpaceX:4} that is being deployed by SpaceX has a Walker Delta pattern, a notation of 53º:1584/22/$F$, an altitude (or $h$) of 550 km, and 72 EOSs in each orbital plane. We should note that the value of the phasing parameter $F$ for this EOSC is not known from SpaceX’s \acrfull{fcc} filings.

\subsubsection{Why VLEO/LEO EOSCs?}
\label{sssec:vleo/leo}

\par EOSCs can be categorized into three types depending on the altitude of their EOSs:

\begin{itemize}
	\item EOSCs in \acrfull{geo}
	\item EOSCs in \acrfull{meo}
	\item EOSCs in \acrfull{leo}
\end{itemize}

\par The altitude of EOSs in LEO and MEO EOSCs is generally below 2,000 km and 20,000 km, respectively. The altitude of EOSs in GEO EOSCs, by contrast, is around 36,000 km \cite{hiriart2009geomeoleo}. LEO EOSCs at altitudes below 500 km are referred to as very low Earth orbit (or VLEO) EOSCs. A 3-EOS GEO EOSC, a 12 to 18 EOS MEO EOSC, and a 40 to 70 EOS LEO EOSC can cover the globe \cite{pelton2006basics}, while a higher number of EOSs is required for global coverage in a VLEO EOSC. In short, the lower the altitude of the EOSC, the more the number of EOSs required to provide minimum global coverage.

\par EOSs in GEO EOSCs have the largest coverage area, while EOSs in VLEO/LEO EOSCs have the smallest. On the one hand, this is a disadvantage, as many more VLEO/LEO EOSs are needed in a VLEO/LEO EOSC to provide the desired coverage. On the other hand, their smaller coverage area results in a much higher frequency reuse of the scarcely available frequency spectrum and thereby higher system capacity. Due to their low altitude, EOSs in VLEO/LEO EOSCs have the least latency, path loss, power requirements, size, and the lowest manufacturing and deployment costs.

\par The lifespan of an EOS at VLEO/LEO altitudes depends to an extent on the amount of fuel it can carry to counter the orbital decay due to atmospheric drag. The low altitude of EOSs in VLEO/LEO EOSCs results in these EOSs facing the highest atmospheric drag and the highest orbital decay. This results in these EOSs frequently burning fuel to stay in their desired orbit, and they have the shortest lifespan compared to EOSs in MEO or GEO EOSCs. 

%
\subsection{Different Architectures for Future SpaceNets}
\label{ssec:DifArch} 


Due to the projected increase in population and connected Internet-of-Things (IoT) devices, it is expected that the next decade will necessitate further technological developments. This will include enhanced latency on the order of hundreds of microseconds, higher data rates on the order of terabits per second, higher frequencies such as mmWave bands, better capacity, massive connectivity, more multimedia services and applications, smart traffic, and immersive reality. In addition, over the next few years, we expect to see a continuation and development of recent space-related trends, including Space Big Data by collecting Earth data from space, Space Internet by providing the internet from space to Earth and space, and space travel to the moon, Mars, or other planets \cite{2020WhiteP5}. 
Accordingly, it will be necessary for existing architectures, multiple access techniques, and technologies to meet these extensive requirements \cite{akhtar2020shift,huang2019novel}. To address issues related to (Internet/network/other) availability and reliability, and to provide high capacity connectivity, a wide range of organizations, including 3GPP and various space companies, are stepping in to help deliver next-generation network services.

Thus, standardization efforts are currently underway to develop the architecture for next-generation networks.
In recent years, we have seen a steady increase in the exploitation of space resources to meet demands for ubiquitous coverage and massive global connectivity. Therefore, the seamless integration of \glspl{ntn} such as SNs, drones, and HAPSs, with terrestrial networks is the envisioned architecture for future wireless communications. To bring this vision about, future architecture will be layered and three-dimensional, involving terrestrial, air, and space networks. These networks will include different types of SN constellations connected through ISNLs with high data rates.
\begin{itemize}
      \item \textbf{Terrestrial network layer:} This layer includes fixed and mobile users and various radio access technologies (RATs) of current networks, including 5G, 4G, 3G, and Wifi. To meet the requirements of Tb/s rate services, such as hologram and full sense digital reality, it is expected that the THz band will be used. Therefore, microwave, mmWave, and THz bands are expected to be deployed by terrestrial networks. However, mmWave and THz bands are subject to very high path loss. Thus, a number of small base stations may be needed to form ultra-dense heterogeneous networks. In \cite{giordani2020satellite}, the authors emphasized the feasibility of employing mmWave frequencies to provide high-capacity SN communications, given the enormous traffic demands and the continuity of service requirements for future applications.
       \item \textbf{Aerial network layer:} This layer includes manned and unmanned flying platforms, such as HAPSs, drones, unmanned aerial vehicles (UAVs), and balloons in order to deliver more reliable and flexible connectivity for temporary or urgent events or in remote areas by using flying base stations. These platforms can provide similar services to a terrestrial base station. Their low-altitude operation, easy deployment, and low cost make these platforms desirable complements for terrestrial networks with their capacity to extend communication speed and efficiency in an agile manner.
        \item \textbf{Space network layer:} This layer consists of SNs comprising GEO and non-GEO constellations, which are expected to enable the delivery of seamless services across the globe using multicast and broadcast streams. Additionally, SpaceNet can provide on-demand connectivity in remote areas or when terrestrial infrastructure is inaccessible. They can also be used for aircraft, ships at sea, trains, and HAPSs, while a minimum of support is required from terrestrial infrastructure \cite{hoyhtya20195g}. This is the layer that will provide internet services for different applications, including space travel. 
\end{itemize}

Three architectural principles are proposed in the literature for the integration of terrestrial networks with NTNs depending on the degree of integration between the different space/air components: 1) transparent NTNs, in which the NTN plays the role of user's signal repeater; 2) regenerative NTNs, where the NTN regenerates the received signals from the Earth while offering inter-SN connectivity; and 3) a multi-connectivity architecture, in which different transparent SNs are involved (LEO or GEO or a combination of both), and the integration of terrestrial and non-terrestrial access is possible \cite{giordani2020non}. The various connectivity links in this integrated space-terrestrial architecture will be generated through the use of radio frequency (RF) and/or optical technologies enabling short-delay connectivity from the ground.\\
Different architectures have recently been proposed in the literature to meet the requirements of the envisaged applications and use cases.
The authors of \cite{niu2020space} introduced a space-air-ground integrated vehicular network (SAGiven) to further improve connectivity between moving automated vehicles and smart transport systems. For more precise positioning of mobile vehicles, they proposed a new UAV-assisted scheme. They also presented a deep reinforcement learning model based on UAV relaying to avoid smart jamming attacks. In \cite {lee2020integrating}, the authors proposed a multi-hop communication framework that involved using a constellation of LEO SNs and a HAPS mobile node to maximize the achievable end-to-end throughput between two distant ground stations. The simulation results revealed that the proposed scheme achieves significant data throughput compared to other fixed relay schemes or when using only an SN network. Furthermore, to provide seamless connectivity to massive access IoT devices, which are not connected to terrestrial cellular networks, a cognitive SN-UAV network was proposed where the SN and UAVs cooperate \cite{liu2020cell}. 
These works all contribute to the use of 3D layered architecture, which we expect will be required by societal demands in the future. In the next subsection, we present different channel impairments and mitigation techniques for this envisaged architecture.

%
\subsection{SN Communication Links}
\label{ssec:CommunicationLinks}
\subsubsection{Space Node (SN) Link with Space Network Terminal}
\label{sssec:UpDownLinks} 
SpaceNet paths are further classified into uplink and downlink communications. Uplink transmission includes ground-to-SN, ground-to-aircraft, and aircraft-to-SN, whereas downlink involves SN-to-ground, aircraft-to-ground, and SN-to-aircraft. Similar to wireless cellular networks, uplink and downlink channels in SpaceNets differ greatly from the other and are subject to different issues. Specifically, in uplink communication, the transmitted beam experiences additional distortion as the main sources of loss are close to the transmitter. Additionally, to avoid channel interference, uplink and downlink transmissions may operate on different frequencies or on the same frequencies while using different multiple access techniques.
\paragraph{Channel impairments}
Most existing SpaceNet applications currently operate on the microwave RF band, which has led to congestion of the RF spectrum. This is compounded by the fact that RF transmission suffers from regulatory restrictions and limited capacity. Moreover, in uplink/downlink communication, the RF wave is dramatically affected by rain, oxygen, and water vapor attenuation, especially at high transmission frequencies. As the distance between the SN and the ground station increases, other phenomena may affect the RF wave resulting in path loss, blockages, and shadowing from multipath effects.

Thus, to meet the high data rates and capacity requirements, FSO communication is the key technology. Due to its promising characteristics, FSO has gained increasing interest from researchers in recent scholarship. However, in FSO uplink/downlink communications, the transmitted optical signal is highly affected by beam scintillation, atmospheric attenuation, pointing errors, and other unpredictable sources of loss. More specifically, when the optical beam propagates through the atmosphere, it may encounter various atmospheric particles, which lead to power loss. Studies have shown that FSO uplink and downlink are highly susceptible to turbulence-induced fading, foggy weather, snow, and clouds. However, the impact of rain is insignificant as the size of rain droplets is much larger than the wavelength used in SN optical communication. On the other hand, an important issue in uplink optical transmission is beam wandering, which is a neglected issue in downlink transmission. This phenomenon occurs when the beam size is much smaller than the size of turbulent eddies, which results in random loss to the received signal.

One common technical challenge for RF and FSO uplink/downlink SpaceNets is the Doppler shift. Due to the high speed of SNs, there are significant Doppler frequency shifts that may occur at the physical layer. One possible solution is to take advantage of existing frequency compensation techniques to exclude at least moderately the Doppler shift. However, a significant residual Doppler frequency shift could compromise the orthogonality of the subcarrier in a multi-carrier system. For SpaceNet design, it is recommended to have a statistical characterization of Doppler variation in the region of service of the SN as given in \cite{khan2020stochastic,ali1998doppler}. The information obtained could be exploited to predict the visibility time function and to enhance the performance of the user’s phase-lock loop. In \cite{ando2013optical}, the authors developed a homodyne receiver based on Doppler frequency shift compensation for LEO-GEO inter-SN optical communication.

\paragraph{Mitigation techniques}
In the literature, several techniques are proposed to mitigate these side effects, guarantee the availability of the transmitted signal, and satisfy the recommended quality of service (QoS). It should be noted that the majority of these techniques are applicable to uplink/downlink, RF, and FSO communications.
\begin{itemize}
    \item \textbf{Hybrid RF/FSO:} RF and FSO downlink/uplink communications are affected by opposite conditions and behave in a complementary way. Therefore, an efficient solution to guarantee reliable communication and improve system availability is a hybrid RF/FSO system. Here, one approach is to use the RF link as a backup link, where the FSO is initially active and the RF is activated only when the FSO is in outage. Another approach is to transmit simultaneously the same data over both links and using combining techniques at the receiver \cite{9446153,yahia2021weather,9678060}.
  \item \textbf{Aperture averaging:} The aperture averaging technique has been extensively studied to alleviate atmospheric turbulence, especially in downlink optical communication \cite{wasiczko2005aperture,andrews1992aperture,barrios2012exponentiated,erdogan2021site}. Increasing the aperture diameter at the receiver to be large enough to handle turbulent eddies improves the performance of the communication by averaging out all random fluctuations over the aperture area and reduces channel fading compared to a point receiver. 
  \item \textbf{Relay-based transmission:} Another effective technique to tackle these challenges is relay-based transmission. It has been shown that cooperative transmission improves the signal-to-noise ratio (SNR) and offers significant performance over single-hop communication. More recently, the authors in \cite{swaminathan2021haps} introduced the use of HAPS-based relaying to improve the reliability of FSO uplink and to mitigate beam-wandering effects. Also, the authors in \cite{9678060} highlighted the use of HAPS-based communication for combating adverse weather effects. However, the optical SN-to-HAPS channel was shown to suffer from stratospheric attenuation caused by polar clouds, molecular absorption, volcanic eruptions, and scattering due to droplets or ice crystals, which are rare phenomenons \cite{giggenbach}. 
  \item \textbf{Power control:} To compensate for power losses caused by atmospheric effects, a ground station can handle uplink power control or an SN can handle downlink power control \cite{panagopoulos2004satellite}. Power is controlled by adjusting the transmit power in order to guarantee a certain level of communication quality; however, is very challenging to control power at the SN due to size and weight limitations \cite{panagopoulos2004satellite,ali2021power}. 
  \item \textbf{Diversity:} Additionally, frequency, time, and space diversity are highly recommended to enhance link availability and to improve the bit-error-rate (BER) performance of communications. In \cite{erdogan2021site}, authors demonstrated that site diversity through the selection of the best ground station enhances the overall performance of downlink SN communication. 
  \item \textbf{Channel coding:} Error control coding or channel coding by adding redundant bits to the transmitted information was also considered to improve channel reliability in SN downlink communication \cite{li2004analysis}. It was shown that, on the one hand, increasing the redundancy decreases the error probability. On the other hand, the required bandwidth will also increase. Therefore, in order to guarantee a certain level of error probability, an error coding technique can be considered as a trade-off between power and bandwidth requirements.
  \item \textbf{Modulation:} In order to improve spectral efficiency and mitigate the adverse effects of turbulence, several modulation schemes can be adopted \cite{mokari2016resource}. The modulation scheme most commonly used in optical communication is the on-off keying (OOK) scheme, which is generally deployed with intensity modulation with a direct detection (IM/DD) technique. This has been shown to be the best fit to overcome the attenuation impact on the signal \cite{ismail2016free}.
\end{itemize}

Table \ref{tab:Imp} summarizes the different channel impairments in FSO and RF communications.

\begin{table*}[t]
\caption{Channel Impairments in FSO and RF communications}
\label{tab:Imp}
\begin{center}
\begin{tabular}{ |m{3cm}| p{6cm} | p{6cm} |  } 
\hline 
\textbf{Technology } & \textbf{Free-space optical (FSO)} & \textbf{Radio frequency (RF)} \\
\hline
\textbf{Impairments } & Misalignment losses (pointing errors), atmospheric losses,  weather conditions (fog, clouds, haze, smog, snow, winds, rain, polar clouds, volcanic activity), noise, beam wandering, scattering, path loss, Doppler shift  & 
Path loss, noise, shadowing, rain, induced fading loss absorption by molecules (oxygen,water vapor, or other gaseous)
blockage and masking effects, Doppler shift \\
\hline
\end{tabular}
\end{center}
\end{table*}

\subsubsection{Inter-Space Node Links}
\label{sssec:ISLs}
\par In this subsection, we discuss the rationale for using \glspl{isnl}, the types of ISNLs that exist, the Iridium NEXT EOSC, which uses ISNLs, and some challenges that have arisen in deploying ISNLs.

\paragraph{Why ISNLs?}
\label{sssec:why}
\par
\begin{figure*}[htbp]
	\centerline{\includegraphics[scale=0.4]{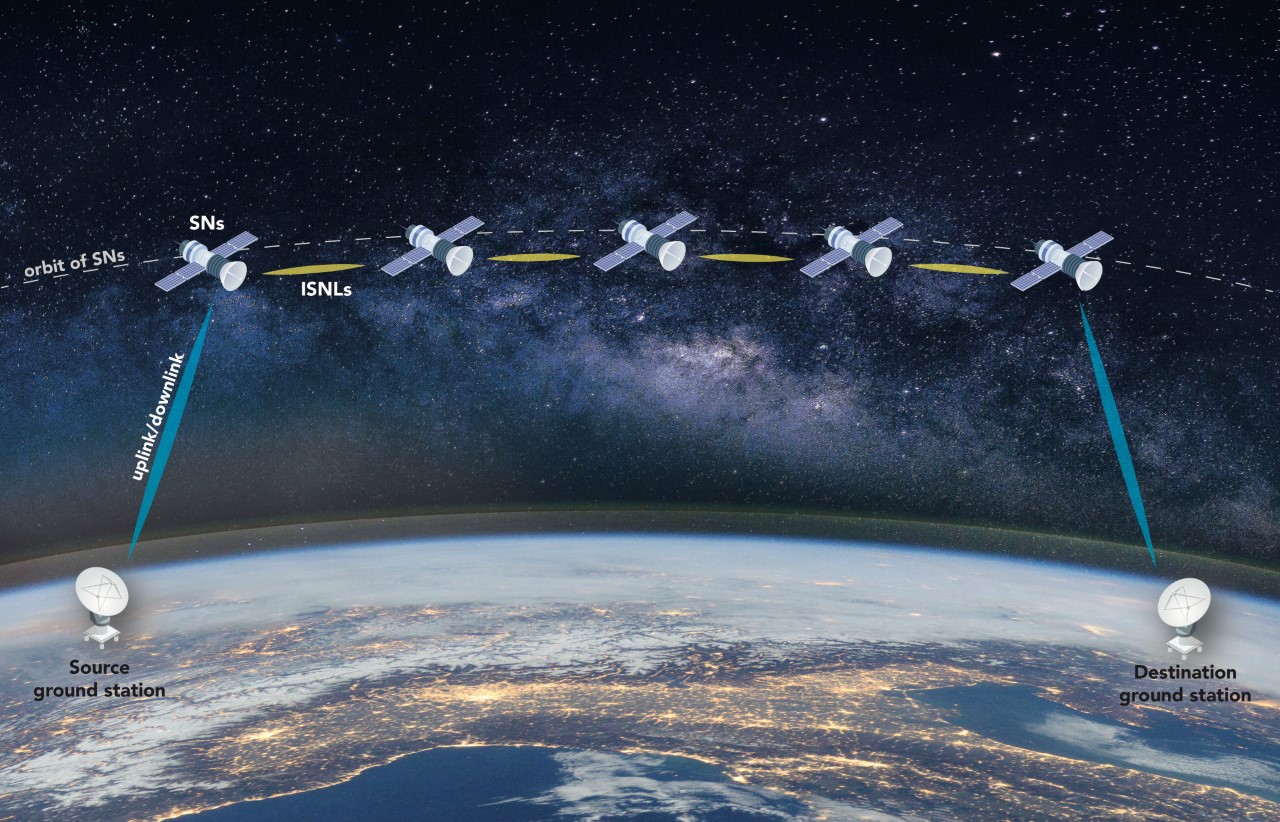}}
	\caption{Data communications with ISNLs.}
	\label{fig:dataComISLs}
\end{figure*}

\begin{figure*}[htbp]
	\centerline{\includegraphics[scale=0.4]{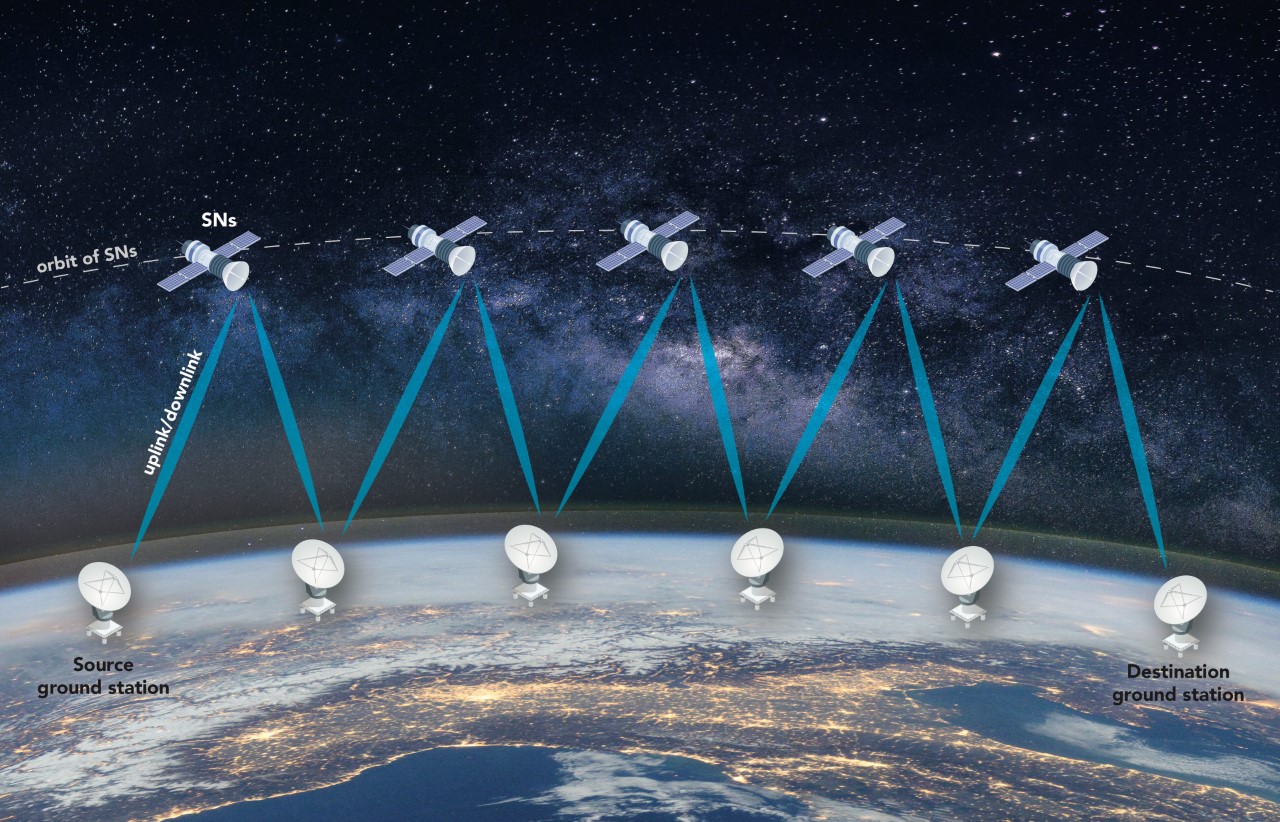}}
	\caption{Data communications without ISNLs.}
	\label{fig:dataComWithoutISLs}
\end{figure*}

 An inter-space node link connects a pair of \glspl{sn}, such as two satellites in a satellite constellation, and ISNLs among SNs result in the formation of an SN network in space. This use case is illustrated in Fig. \ref{fig:dataComISLs}, where the SNs operate as routers and the source and destination ground stations can communicate over the SN network without any intermediate ground station. In the context of a satellite constellation, by SNs we mean satellites in that constellation. Without inter-space node links, the SNs in space function in a bent-pipe mode, where their main function can be to relay data traffic to and from ground stations on the surface of a planet. This mode of operation is shown in Fig. \ref{fig:dataComWithoutISLs}, where data traffic between the source ground station and the destination ground station has to go through multiple intermediate ground stations. 
	
\par Inter-space node links are seen as the key enabler for the creation of a communications network in space \cite{handley2019isls},\cite{hauri2020isls}. They offer significant advantages, such as a reduction in propagation distance, and a reduction in the number of ground stations, which is due to the elimination of communications between SNs and intermediate ground stations. Without ISNLs, an inter-continental long-distance connection that uses SNs for data communications between two cities, such as Toronto and Istanbul, would require several intermediate ground stations, and the data traffic would have to go back and forth several times between these intermediate ground stations on Earth and the SNs in space. This would result in an increase in the propagation distance and, by extension, the latency of the data communications between the source and destination ground stations.

\paragraph{Types of ISNLs}
\label{sssec:types}

\par Based on the technology of the link used for communications between two SNs, there can be two types of ISNLs: \glspl{rfisnl} and \glspl{fsoisnl}.
\par
\begin{figure*}[htbp]
	\centerline{\includegraphics[scale=0.1]{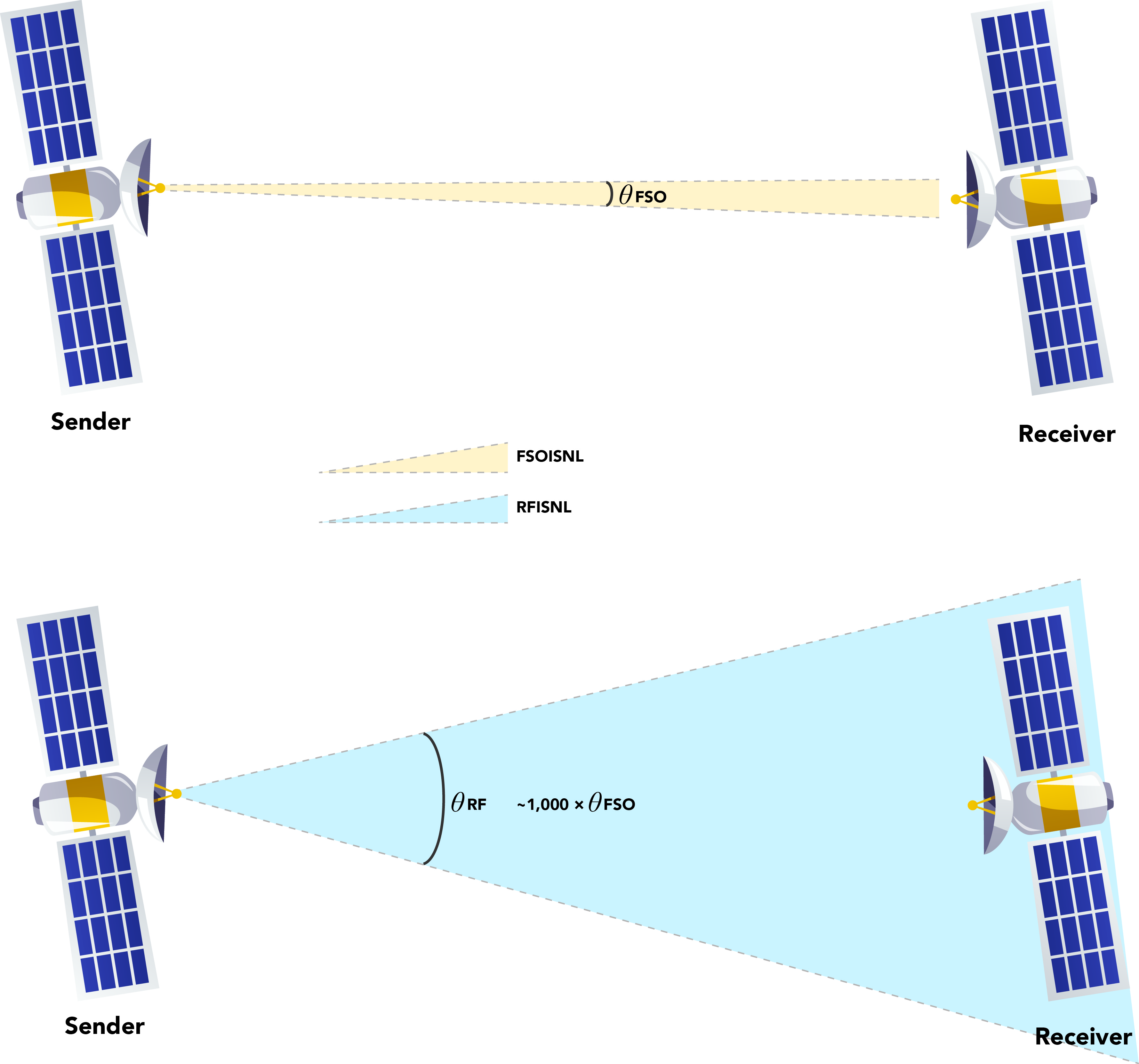}}
	\caption{Beam divergence – RFISNL vs. FSOISNL.}
	\label{fig:beamDiv}
\end{figure*}

 The Iridium NEXT EOSC is currently the only one that has demonstrated the feasibility of using Ka-band RFISNLs to interconnect SNs.  However, for point-to-point wireless communication in space, especially SN-to-SN links, FSO communications have emerged as an attractive alternative to RF communications. The pros of an \acrshort{fsoisnl} compared to an RFISNL heavily outweigh its cons, and a detailed comparison of the two is provided later in Section \ref{ssec:LaserISLs}. The beam divergence of an RF signal is typically 1,000 times greater than that of an FSO signal, and Fig. \ref{fig:beamDiv} provides an illustration of this difference, where $\theta_\text{FSO}$ and $\theta_\text{RF}$ are beam divergences of FSOISNL and RFISNL, respectively \cite{franz2000opticalComm}.

\par Free-space optical ISNLs are also commonly referred to as optical ISNLs (OISNLs) or laser ISNLs (LISNLs). A comparison of RFISNL and LISNL has been examined in \cite{toyoshima2005fso}, where RFISNLs are considered to operate in Ka and mmWave bands. The transmit power for this analysis was set at 10, 20, and 50 W for optical, mmWave, and Ka ISNLs, respectively. It was concluded that a Ka or mmWave ISNL needs an antenna that is at least 19 times greater in diameter and more than twice the onboard power and mass compared to an LISNL when the data rate for the link is 2.5 Gbps and the inter-SN distance is 5,000 km.

\paragraph{ Iridium NEXT -- an EOSC using ISNLs}
\label{sssec:iridium}

 \par The Iridium EOSC system was designed by engineers at Motorola in the early 1990s \cite{leopold1991iridium},\cite{leopold1993iridium}. The deployment of this EOSC began in 1997, and it started offering commercial services (including voice and low-speed data) in 1998. Its original design consisted of 77 SNs with six ISNLs per SN, although this was later reduced to 66 SNs and four ISNLs per SN in the deployed EOSC \cite{wood2001constellations}. There are 77 electrons that orbit the nucleus of the element Iridium, and this led to the name Iridium for this EOSC. The Iridium EOSC was not able to compete with the rapid development of terrestrial mobile communication networks in the 1990s, and its operating company went bankrupt in 2000. In 2007, a new company, Iridium Satellite LLC, was created to revive this EOSC, and it announced plans for a new Iridium NEXT EOSC that would replace the original Iridium SNs with new SNs to provide higher data rates.

\par In its 2013 FCC filing \cite{iridium}, Iridium Satellite LLC proposed to replace its existing EOSC with a second-generation Iridium NEXT EOSC system that would use the same orbital parameters, provide the same coverage, and transmit on the same frequency bands. For continuity of service during the transition from old to new EOSC, the existing SNs were replaced one-for-one with new ones as they were launched. The upgrade to Iridium NEXT SNs was completed in 2019. Similar to the old Iridium SNs, each Iridium NEXT SN is capable of establishing four ISNLs with other Iridium NEXT SNs. These include two ISNLs with SNs in the same orbital plane that are at the front and aft and two ISNLs with the nearest SNs in adjacent orbital planes. These ISNLs operate on RF frequencies in the 23.18–23.38 GHz range of the Ka band at a capacity of 12.5 Mbps \cite{yoon2019iridium}.

\paragraph{Challenges for ISNLs}
\label{sssec:challenges}

\par In the following, we highlight major challenges that are encountered in establishing ISNLs between SNs in a constellation.
\par {\bfseries Cross-seam communication:} In Walker Star (or Polar) constellations, SNs generally move from north to south in the first orbital plane and from south to north in the last orbital plane. As a result, SNs in the first and last orbital planes move in opposite or counter-rotating directions, and the edge where these orbital planes meet is called orbital seam. The ISNLs between SNs in these counter-rotating orbital planes need to be established between these SNs when they are moving at high speeds in opposite directions. In the Iridium NEXT constellation, such ISNLs are not supported.

\par Similarly, in Walker Delta (or inclined) constellations, half of the SNs typically move in a northeasterly direction in descending orbital planes and the other half move in a southeasterly direction in ascending orbital planes. Consequently, SNs in these orbital planes cross each other while orbiting, and it is difficult for a northeasterly moving SN to establish an ISNL with a southeasterly moving SN. The Starlink Phase 1 EOSC currently does not support ISNLs between crossing SNs. The ISNLs between counter-rotating or crossing SNs are generally referred to as cross-seam ISNLs.

\par	{\bfseries Pointing, acquisition, and tracking:} The main challenge in establishing cross-seam ISNLs between counter-rotating or crossing SNs moving at high speeds in different directions is the need for a highly sophisticated and accurate \acrfull{pat} system \cite{kaushal2017pat} (also sometimes known as an \acrfull{atp} system) on board these SNs to establish and maintain such ISNLs. For example, the methodology of the PAT mechanism for setting up an LISNL between SNs involves pointing the laser beam of the \acrfull{lct} of the transmitting SN in the direction of the LCT of the receiving SN, acquiring the incoming optical signal from the transmitting LCT, and then tracking the position of the LCT in the remote SN to maintain the LISNL \cite{kaymak2018pat}.

\par Although the small beam divergence is an advantage of FSO communications, it is a disadvantage in LISNLs. The establishment of an LISNL between a pair of SNs involves pointing a laser beam at a moving SN from another moving SN, and this is a major challenge due to the different relative velocities of SNs and the small divergence of their laser beams. SpaceX initially planned to equip its Starlink SNs with five LCTs to interconnect the SNs within the EOSC via LISNLs to realize an optical wireless satellite network \cite{SpaceX:1}. Four of these LCTs were meant for connecting to two neighbors in the same orbital plane and two neighbors in adjacent orbital planes, while the fifth LCT was intended to connect to a crossing SN. However, SpaceX revised the number of LCTs per SN to four in a later \acrfull{fcc} filing \cite{SpaceX:3}. This was likely due to the difficulty of developing an appropriate PAT system suitable for the fifth LCT meant for creating the LISNL between crossing SNs.

\par	{\bfseries Doppler shift:} Doppler shift is the change in frequency of the received signal due to the difference in the relative velocity of the sender and the receiver. A Doppler shift exists between SNs moving in adjacent orbital planes and is significantly higher between counter-rotating or crossing SNs \cite{yang2009doppler}. It needs to be addressed to establish reliable ISNLs between SNs, and it is another major challenge in the implementation of cross-seam ISNLs in satellite constellations.

%
\subsection{Artificial Intelligence and Machine Learning for Future SpaceNets}
\label{ssec:AI_ML}
Mimicking human intelligence, artificial intelligence (AI) is the ability of a system, or an ``intelligent agent," to take actions toward achieving certain goals while being aware of its surrounding environment. To do this, the system requires capabilities such as learning, reasoning, planning, and perception, among others. Machine learning (ML) is one of the major functions that reflect machine intelligence. ML involves building a mathematical model to perform a certain task (e.g., classification, regression, clustering) without being explicitly programmed to accomplish it. This is done with the aid of data. The main categories of ML are as follows:
\begin{itemize}
\item Supervised learning utilizes a labeled data set (i.e., containing both the input and the desired output) to learn a certain unknown mapping function that maps the input to the output. That is, the model is trained using solved examples, and then it can provide predictions for the output based on new inputs. The primary models that are used in supervised learning are artificial neural networks (ANNs), support vector machines (SVMs), k-nearest neighbours (KNN), and logistic regression.
\item In unsupervised learning, the data is unlabeled (i.e., contains inputs only) but has a certain hidden structure that needs to be identified by the model. A good example of unsupervised learning is the clustering of data into groups with some common features. Examples of models that utilize this learning approach are k-means clustering, principal component analysis (PCA), and Gaussian mixture model (GMM).
\item In reinforcement learning, the agent assigns positive and negative rewards for the actions it takes while interacting with the environment and tries to maximize the positive rewards from its experience. Therefore, this learning technique does not require a dataset for training. Examples of reinforcement learning algorithms include Q-learning, actor-critic learning, and multi-armed bandit learning.
\end{itemize}

\subsubsection{Benefits of Utilizing AI/ML in SpaceNets}
Incorporating AI/ML techniques for different functionalities in future SpaceNets will have many advantages. Below, we summarize the main benefits that AI/ML can bring to SpaceNets:
\begin{itemize}
\item \textbf{Overcoming modeling complications:} Due to the high mobility of most SNs, the dynamic environment of space, the use of wireless links for most segments, the modeling of channels, interference, and mobility (among other aspects) is very complicated in SpaceNets. The accuracy of these models has a significant impact on network performance and user satisfaction. However, ML techniques do not require accurate modeling. Instead, they can automatically extract the required information by learning from the data and prior knowledge, without going into the details of the model behind this behaviour, and provide efficient predictions for the decisions.
\item \textbf{Low computational and time complexity:} The operation of ML techniques can be divided into two stages: training and testing. In the training phase, the model parameters are optimized to learn the structure of the data. This learning process is computationally complex. However, it can be implemented offline utilizing available data. During the testing phase, the trained model operates online with low complexity and a short response time. For instance, a trained ANN predicts the output for new inputs mainly by matrix multiplications. This reduces the computational complexity and response time during online operation. For example, according to \cite{sun2018learning}, it was shown that a trained deep neural network (DNN) could predict the power allocation for a network with $15$ users in a response time of just $0.0149$ ms.
\item \textbf{Adaptability to dynamic environments: } SpaceNets are characterized by their highly dynamic environment. This includes a network topology that changes with time due to the movement of SNs, channel conditions that depend on the weather and location of SNTs and SNs, and traffic loads that differs from one spot to another. The development, configuration, and management of the network need to react to these changing conditions quickly and efficiently. In this regard, AI/ML can adapt to these changes smoothly and even predict them and act in a proactive manner.
\item \textbf{Scalability:} Since AI/ML techniques can operate with low computational and time complexity, the increase in complexity when the network parameters scale up is lower than that of traditional approaches. This is suitable for SpaceNets in general, and EOSNs in particular, which are required to serve a massive number of users per EOS as its footprint covers a wide area on the ground.
\end{itemize}

\subsubsection{Challenges of Using AI/ML in SpaceNets}
Several ML techniques have been closely studied in the context of terrestrial networks. However, more studies are needed if they are to be exploited in next-generation SpaceNets, which have different characteristics in terms of propagation delay, computational capabilities, scalability, mobility, and data exchange between network entities. In this respect, the following challenges need to be addressed to leverage the intelligence of AI/ML in future SpaceNets:
\begin{itemize}
\item \textbf{Dataset availability:}  High-quality data is essential for ML models, and their accuracy depends greatly on the amount of data used for training. Therefore, the availability of data is necessary for AI-driven SpaceNets. In this regard, data for SpaceNets might not be available for all use cases due to cost or computational resource limits.
\item \textbf{Suitable techniques:} There are numerous ML approaches that can be grouped under the three categories discussed in this section. Each technique has its own advantages and shortcomings. For example, supervised learning-based models can provide accurate predictions for decision making. However, labeled data sets are required to train those models, which might not be available. Therefore, there is a need to identify suitable techniques for each use case and data availability scenario. 
\item \textbf{Distributed versus centralized:} To address scalability issues and avoid centralized processing shortcomings (e.g., single point of failure), distributed ML techniques, such as federated learning \cite{ramzi2022ground}, can be utilized to allow the SNs to share the parameters of their local models to construct a global one without needing to exchange any actual data. However, this impacts the global optimization of the models. Therefore, there is a trade-off between centralized and distributed ML techniques.
\item \textbf{Standardization:} Despite the great interest that AI/ML has received, it is still in the study phase and is not widely standardized in wireless networks. For example, to automate mobile networks, 3GPP has studied what is called a network data analytics function (NWDAF) and big-data-driven network architecture for 5G networks \cite{3GPPTR23-791}. In addition, AI/ML is included in the work items to be studied or specified in Release 18 \cite{rel8News}.  This means that AI/ML is beginning to find a place in wireless network standards. However, for SpaceNets, the technology is still in its infancy.
\item \textbf{Unique characteristics of SpaceNets:} Compared with terrestrial networks, SpaceNets have unique characteristics that can impact the performance of AI/ML models and techniques. For example, the propagation delay in SpaceNets is high compared to terrestrial networks due to the long distance between the SNs and SNTs (e.g., connecting to an EOS from the ground). This makes it challenging for AI/ML models to utilize such kind of data that could be outdated due to that delay. Another example is the highly dynamic topology due to the motion of SNs, which may make AI/ML models need to be retrained for the highly different network topology. Furthermore, the communications environment in SpaceNets makes the data susceptible to high errors due to pointing and polarization errors, weather conditions (for communications to/from the ground), radiation, and high interference. Consequently, such factors associated with SpaceNets pose several challenges for data-driven techniques.
\end{itemize}

\subsubsection{Potential Roles of AI/ML in Future SpaceNets}
Advances in AI technology have begun to revolutionize numerous industries. Utilizing AI/ML to improve wireless networks has received significant interest from the industry and the research community. In this regard, several applications of AI/ML in wireless communications and networking have been investigated in the literature \cite{sun2019application,abdelsadek2020resource,chen2019artificial,tang2021survey}. Most of these applications (and others too) can be used in SpaceNets \cite{fourati2021artificial}. Such applications include channel modeling and estimation, resource allocation, handover prediction, mobility management, routing, localization, and security, among others. The following points highlight AI/ML applications in the literature as they pertain to SpaceNets, in general, and EOSNs, in particular:
\begin{itemize}
\item \textbf{Handover prediction and  management:}
In \cite{xu2020qoe}, the authors used reinforcement learning to design a user-centric handover technique in which the ground user terminal makes decisions about the handover process based on predicted handover factors, such as service time, channel resources, and relay overhead. For this purpose, the environment state is selected to indicate the features related to the candidate LEO satellites (e.g., relay overhead, estimated available channels, and spatial relationship value). The action is set to be the decision of handover to each satellite. The quality of experience (\acrshort{qoe}) of the user is set as the reward. Therefore, this reinforcement learning problem aims to maximize the QoE of users. Similarly, in \cite{he2020load}, the authors leveraged multi-agent Q-learning to minimize the average number of handovers in an EOSN while reducing the blockage rate of the users. To achieve this, the state incorporated the covering LEO satellites, the available channels per satellite, and the period of service by each satellite before the handover was required. The reward was designed to indicate the remaining visible time and the channel budget of the LEO satellite.

\item \textbf{Routing:} In \cite{cigliano2020machine}, data-driven solutions were used for packet routing in EOS megaconstellations. In this respect, the authors adopted an unsupervised learning technique to define clusters of traffic based on their data type while taking into account the geographical traffic distribution density. Then, a reinforcement learning approach was applied to learn following the replays and selecting the route with the lowest round-trip time.  The authors demonstrated that this approach outperformed the classical shortest path routing approach. In \cite{liu2020drl}, the authors leveraged deep reinforcement
learning to design an energy-aware routing technique for EOSNs. This routing scheme considered the battery level and path delay of EOSs without calculating the propagation delay. Therefore, the packets were routed to EOSs with a high level of battery power while taking into consideration a guaranteed end-to-end delay constraint. The authors proved that this ML-driven technique decreased the average energy consumption of EOSs by more than $ 55\%$ relative to conventional routing approaches.

\item \textbf{Channel modeling and estimation:}
In \cite{zhang2021deep}, the authors proposed a deep learning-based channel state information predictor for collocated massive multiple-input multiple-output (MIMO)-based LEO satellite communications. This channel condition predictor exploited the channel correlation to overcome channel aging issues related to satellite and space communications. This is because the long propagation delay caused the channel state information (CSI) estimated by LEO satellites using traditional techniques, such as maximum-likelihood and minimum-mean-square error (MMSE), to be out of date. The deep learning model was based on long short term with memory (LSTM), which is an upgraded version of a recurrent neural network (RNN) suitable for time series problems. 

\item \textbf{Resource allocation:}
The authors in \cite{sun2019deep} utilized ML in a resource allocation problem in EOS IoT applications based on non-orthogonal multiple access (NOMA). In so doing, they used deep learning to study the relationship between the queue and channel states and overcome the non-convexity of the power allocation problem. This resulted in a successive interference cancellation (SIC) decoding order that was more accurate than conventional approaches. In \cite{qiu2019deep}, the authors jointly optimized the networking, caching, and computing resource allocation in software-defined EOS-terrestrial networks. After modeling the problem as a Markov decision process, they used deep Q-learning to solve the problem in an efficient manner. 
\end{itemize}
%
\subsection{Softwarized Network Management Enablers for Future SpaceNets}\label{ssec:NetworkManagement}
\subsubsection{Software Defined Space Networks (SDSN)}

 \acrfull{sdn} is a concept that separates the data plane from the control plane, where controllers perform the intelligent network management functionalities. SDN switches are located in the data plane and their role is to forward data packets. Actions on how to treat the received data packets are predefined in the switch flow tables. Through the northbound interfaces, the controllers interact with applications to decide on how to create and update flow tables saved in the SDN switches. Westbound and eastbound interfaces are used to communicate among SDN controllers. Once a new packet is received at the SDN switch, the flow table is checked. If the packet destination can be found in the flow table, then the predefined actions are performed. Otherwise, the packet will be forwarded to the controller through the southbound interface \cite{Kaloxylos2020}. The main advantage of implementing the SDN concept in space networks is the flexibility that it adds in managing the network topology and routing data traffic. This makes it a good solution for the dynamic network topology of future SpaceNets.

 Topology discovery (\acrshort{td}) is a key component to support the logically centralized control and network management principle of an SDN. TD provides global visibility of the complete network for a controller. Discovering the network topology includes the discovery of switches, hosts, and interconnected switches. In the TD process, each entity in the network can collect information about the network topology. The information can be collected at different levels and in many ways with the goal of delivering the topology information to the controllers. 
To avoid flooding the network and controllers with unnecessary information, the TD process must be efficient in terms of sending topology information only when changes happen \cite{Kipongo2018}. However, the fast mobility of satellites, which results in frequent topology changes in densely deployed megaconstellations, results in more packets being sent to the controller to update the topology and flow tables. Such overhead traffic could negatively affect the efficiency of network resource utilization.

In an SDN, the controller is responsible for updating the forwarding rules of the switches in the data plane. The time required for a rule to be installed is referred to as the flow setup time. In a large-scale network, such as an EOSN or SpaceNet, a single controller with limited resources would not be able to handle all the update requests originating from the data plane, and the controller might encounter bottlenecks. Furthermore, due to the large distances between the SNs and the controller, there is no guarantee of meeting the acceptable control plane latency. Therefore, having a distributed control plane becomes mandatory. However, it is very important to choose the optimum number of controllers and their locations based on the traffic load distribution and topology changes~\cite{Papa2018}.

\subsubsection{Network Function Virtualization (\acrshort{nfv}) in Future SpaceNets}

Virtualization simplifies network management and facilitates resource sharing, aggregation, and dynamic allocation, which improves network scalability and service agility. Network function virtualization (NFV) is viewed as a new paradigm in allocating network resources on-demand. The main ideas of NFV are (i) the separation of a network function and physical device, (ii) the flexible deployment and management of network functionalities, and (iii) the provision of reconfigurable services. The means of implementing NFV involves (i) decoupling a network function from the dedicated physical device, (ii) implementing virtual network functions (\acrshort{vnf}s) on virtual machines, and (iii) assembling and chaining VNFs to create services \cite{bertaux2015software}. The major benefits of NFV are (i) running and creating network services
with high flexibility by adaptively assembling and chaining software-based network functions without changing network
architecture, (ii) lowering capital expenditures (\acrshort{capex}) and operational expenditures (\acrshort{opex}) by using the centralized
servers instead of installing specialized hardware equipment for new services \cite{zhang2019satellite}, and (iii) facilitating the global and optimal control and management of the network by implementing the software-based network functions in centralized
network servers \cite{xu2018software}.

However, implementing NFV in SpaceNets or EOSNs is challenging due 
to large-scale and rapid topology changes in the network.
In \cite{xu2018software}, a software-defined architecture was presented for next-generation satellite networks, called SoftSpace, which exploited NFV, network virtualization, and software-defined radio. This was done to facilitate the incorporation of new applications, services, and satellite communication technologies. It was shown that this was capable of not only reducing capital and operational expenditures but also of integrating EOSNs with terrestrial networks seamlessly, and it was shown that it could improve the interoperability of satellite network devices. The author of \cite{zhang2019satellite} proposed a dynamic virtualization system containing three layers: a virtualized infrastructure layer, a virtualized network function (VNF) layer, and an orchestrator layer. To adapt to the mobility of LEO satellites, a dynamic resource monitor was added to the satellite orchestrator. The monitor would follow a VNF to know whether it was connected or not. The resource registration and deletion operations could be quickly done in the orchestrator. In addition, when the available resources changed, the information would be sent to the users and edge computing servers to
adjust their policies. The author in \cite{gao2020virtual} proposed a potential game approach for virtual network function (VNF) placement in EOSN edge computing. The author formulated the VNF placement problem with maximum network payoff as a potential game and analyzed the problem by a game-theoretical approach. Integrating virtualization technologies and machine learning techniques can pave the way for automated and intelligent network design, operation, and management.

\subsubsection{Network Slicing in Future SpaceNets}
Virtualization enables the support of logical sub-networks which are referred to as network slices. Each slice provides a configurable custom-fit solution for the target use-cases, which may have different requirements in terms of reliability, security, latency, and capacity. Due to the diverse range of envisioned applications in future SpaceNets with varying requirements, network slicing emerges as a key enabler with the potential for autonomous management in terms of slice customization and prioritization.

The potential of network slicing in satellite mega-constellation is already noted in the literature. The problem for on-demand network slicing for resource optimization in integrated 5G satellite networks is proposed in  \cite{ahmed2018demand}, which targets the minimization of resource consumption while meeting the target service specifications.  The authors in \cite{bisio2019network}  use the network slicing paradigm to reduce the end-to-end packet delay for different services focusing on an exemplary use case of providing connectivity to railway communications building over an integrated 5G-satellite network, where both queuing theory and neural network based approaches are shown to improve the target QoS levels. The authors in \cite{de2020qos} extended the notion from QoS to QoE, and numerically showed that it is possible to satisfy different requirements of the users’ and verticals’ requirements with the use of network slicing. Neural networks are used to determine the  optimal resource allocation and through extensive numerical studies that the proposed  neural network based solution can  provide better solutions than the optimised solutions based on queuing theory
paradigms. 

In \cite{drif2021extensible}, the authors propose a satellite slicing framework for satellite network operators with the goal of facilitating the integration of satellite services and terrestrial networks, specifically focusing on 5G networks. Slicing of the satellite network is proposed in accordance with the 3GPP network slicing guidelines, specifically addressing the requirements from the slice isolation and architectural differences perspectives. The lack of a clear difference between the core network and the radio access network is cleverly utilized to take the advantage of NFV/SDN paradigms to orchestrate the flows. 

The extension of the network slicing to the satellite mega-constellations, making use of the inter-connectivity between the satellites is investigated in  \cite{kak2021towards}, where the authors introduce automatic network
slicing framework for the Internet of Space Things. The analytical formulation of the optimization problem addressing the dual objectives of route computation and resource allocation with minimal service level agreement violation is presented and the efficacy of the proposed approach is shown numerically. These works solidify the expected role of network slicing for future EOSNs networks, where user requirements will be even more demanding and the communication environments will be more challenging. 

Softwarized network management is essential to support satellite network use cases in space, air and ground. Utilizing SDN concepts adds flexibility and automation to satellite network management. Also, it creates a suitable environment for intelligent automated satellite network management, which turns satellite networks into a self-organizing and even a self-evolving network of networks.
%
\subsection{Edge Computing and Caching for Future SpaceNets} \label{ssec:cachingMEC}
EOSNs provide coverage in rural and remote areas and are expected to complement terrestrial networks in urban areas. Future EOSNs  will face the challenge of growing user demands for a higher quality of service (QoS), including higher data rates, lower communication latency, and lower energy costs on data communication and processing. In addition, many new computation and energy intensive
applications are emerging. Relying on cloud computing to process these tasks faces the disadvantages of high delays and the consumption of communication resources. Mobile edge computing (\acrshort{mec}) has been
proposed to transfer computation and storage
capacity from cloud computing to local servers near users to improve QoS by reducing user-perceived latency. In addition,
MEC can provide content caching and storage
services, which are effective in reducing traffic in the core network \cite{zhang2019satellite}. 

Equipping LEO satellites with MEC supports computation offloading, especially in rural and remote areas (e.g., IoT applications); and provides content caching and storage to reduce repeat transmissions from remote Clouds \cite{zhang2019satellite}. In addition EOS MEC is an effective way to support delay-sensitive and resource-hungry wide-area IoT applications in remote areas. By processing offloaded data (or part of it) directly by EOSs, loads can be considerably reduced on satellite-to-satellite links and satellite-to-ground links, and processing delays can be reduced \cite{li2021integrating}. Satellites with software defined payloads, such as the “Eutelsat-quantum,” designed by the European Space Agency (ESA) and Eutelsat \cite{fenech2015eutelsat}, have a potential to serve various applications and operate as an MEC node in space. However, satellite computing power is still considered limited, even though satellites can be equipped with CPUs, GPUs, and FPGAs \cite{wei2019application}.

Some recent studies have proposed solutions to enable MEC in EOSNs. In \cite{wang2018computation}, a satellite terrestrial network architecture with double edge computing was proposed to provide computing service for remote areas. The edge servers were located both on EOSs and in terrestrial gateways. In addition, the author introduced a strategy to efficiently schedule the edge servers distributed in the satellite terrestrial networks. The strategy employed a minimum cost matching algorithm to optimize energy consumption and reduce latency by assigning tasks to edge servers with minimal cost. When a large number of burst offloading tasks were stacked in the terrestrial MEC sever, tasks were offloaded to the EOSN edge servers when the computation load of the MEC terrestrial edge servers exceeded a given threshold.
 
The authors in \cite{wei2019application} proposed using satellite edge computing to enable onboard intelligent processing to support image target detection of satellite IoT. The EOSs act as edge nodes that can communicate with each other and with the upper layer of EOS cloud nodes. The data can be offloaded from the EOS edge nodes to the EOS cloud nodes or to the ground cloud center. A service cluster can be created by a set of cooperative EOS edge nodes. The EOS edge nodes and EOS cloud nodes can implement network slicing through SDN/NFV. To do this, a deep learning model can be placed onboard, and a training model can be prepared in advance and then uploaded to the EOS. Online optimization training can be performed on EOS. Computing with low complexity can be done on the EOS IoT edge node. Computing with high complexity and high requirements for real-time is suitable for completion in the EOS IoT cloud node. Computing with high complexity and low real-time processing requirements
should be placed in the ground cloud center on the ground.  

An EOS MEC was proposed in \cite{zhang2019satellite}, where user 
 equipment without access to a terrestrial MEC server
could utilize  MEC services via satellite links. The author proposed a cooperative computation offloading model to provide parallel computation
in EOSNs and terrestrial networks. Moreover, a dynamic network virtualization technique was designed to integrate the network resources. In \cite{wang2019satellite}, EOS play the role of a space edge computing node to serve the remote IoT devices that can not access terrestrial networks. The proposed space edge computing system architecture virtualizes onboard resources of satellites and form a resource pool. Resources are allocated on demand to meet the processing requirements of the tasks. The author provided an analysis of the execution time and energy consumption of the task. 

\subsubsection{Challenges Facing Edge Computing in EOSNs}

The integration of MEC in EOSNs faces many technical challenges that are different from edge computing in terrestrial networks. 
\begin{itemize}
    \item The storage and processing resource of a single satellite is considered relatively scarce. This is because the power, weight, and size of a satellite are limited and the environment of space is considered harsh. 
    \item The mobility of LEO satellites at high speeds results in frequent handovers from one satellite to another for devices located on Earth \cite{li2021integrating}.
    \item The propagation delay of satellite-ground links and inter-satellite links are longer than the wireline links in terrestrial networks. Therefore, such delays need to be considered while making data offloading decisions \cite{li2021integrating}. 
    \item Providing edge computing services by deploying a domain server in an EOS might not be economically feasible when EOSNs serve sparsely distributed users in remote or rural areas. However, future EOSNs are expected to serve urban areas as well, where there is high density of users.
\end{itemize}
\subsubsection{LEO Satellite Edge Computing System Requirements}
To provide efficient edge computing services in LEO satellite networks the following points should be taken into consideration: 
\begin{itemize}
    \item Service mobility capability: When a service is provided by a certain LEO satellite, the hop count between the user device
    and the service will increase rapidly due to the very high speed of the satellite. Therefore, enabling service migration from one satellite to another is necessary \cite{li2021integrating}.
    \item Pooling satellite resources: It is important to provide solutions that integrate and manage satellite resources efficiently. This is because the resources of a single satellite are scarce and may not be sufficient to process computation-intensive tasks \cite{li2021integrating}.
    \item Efficient scheduling in satellite edge computing: To fully utilize the capabilities of an LEO satellite edge computing system, task scheduling is necessary to assign proper processors for pending tasks. In addition, the time and energy costs of the space-terrestrial and inter-satellite links also need to be considered, which makes the scheduling problem more complex compared to terrestrial MEC.
    \item Service discovery: Since LEO satellites have scarce resource, placing every service on all the satellites is not possible. So when a user request for a service, a service discovery procedure should be performed to find the satellite(s) which can serve the user \cite{li2021integrating}.
    \item Service migration decision: Although service migration is required in LEO satellite edge computing, frequent migration will consume the inter-satellite link resources and generate unnecessary traffic overhead in the network. Therefore, it is important to make intelligent service migration decisions that take into consideration the trade off between the costs using a satellite service located multiple hops away relative to a service offered by a closer satellite \cite{li2021integrating}. 
    \item Distributed computing: As some computational tasks may not be processed by one satellite, efficient distributed computing algorithms are required which can adapt to the dynamic topology of satellite servers. 
    \item Laser inter-satellite links: Although laser inter-satellite links will provide a fast way to communicate between different satellite edge computing servers, such links should not be overused to save the communication and energy resources. 
\end{itemize}

%
\subsection{Antenna Technologies for Future SpaceNets}
\label{ssec:AntennaTechnologies}
To support new frequency bands, terabit high-speed links, and to ensure beam steering, new antenna technologies are needed. To this end, space companies are investigating the use of modern reconfigurable phased antennas, which are lighter, smaller, and use less energy than current antennas. Using such antennas will enable the integration of multibeam architectures, which can operate in mmWave bands to allow simultaneous transmission of information to different places on Earth, while offering better spectrum efficiency.

\subsubsection{Broadband RF Links between EOSs and UEs}
\label{sssec:spacemobile}

\par To connect to EOSs in AST’s SpaceMobile network via broadband RF links, the UEs will operate on frequencies already licensed to AST partner MNOs. For the gateway (or ground station) to EOS to UE link, the gateway uplink carriers in the V band will uplink the UE forward link signals for each active cell to a SpaceMobile EOS, and a payload processer on the EOS will demultiplex the V band uplink signals and map them to the phased antenna array’s downlink beams covering the cells in the assigned UE channels. At the return link side between the UE, EOS, and gateway, the UE uplink signals from different cells in the assigned UE channels will be received by the beams of the phased array antenna on a SpaceMobile EOS, and the received signals will be multiplexed in frequency domain, upconverted to the V band downlink frequencies, and transmitted to a gateway \cite{ast:1}.
	
\par The AST SpaceMobile EOSs will have beamforming technology with MIMO functions. Each AST SpaceMobile EOS will have up to 2,800 user beams and two or three gateway beams. Each user beam will be steered through a large phased array antenna (900 m$^2$ in size \cite{ast:3}) on the EOS and will have the ability of being pointed anywhere within the \acrfull{fov} of 20º elevation angle. An EOS will be able to transmit all of its active beams on the same frequency or on different frequencies. Each active user beam will be able to track a fixed cell on the ground within its FoV, and a cell on the ground will be illuminated by a single beam or multiple beams from the same EOS or by multiple beams from different EOSs that are within view of the cell to improve the user experience.
	
\par As the beams of a SpaceMobile EOS track specific ground cells in the EOS’s FoV, the elevation angle of a given cell viewing the EOS will vary as the EOS passes over its service area. At different elevation angles, the digital beam former on the SpaceMobile EOS will select different sections of the EOS’s phased array antenna aperture to form the beam to track the cell. The smaller the elevation angle, the larger the size of the aperture selected until the maximum aperture size of the phased array antenna is reached. This approach will result in an almost consistent carrier-to-noise ratio regardless of the cell elevation angle as long as the cell is within the FoV of the EOS \cite{ast:3}.

LEO EOSs with very large antennas are being planned by AST for its SpaceMobile EOSC so that its EOSs are able to directly communicate with off-the-shelf mobile phones/smartphones on ground. However, very large antenna sizes on EOSs mean a higher probability of collisions with other spacecrafts at LEO altitudes, especially if such an EOS fails and has no propulsive capability to avoid collisions.

\begin{figure*}[!htb]
	\centerline{\includegraphics[scale=0.75]{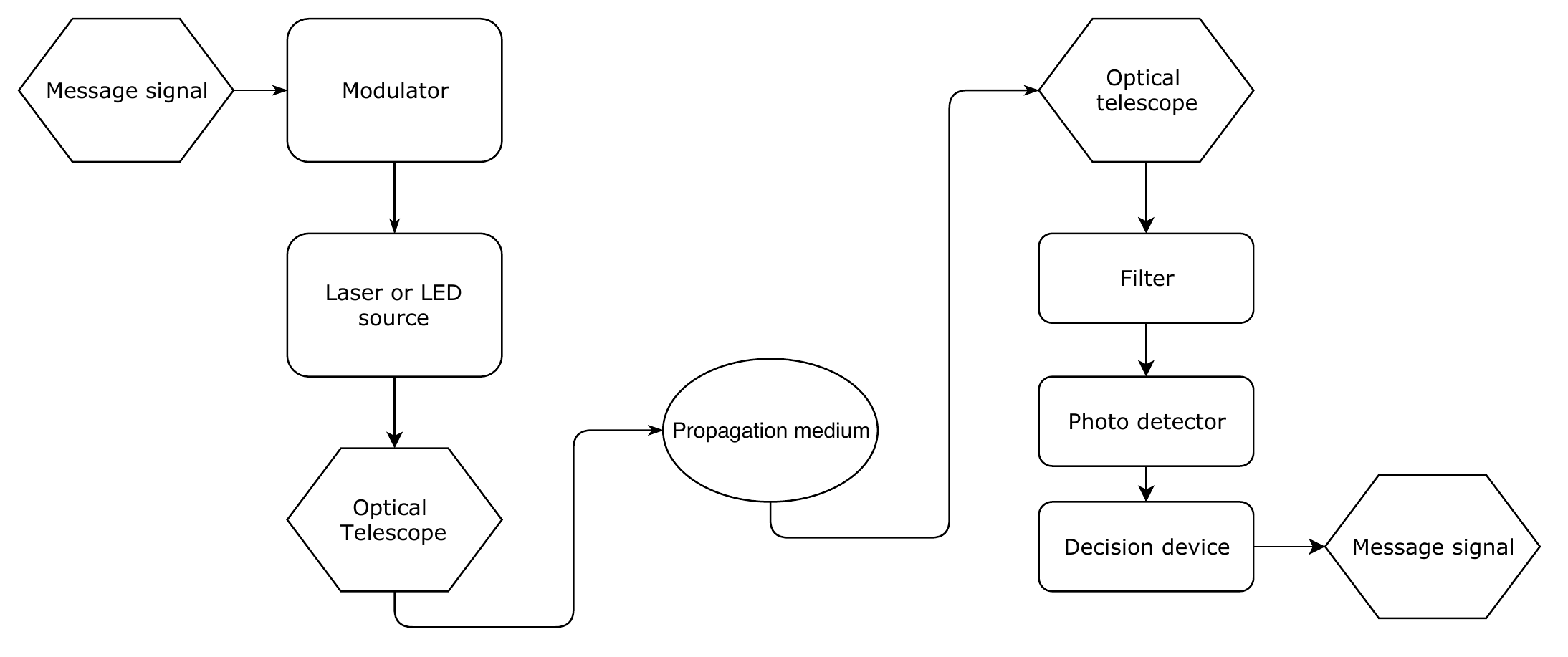}}
	\caption{Block diagram of optical wireless communication system.}
	\label{fig:OptWireSystem22}
\end{figure*}

\subsubsection{Optical Telescopes for LISNLs}

As discussed in the previous section, using RF antennas in ISNLs may create important problems as satellites are limited with size and power. For these reasons, LISNLs have attracted considerable attention from the industry. In Fig. (\ref{fig:OptWireSystem22}), the block diagram of an optical wireless system that can be used in LISNLs is drawn. This schematic view shows the components of an optical transmitter and receiver. As can be seen, the transmitter is composed of a modulator, a laser or LED source and an optical telescope. The receiver on the other hand, is comprised of an optical telescope, filter, photo detector and decision device to obtain the transmitted message signal. The most important part of a laser transmitter is the optical telescope (also known as optical antenna). An optical telescope can collect light and bring it to a point of focus so that it may be magnified and examined through an eyepiece. With this magnification, the beam's width increases and its divergence decreases equally, so that the telescope can protect the beam to spread out. The receiver telescope on the other hand, increases the area of the photo-detector which is responsible of converting light into electrical pulses. The aperture size of a telescope varies from $10$ mm to $1.5$ m depending on the communication distance and infrastructure. For LISNLs, $80$ to $135$ mm apertures can be sufficient for data reception up to $2000$ km. So far, Cassegrain telescopes have been used in LISNLs. A Cassegrain telescope is comprised of a concave parabolic main dish and a convex hyperbolic secondary dish. Cassegrain telescope is more compact than conventional telescopes and it can provide much higher gains which is very important for long-distance communications. Furthermore, new technologies enable transportable optical telescopes which are lighter and more compact. Overall, an optical telescope should provide the following features:
\begin{itemize}
    \item A laser telescope should create a thin beam width which is essential for long range communications. 
    \item The energy consumption of a laser telescope should be tolerable as the satellites are limited with power.
    \item The aperture of the telescope should be larger enough to provide reliable communications.
\end{itemize}

%
\section{The Road to Future SpaceNets: Activities and Projects}
\label{sec:Projects}
In connection with the SpaceNets we envision for the future, several activities and projects have been carried out by different organizations. In this section, we highlight the major activities of those organizations. More specifically, we start by discussing those of standardization bodies, such as the CCSDS, 3GPP, and IETF, in Section \ref{ssec:Standardization}. Then, in Section \ref{ssec:SatOps}, we provide an in-depth discussion on the four largest upcoming commercial EOSCs, namely Starlink, Kuiper, Lightspeed, and OneWeb. We discuss Blackjack and SpaceMobile EOSCs, which are being developed for military purposes and for direct broadband RF communications between EOSs and off-the-shelf mobile devices, respectively. We provide an overview of upcoming CubeSat EOSCs that are being planned for IoT applications. We examine EOS projects related to very high throughput EOSs. Also, in Section \ref{ssec:IandP}, we give a detailed review of the significant near-Earth, Lunar, and Martian projects led by the international space agencies, including the European Space Agency’s (ESA) European data relay system (EDRS), ESA’s SeCure and Laser communication Technology (ScyLight), ESA’s high throughput optical network (HydRON), ESA’s Moonlight and the National Aeronautics and Space Administration’s (NASA) LunaNet, and NASA’s Mars Cube One (MarCO). We also provide an overview of the activities under Canada’s optical satellite communications consortium (OSC) initiative.

%
\subsection{Standardization Activities}
\label{ssec:Standardization}

\subsubsection{CCSDS Activities}
\label{sssec:ccsds}


\paragraph{Overview of the CCSDS}
\label{sssec:ccsds_overview}

The \acrfull{ccsds} was founded in 1982 by the world's leading space agencies to offer a platform for the exchange of ideas on common issues in the development and management of space data systems. It now has 11 member agencies, 32 observer agencies, and more than 119 industrial collaborators. It has been actively establishing standards for data and information systems since its inception in order to foster interoperability, cross-support, cooperation, and new capabilities for future missions. The activities of the \acrshort{ccsds} are organized around six topic areas: Systems engineering (SE), Spacecraft onboard interface services (SOIS), Space link services (SLS), Cross support services (CSS), 
Space internetworking services (SIS), Mission operations and information management services (MOIMS).

\paragraph{Standards for space communication protocols}
\label{sssec:ccsds_standards}

The 130.0-G-3 informative report \cite{ccsds2014commprot} gives an overview of the \acrshort{ccsds} recommended space communications protocols. The following is a summary of the protocols categorized according to the layers of the open systems interconnection (OSI) model:

\begin{itemize}
\item \textbf{Physical layer:} 
The standard for radio frequency and modulation systems \cite{ccsds2013rfm} offer recommendations for their use in communications over space links between spacecraft and ground stations. 
In \cite{ccsds2013proxpl}, recommendations are given for the physical layer of proximity space links.

\item \textbf{Data link layer:}
The TM Space Data Link Protocol \cite{ccsds2003tm} is used in the return link for sending telemetry from a spacecraft to a ground station. 
The TC Space Data Link Protocol \cite{ccsds2010tc} is used in the forward link for sending commands from a ground station to a spacecraft. 
The AOS Space Data Link Protocol \cite{ccsds2006aos} may be used on the return link alone, or on both if higher speed communications are needed.
The Proximity-1 Space Link Protocol \cite{ccsds2013proxll} is used over short range, bi-directional, fixed, or mobile radio links, to communicate among orbiting constellations, relays, probes, landers, or rovers.

Furthermore, three standards were developed for synchronization and channel coding:
TM synchronization and channel coding \cite{ccsds2011tmscc} is used with the TM or AOS Space Data Link Protocol.
TC synchronization and channel coding \cite{ccsds2010tcscc} is used with the TC space data link.
Proximity-1 Space Link Protocol—Coding and Synchronization Layer \cite{ccsds2013proxcs} is used with Proximity-1 Space Link Protocol—Data Link Layer.

\item \textbf{Network layer:}
The Space Packet Protocol \cite{ccsds2003spp} provides the capability to transfer space application data over a path that involves a ground-to-space or a space-to-space communications link.
The Encapsulation Service \cite{ccsds2009es} enables the use of other CCSDS-recognized network protocols such as \acrshort{dtn} and Internet Protocol (IP) over space links.
IP over CCSDS \cite{ccsds2012ip} specifies how CCSDS-recognized IP datagrams are transferred over the link.
 
\item \textbf{Transport layer:} 
SCPS Transport Protocol (SCPS-TP) \cite{ccsds2006scps} supports end-to-end communications between applications. It defines extensions to transmission control protocol (TCP), incorporates  user datagram protocol (UDP) by reference, and may be used on top of the space packet, encapsulation packet, or IP over CCSDS.
The CCSDS File Delivery Protocol (CFDP) \cite{ccsds2007cfdp} provides functionality of the transport and application layers.
Transport protocols used in the Internet (such as TCP and UDP) can also be used on top of the encapsulation packet, or IP over CCSDS space links.
The \acrfull{bp} specification \cite{ccsds2015bp}, based on \cite{rfc9171}, defines procedures for forwarding data bundles through a delay-tolerant network.
The \acrshort{bp} can be used on top of other heterogeneous transport or network layer protocols and provides the ability to cope with intermittent connectivity including scheduled, predicted, opportunistic, and continuous connectivity. Furthermore, it provides custody-based retransmission and notional data accountability with built-in status reporting. The procedures for routing in space flight missions are specified in an additional document called schedure-aware bundle routing \cite{ccsds2019sabr}. The Licklider Transmission Protocol (LTP) \cite{ccsds2015ltp} provides optional reliability mechanisms on top of an underlying data link communication service.
It is intended for use over the current and envisaged packet delivery services, including packet telecommand and packet telemetry.

\item \textbf{Application layer:} 
The asynchronous messaging service (AMS) \cite{ccsds2011ams} implements a protocol under which the different mission modules or processes may be designed without explicit awareness of which other modules are currently operating nor of where they are deployed.
The CCSDS File Delivery Protocol (CFDP) \cite{ccsds2007cfdp} is a file transfer protocol, but it also provides services typically found in the transport layer (i.e, complete, in-order, without duplicate data delivery).
Lossless data compression \cite{ccsds2012ldp} promotes greater science returns and reduces the requirements for onboard memory, station contact time, and data archival volume.
Image data compression \cite{ccsds2005idc} was developed to establish a standard for a data compression algorithm applied to digital image two-dimensional spatial data from payload instruments.
Lossless multispectral and Hyperspectral image compression \cite{ccsds2012lmhic} provides a data compression algorithm applied to digital three-dimensional image data from payload instruments, such as multispectral and hyperspectral imagers.

\end{itemize}

\paragraph{Solar system Internet architecture}
\label{sssec:ssi_architecture}
The 730.1-G-1 informational report \cite{ccsds2014ssi} describes the \acrfull{ssi} as an architecture that would facilitate automated communication for space agencies and ventures through the evolution of the familiar CCSDS communication standards and the use of shared network infrastructure in more complex configurations. To ensure successful operation of the protocols and network administration, the SSI requires investments in ground systems and space assets with computational power. The return on that investment will include support for enhanced functionality in space exploration missions, including EOSNs, DSNs, and relay-enabled missions.
The full implementation of the SSI architecture is planned as a three-stage process: Mission functionality, Internetwork functionality, and Advanced functionality.

\subsubsection{3GPP Activities}
\label{sssec:3GPP}
EOSNs and terrestrial networks have always been considered two independent ecosystems  and their standardization works have been carried out independently. However, as integrating EOSNs into 5G has tremendous market potential, the satellite communication industry has shown increasing interest in participating in the 3GPP standardization effort for 5G. The 3GPP community considers satellites as a main component in NTNs, which will  complement terrestrial networks of the future. 3GPP documents define an NTN as a network where spaceborne (i.e., GEO, MEO, LEO satellites) or airborne (i.e., HAPS \cite{kurt2021vision})
node acts either as a relay node or as a base station. 

The first consideration of EOS communications in 3GPP standardization was in Release 14, where scenarios and requirements for next-generation access technologies were presented \cite{3gpp38913}. In subsequent releases (i.e.,
Releases 15, 16, 17, and 18), 3GPP considered EOS communication
networks from several aspects, such as architectures, use cases and scenarios, New Radio (\acrshort{nr}), and management and orchestration aspects.

%
\paragraph{Overview of EOS access network 3GPP standardization}
\label{sssec:Studies} 

3GPP initiated several study items (\acrshort{si}s) and work items (\acrshort{wi}s) within the Technical Specification Group (\acrshort{tsg}) of the radio access network (\acrshort{ran}), the service and system aspects (\acrshort{sa}), and the core network and terminals (\acrshort{ct}).
The goal of 3GPP  standardization activities is to support the integration of EOSNs and 5G terrestrial networks. The following is a summary of the EOSN-related 3GPP standardization
activities:
\begin{itemize}
    \item \textbf{Release 15:} In 2017, two SIs were initiated: (1) 3GPP TR 38.811 ``Study on NR to support Non Terrestrial Networks" under the RAN TSG; and (2) 3GPP TR 22.822 ``Study on using Satellite Access in 5G" under the SA TSG. The objective of the first SI was to define the NTN deployment scenarios and their system parameters, to identify the NR adaptation required to accommodate NTN, and to propose preliminary solutions to address the impacted areas of NR. Although the second SI was initiated in 2017, it was shifted to Release 16.

\item \textbf{Release 16:}  The SA TSG  initiated four activities: (1) an SI called ``Study on using satellite access in 5G" \cite{3gpp22822}; (2) an SI called ``Study on architecture aspects for using satellite access in 5G" \cite{3gpp23737}; (3) a WI called ``Integration of Satellite Access in 5G" (WI\#800010-5GSAT, Release16); and (4) an SI called ``Study on management and orchestration aspects with integrated satellite components in a 5G network" \cite{3gpp28808}. A number of use cases were introduced in the first and second SIs with the objective of supporting  services in the integrated satellite-based access components in 5G. As a result, modified or new requirements were identified in the aspects of connectivity, roaming, QoS, UE, security, and regulatory. To support the management and orchestration of the integrated EOSNs and 5G networks, some critical issues and possible solutions were presented in \cite{3gpp28808}. With respect to NR 3GPP activities, the SI ``Solutions for NR to support Non-Terrestrial Networks (NTN)" \cite{3gpp38821} was completed at the end of 2019, where several important adaptations were introduced to enable NR technologies and operations in EOSNs. This SI, was to complete the work that was started in Release 15 \cite{3gpp38811}.

\item \textbf{Release 17:}  By the end of 2019, two WIs for NTN were introduced: (i) ``Solutions for NR to support NTN" \cite{3gpp38821}, under RAN activities; and (ii) ``Integration of satellite components in the 5G architecture" \cite{5gsatarch860005}, under SA. The former was last updated on June 2021 and its activities are in the final stage. The objectives of \cite{3gpp38821} were stated as follows: (i) consolidation of the  impacts on the physical layer  and definition of potential solutions; (ii) evaluation of the performance of NR in selected deployment scenarios (LEO-based satellite access, GEO-based satellite access) through link level (radio link) and system level (cell) simulations; and (iii) identification of the potential requirements for the upper layers based on the considered architectures. The goal of the latter WI was to extend the analysis provided in \cite{3gpp23737} through the following: (i) identification of impacted areas in NR system due to  the integration of satellite components  in 5G; (ii) analysis of the issues related to the interaction between the core network and the RAN; and (iii) identification of solutions for the two highlighted use cases  (terrestrial/satellite network roaming and 5G fixed backhaul). Work on the SI ``Study on architecture aspects for using satellite access in 5G" \cite{3gpp23737} continued and was last updated in March 2021.

\item \textbf{Release 18:}  The workshop held on June 28 – July 2, 2021 was the start of 5G-Advanced (Release 18). It specified the topics of Release 18 with submissions divided into three areas: \acrfull{embb} driven work, non-eMBB driven functionality and cross-functionality for both. This 3GPP workshop focused on the radio related content of Release 18. It reviewed over 500 company and partner organization presentations to identify topics for immediate and longer-term commercial needs. The detailed discussions on how to consolidate topics into work items and study items was made before the RAN\#93-e meeting in September 2021. This meeting discussed progress on ``high-level descriptions" of the objectives for each topic \cite{rel8News}.
\end{itemize}

%
\paragraph{3GPP architectures for EOS access in 5G}
\label{sssec:Architectures} 

Two types of EOS access networks were introduced by 3GPP in TR 38.811 \cite{3gpp38811}:
\begin{itemize}
    \item \color{black}{A broadband access network serving \acrfull{vsat} that can be mounted on a moving platform (e.g. aircraft, trains, buses, ships). In this context, broadband refers to a data rate of at least 50 Mbps and of up to several hundred Mbps for downlink. The service links operate in frequency bands allocated to satellite and aerial services (fixed, mobile) above 6 GHz.}
    
    \item \color{black}{Narrow- or wide-band access network serving terminals equipped with omni- or semi-directional antennas (e.g., handheld terminal). In this context, narrow-band  refers to less than 1 or 2 Mbps of data for downlink. Typically, the service links operate in frequency bands allocated to mobile satellite services below 6 GHz.}
\end{itemize}

\color{black}{ To integrate EOS access networks in 5G, 3GPP TR 38.821 introduced the following three types of satellite-based NG-RAN architectures:}

\begin{itemize}
    \item \color{black}{\textbf{Transparent satellite-based NG-RAN architecture:} The satellite payload implements frequency conversion and a radio frequency amplifier in both uplink and downlink directions. Several transparent EOSs may be connected to the same gNB on the ground.}
\item \color{black}{\textbf{Regenerative satellite-based NG-RAN architectures:} The satellite payload regenerates the signals received from Earth. The satellite payload also provides inter-satellite links (\acrshort{isl}s). An ISL may be a radio interface or an optical interface that may be 3GPP or non-3GPP defined. The regenerative satellite-based NG-RAN architecture has two types: gNB processed payload and gNB-DU processed payload.}

\item \color{black}{ \textbf{Multi-connectivity involving satellite-based NG-RAN architecture:} This may apply to transparent EOSs as well as regenerative EOSs with gNB or gNB-DU functions on board.}
\end{itemize}


\paragraph{\color{black}{Required NR adaptation for the integration of EOS components in 5G}}
\label{sssec:Integration} 

\color{black}{Channel modeling for satellites was discussed in the technical reports 3GPP TR 38.811 \cite{3gpp38811} and 3GPP TR 38.821 \cite{3gpp38821}, where channel model parameters were provided that took different atmospheric conditions and user environments into consideration. Due to  EOS related design constraints, some areas of NR are impacted. The documents identified the following impacted areas of NR.}

\begin{itemize}
\item \color{black}{\textbf{Handover paging:} UEs are only kept within an LEO satellite beam for a few minutes due to the fast movement of LEO satellites and their many beams. This rapid change creates problems for handovers as well as for paging for both moving UEs and stationary UEs. Without a quick handover procedure, the UE may not efficiently utilize the LEO satellite resources and may experience  data loss. With fixed tracking areas on the ground, there is no one-to-one correspondence between moving beams and registration areas or fixed tracking areas, which is essential for the paging process.}

\item \color{black}{\textbf{Tracking area (\acrshort{ta}) adjustment:}  The movement of LEO satellites creates rapid  changes  in the overall distance of the radio link between the UE and BS, which leads to strong delay variations. This delay largely exceeds the \acrfull{tti} of NR, which is equal to or less than 1 ms. Hence, the TA alignment of NR needs to be modified to adapt to the introduction of satellites in 5G and to ensure that all uplink transmissions are synchronized at a gNB reception point.}

\item \color{black}{\textbf{Synchronization in downlink:} In order to access the 5G network, the UE has to detect the primary and secondary synchronization signals. Even though the SNR level of EOS systems is typically in the range of -3 to 13 dB SNR, the movement of the LEO satellite creates a higher Doppler shift, depending on the frequency band and the velocity of the LEO satellite relative to the UE.}  

\item \color{black}{\textbf{Hybrid automatic repeat request (\acrshort{harq}):} In EOSNs, the \acrfull{rtt} normally exceeds the maximum conventional HARQ timers and the maximum possible number of parallel HARQ processes. Due to the memory restrictions of some UEs, extending the number of HARQ processes in proportion to RTT might not be feasible. Also, the impact of this delay has to be considered by the gNBs on all of their active HARQ processes. Despite the increment of the number of HARQ processes in Rel. 15 to be 16 processes, EOSN NR requires further extension of the number of HARQ processes to flexibly adapt to the induced RTT delay.}

\item \color{black}{\textbf{MAC/Radio link control (\acrshort{rlc}) procedure:} For LEO satellite systems, the one-way propagation delay changes continuously (e.g., 2-7 ms for 600 km orbit). The \acrfull{arq} requires a buffering of the transmitted packets until there is the successful receipt of an acknowledgement or a time out. A larger transmission buffer is therefore required to suit the long RTT, and this limits the number of retransmissions allowed for each transmitted packet. The ARQ transmit buffer size and retransmission mechanism must be designed for the longest possible delay of an EOSs. Scheduling mechanisms must be modified to adapt to the long RTT.}

\item \color{black}{\textbf{Physical layer procedure (automatic coding and modulation [\acrshort{acm}], power control):} For EOS links, only a limited amount of power control is available. This is due to the large free-space loss and limited power available at the UE and EOS. In addition, the long delay in the loop, the power control might be efficient only in tracking slower power variations. The slow reaction time may impact the performance of the physical layer procedures which has close control loops, such as power control and ACM. However, most control loops require some adjustments in implementation, but not major modifications in design.}

\item \color{black}{\textbf{Time advance in random access response (\acrshort{rar}) message:} Time advance mechanisms ensure the synchronization of the transmissions from  UEs operating in the same cell when received by the same gNB. In the RAR message, a time advance command is provided to the UE during initial access and later to adjust the uplink transmission timing. The maximum value of the time advance command limits the maximum allowed distance between UEs and the base station, and this defines the cell size.}

\item \color{black}{\textbf{Physical random access channel (\acrshort{prach}):} Considering the long RTT impact on PRACH is important in EOSNs. For a given beam covering a cell, there is one relative propagation delay for each UE served, and one common propagation delay for all UEs served. If the common propagation delay can be compensated for, then the EOS PRACH signal design will depend on the relative propagation delay, which is designed for a maximum TA size of 200 km in current NR specifications. However, an EOS TA is thousands of kilometers in size, so this requires modifying the satellite PRACH signal and procedure design.}

\item \color{black}{\textbf{Access scheme (time division duplex [\acrshort{tdd}]/ frequency division duplex [\acrshort{fdd}]):} Although most existing EOS systems operate in the frequency bands designated for the FDD mode, the TDD mode is possible in some frequency bands. When considering the TDD mode, a guard time is required to prevent the UE from simultaneously transmitting and receiving. This guard time depends directly on the propagation delay between the UE and gNB. This guard time will directly impact the useful throughput and, by extension, spectral efficiency.}

\item  \color{black}{\textbf{Phase tracking reference signal (PT-RS):} In NR, \acrshort{ptrs} has been introduced to compensate for phase errors. The flexibility of the PT-RS configuration in NR allows user-specific configurations depending on demodulation reference signal configuration, UE RF characteristics, scheduled MCS/bandwidth, waveform, etc. PT-RS configuration flexibility is beneficial for EOSNs.}

\item \color{black}{\textbf{Peak-to-average power ratio (PAPR):} The power amplifier in a satellite payload exhibits nonlinear behavior when operating near saturation in an effort to increase power efficiency. Nonlinear distortion causes constellation warping and clustering, thus complicating signal reception. \acrshort{papr} measures the vulnerability of the transmitted signal to nonlinear distortion, where high values indicate a negative impact. Cyclic prefix – orthogonal
frequency division multiplexing (\acrshort{cpofdm}) is used in the NR downlink, which results in higher PAPR values compared with the underlying modulation in a single carrier. By increasing the backoff of the amplifier operating point, the distortion can be reduced. However, this reduces the amplifier efficiency accordingly.}

\item \color{black}{\textbf{Protocols:} Mapping is needed between the NG-RAN logical architecture and the EOSN architecture. Several mobility scenarios should be considered, specifically the mobility induced by the motion of LEO satellites, the motion of UEs from one  beam to another beam  generated by the same EOS, the motion of UEs between beams generated by different EOSs, and the motion of UEs between an EOS and cellular access. Location updating, paging, and handover RAN related protocols need to accommodate the extended delay of intra-satellite access mobility, the differential delay when mobility is between an EOSN and a cellular network, and the mobility of the cell pattern generated by LEO satellites.}

\end{itemize}
%
\paragraph{Vision}
\label{sssec:Vision} 
Although many aspects of NTN integration in terrestrial networks have been studied in 3GPP Releases 15, 16, and 17, the standardization of NR-NTN in Release 17 focused on a few aspects and left others for the subsequent releases and future studies. The major future planned work of 3GPP can be summarized as follows:
\begin{itemize}
\item \textbf{Regenerative payload:} As previously discussed, two types of EOS payload were studied in 3GPP Releases 15, 16, and 17. The first one is transparent, with the EOS acting as a repeater for the user-link signals  directing them to the ground gateways after amplification and frequency conversion. On the other hand, a regenerative EOS can operate as an onboard-gNB. That is, the user's signal can be demodulated, decoded and routed to ground gateways with or without using ISNLs. The standardization efforts in Release 17 considered EOSs with transparent payloads, whereas regenerative payloads will be considered in Release 18+.
\item \textbf{Unmanned aircraft systems (UASs):} UASs (e.g., HAPSs) were included in the 3GPP study items as a special case of EOSs with lower altitudes and delays. However, no specific analyses were undertaken for NTN using UASs \cite{3gpp38821}. Therefore, the specification in Release 17 focused on LEO/GEO satellites, and the consideration of HAS has been postponed for subsequent releases.
\item \textbf{IoT-NTN:} A study for IoT service via NTN was carried out in Release 17 in \cite{3GPPTR36-763}. In this study item, the support of narrowband IoT (NB-IoT) and enhanced machine-type communication (eMTC) technologies for IoT applications using GEO/LEO satellites was investigated. The scenarios studied involved transparent payloads, moving or fixed beam footprint on the ground, and sub-6 GHz bands. The specification of IoT-NTN was not considered in Release 17 and will be taken into consideration in Release 18+.
\item \textbf{Dual connectivity and MIMO:} Dual connectivity of ground user terminals to non-terrestrial and terrestrial networks or LEO and GEO satellites simultaneously was highlighted in Release 16 in \cite{3gpp38821}. This dual connectivity concept was not considered in the specification of Release 17. In addition, the use of MIMO techniques has not yet been investigated in NTN; however, we expect them to be included in Release 19+. 
\end{itemize}

Furthermore, some other aspects and technologies need to be considered in standardization works to fully integrate SpaceNets and terrestrial 5G/6G networks.  The most important aspects are the following:

\begin{itemize}
\item \textbf{Routing standards:} A notable characteristic of future SpaceNets will be their ability to form networks among SNs using ISLs \cite{chen2021analysis}. However, ISLs have limited lifetimes due to the frequent topology changes in SpaceNets. In addition, the high traffic loads in certain parts of SpaceNets may create congestion for some ISLs. Moreover, it is expected that EOSNs will be used by different types of applications with different QoS requirements (e.g., packet delivery ratio, packet delivery delays). Therefore, it is necessary to have robust routing schemes that can deliver data successfully while meeting the QoS requirements of each application type and adapting to the dynamic environment of EOSNs. For example, applications with high bandwidth requirements will be served well through multi-path routing, while delay-tolerant routing will be adequate for delay-tolerant applications. Thus, there is an urgent need for developing standard routing protocols that adapt to the dynamic environment of EOSNs and fulfill the various QoS requirements of different user applications. To achieve efficient routing,  aspects of network monitoring, resource allocation, and congestion
control should be taken into consideration by standardization bodies. Moreover, standards should support cross network routing (i.e., across terrestrial, aerial, EOSNs, and SpaceNets) to achieve  full integration in 6G. Interoperability among the different EOSNs and operators is crucial.

\item \textbf{SDN/NFV standards:} The SDN/NFV paradigms will play a key role in future integrated networks. To provide interoperability and compatibility  among integrated network components, different service providers, and operators, SDN-based solutions for SpaceNets
should be considered in standardization works. For instance, developing standardized onboard SDN-compatible payloads to operate on LEO satellites will provide softwareized routing functions that can adapt to the frequent changes of the LEO satellite network environment. NFV will be particularly necessary to hide the complications of integrated networks, and to reduce service and product introduction times as well as capital and operational expenditures. According to ETSI, efficient control of the NFV environment can be achieved through automation and orchestration. ETSI introduced a full set of standards to enable an open ecosystem where a Virtualized Network Function (VNF) can be interoperable with independently developed management and orchestration systems. However, the adoption of SDN/NFV concepts and technologies is still in its infancy in SpaceNets. Further research is required to identify the requirements needed to adopt SDN/NFV in SpaceNets. In addition, the support for SDN/NFV should be considered in the design of SpaceNet components.

\item \textbf{Standardized management and orchestration:} AI and ML will play a significant role in SpaceNet management and orchestration.
ETSI launched the Industry Specification Group (ISG) on experiential networked intelligence (ENI) in February 2017 \cite{ETSIENI}. ENI is an entity that provides intelligent network operation and management recommendations and/or commands to an assisted system. In another effort, 3GPP introduced the concept of SON \cite{3gpp32500}, where AI/ML can be applied to automate several network management functions. However, both  the ENI and SON concepts are still limited to the 5G context and may not be sufficiently agile in coping with the immense levels of complexity, heterogeneity, and mobility in the envisioned beyond-5G integrated networks. To support the intelligence and autonomous nature of 6G, the concept of a Self-Evolving Network (SEN) was presented in \cite{darwish2020vision} \cite{farajzadeh2021self}. A SEN utilizes AI/ML to make future integrated networks fully automated, and it intelligently evolves with respect to the provision, adaptation, optimization, and management aspects of networking, communications, computation, and infrastructure node mobility. A SEN can be adopted to support real-time decisions, seamless control, and intelligent management in SpaceNets to achieve high-level autonomous operations. Nevertheless, a SEN is quite a recent concept and has not yet been considered by standardization organizations.

\item \textbf{Standardized mobility management:}
EOSNs have the disadvantages of frequent handover and topology changes. For example, there are different types of handovers in LEO satellite networks, such as intra-satellite handovers, inter-satellite handovers, and inter-access network handovers (also known as vertical
handovers). In addition, in the 6G era EOSNs will not only serve rural or remote areas but will also provide communication services and coverage in urban and highly populated areas.
Such a scenario will lead to thousands of UTs being connected
to an LEO satellite, and this large group of users will need to go through a frequent handover process at almost the same time. Managing handovers in LEO satellite networks with existing mobility management standardized protocols may not be feasible. This is because such protocols were not
designed to deal with the high topology change rate in EOSNs, where everything is moving including the gNB (e.g., LEO satellite base station). A number of approaches have been proposed
to address this problem \cite{darwish2021location}. Nevertheless, the concept of
separating the control plane and data plane of a Software Defined
Network is a promising approach to efficiently manage EOSN and SpaceNet topology.

\end{itemize}


\subsubsection{IETF Activities}
\label{sssec:ietf}


\paragraph{Overview of the IETF}
\label{sssec:ietf_overview}
The goal of the \acrfull{ietf} is to produce high quality, relevant technical and engineering documentation that influences how people design, utilize, and manage the Internet so as to make the Internet work better. The documentation includes protocol standards, current best practices, and informative documents of various kinds \cite{alvestrand2004rfc}. The \acrshort{ietf} divides its work into the following areas: Internet, Routing, Transport, Applications and real-time, Security, Operations and management.

\paragraph{Working groups addressing satellite communication related issues}
\label{sssec:wg_satcom}
As satellite service providers have become increasingly interested in extending the Internet globally, the following working groups have emerged and contributed a variety of Internet drafts and RFC (request for comments) documents.

\begin{itemize}
\item \textbf{TCP over satellite:}
It produced informational RFCs which describe the issues that may affect the TCP throughput over satellite links, and identified domains in which each issue applies, including
link rates, network topology and satellite orbit. In addition it proposed fixes and areas for further research.

\item \textbf{Performance implications of link characteristics:}
It produced informational documents that discuss the capabilities and limitations of performance enhancing proxies (PEPs). PEPs are active network elements that modify or split
end-to-end flows in order to enhance the performance they obtain when dealing with particular link characteristics.

\item \textbf{Robust header compression:}
It developed new header compression protocols to suit the needs presented by various wireless link technologies including WCDMA, EDGE, CDMA-2000 and others.

\item \textbf{IP routing for wireless/mobile hosts:}
It developed routing support to allow IP nodes to seamlessly roam among IP subnetworks. 
It focuses on mobile IP deployment challenges and provides appropriate protocol solutions to overcome known issues and limitations.

\item \textbf{Mobile ad-hoc networks:}
Focuses on IP routing protocol capability that is appropriate for wireless routing applications in both static and dynamic topologies with higher dynamics owing to node mobility or other considerations.

\item \textbf{Delay/disruption tolerant networking (DTN):}
It defines an architecture and mechanisms for data communications in intermittently connected networks that may suffer from high delay, frequent partitions, variable error rates, and that may be comprised of more than one divergent set of protocols. DTN protocols have been the subject of extensive research and development in the delay-tolerant networking research group (DTNRG) of the \acrshort{irtf} since 2002 \cite{cerf2007delay}, and their development and revision is currently ongoing in the \acrshort{ietf}. The latest key documents are ``Bundle Protocol Version 7 (BPv7)'' \cite{rfc9171}, ``Bundle Protocol Security Specification (BPSec)'' \cite{rfc9172}, the ``TCP Convergence-Layer Protocol Version 4 (TCPCLv4)'' \cite{rfc9174}, and the ``DTN Management Architecture'' \cite{draft-ietf-dtn-dtnma-00}. Multiple independent implementations exist for these technologies in space and terrestrial environments, and the technology is being used by commercial organizations and governments around the world.

\end{itemize}

\subsection{Satellite Operator Activities}
\label{ssec:SatOps} 
 \par In this subsection, we present a discussion of the four largest upcoming commercial EOSCs, including Starlink, Kuiper, Lightspeed, and OneWeb. These EOSCs expect to provide commercial broadband services either to individual users or enterprise customers or both. For example, Starlink aims to provide broadband and communications services to residential, commercial, institutional, governmental, and professional users worldwide. Kuiper’s focus is on providing consumer and enterprise broadband services, IP transit, carrier grade Ethernet, and wireless backhaul traffic services.
\par Telesat’s Lightspeed will allow the provision of broadband services to unserved and underserved communities and individuals, provide the ability to connect any two points on Earth and deliver unique connectivity capabilities to governments and enterprises. OneWeb plans on providing services comparable to the broadband terrestrial services that are currently available in the densely populated areas of the USA.  
\par In addition, we discuss the Blackjack and SpaceMobile EOSCs. Blackjack is being developed for purely military purposes by the USA’s \acrfull{sda} with the support of the \acrfull{darpa}, while SpaceMobile is being planned by AST as the first EOSC to provide broadband services directly to off-the-shelf mobile phones/smartphones. We also discuss some upcoming CubeSat EOSCs that are being planned for IoT applications. Finally, we examine some EOS projects related to \glspl{vhts} in \acrfull{geo}.
%
\subsubsection{SpaceX's Starlink}
\label{sssec:Starlink}
{\color{black} \par SpaceX’s Starlink is currently the largest EOSC and consists of nearly 12,000 EOSs in \acrfull{leo} or \acrfull{vleo} as per SpaceX’s \acrfull{fcc} filings \cite{SpaceX:1,SpaceX:2,SpaceX:3,SpaceX:4}. It is also known as Starlink mega-EOSC since it will be composed of several sub-EOSCs (or shells) of EOSs at different altitudes. The original plan for SpaceX's Starlink LEO sub-EOSCs consisted of 4,425 EOSs in five different LEO orbits \cite{SpaceX:1}. Later SpaceX proposed a plan for three different VLEO sub-EOSCs consisting of 7,518 EOSs \cite{SpaceX:2}.}

{\color{black} \par The sub-EOSC for the Phase I of Starlink is already in the process of being deployed. The Phase I (or initial phase) of Starlink consisted of 1,600 EOSs at 1,150 km altitude in 32 orbital planes at an inclination of 53º relative to the equator. Later SpaceX proposed a modification to Phase I in its November 2018 FCC filing \cite{SpaceX:3}. As per the new plan for Phase I, its sub-EOSC now consisted of 1,584 EOSs in 24 orbital planes at a lower altitude of 550 km but at the same inclination of 53º.}

{\color{black} \par More recently, in April 2020, SpaceX proposed a modification to its plan for its five LEO sub-EOSCs \cite{SpaceX:4}. Instead of the 4,425 EOSs in its original plan, the new plan consisted of 4,408 EOSs in five different LEO sub-EOSCs at lower altitudes. It should be noted that the last two sub-EOSCs in this plan have the same altitude and inclination but differ in the number of orbital planes and EOSs per plane. After this modification, the Phase I sub-EOSC changed again, and it now consists of 1,584 EOSs in 22 orbital planes at 550 km altitude and 53º inclination. This indicates that SpaceX’s plan continues to evolve and that SpaceX is continuously modifying and optimizing its Starlink Phase I sub-EOSC as well as its other LEO sub-EOSCs. Starlink can be seen as the biggest ongoing engineering experiment in space and further modifications to it over time cannot be ruled out. Once fully deployed, the nearly 12,000 Starlink EOSs in multiple near-polar and inclined orbits will cover virtually the entire surface of the Earth and will have the ability to provide ubiquitous global coverage.}

{\color{black} \par In May 2020, SpaceX submitted another ambitious plan to the FCC for approval \cite{SpaceX:5}. This Gen2 (or second-generation) EOSC system will consist of another 30,000 EOSs in eight different LEO and VLEO orbits. This plan is also likely to change, as 7,178 EOSs in a single orbital plane does not seem to be a practical number. When combined with Gen1 (or first-generation EOSC system), SpaceX plans to deploy around 42,000 EOSs as part of its Starlink mega-EOSC.}

%
\subsubsection{Amazon's Kuiper}
\label{sssec:Kuiper} 
{\color{black} \par Amazon’s Kuiper EOSC will consist of a fleet of 3,236 EOSs in three shells (or sub-EOSCs) at different LEO altitudes and inclinations \cite{Kuiper:1}. Amazon’s deployment plan for Kuiper consists of five phases. Amazon will commence commercial operations of its Kuiper EOSC after the launch of the first phase, which will consist of 578 EOSs. When fully deployed, Kuiper is expected to provide continuous coverage throughout the 56ºN to 56ºS latitude range, which includes the USA and its territories with the exception of Alaska.
}

%
\subsubsection{Telesat's Lightspeed}
\label{sssec:Lightspeed} 
{\color{black} \par As per Telesat’s initial FCC filing in November 2016 \cite{Telesat:1}, its Lightspeed EOSC consisted of 117 EOSs in two shells, one polar and one inclined. The polar shell comprised 72 EOSs in six orbital planes at 1,000 km with 12 EOSs in each plane at an inclination of 99.5º. The inclined shell consisted of 45 EOSs in five planes at 1,248 km altitude with nine EOSs per plane at 37.4º inclination. 
\par The polar shell was meant to provide global coverage and had a concentration of EOSs in the polar regions. On the other hand, the inclined shell concentrated EOSs over equatorial and mid-latitudes, where the population density was higher and the demand for communications services was greater. By using two shells (or sub-EOSCs) in complementary orbits, Telesat’s Lightspeed is aiming to achieve global coverage.
\par In May 2020, Telesat filed an application with the FCC seeking to modify its original plan for the Lightspeed EOSC \cite{Telesat:2}. The modified Lightspeed EOSC plan consists of two phases. In Phase 1, Telesat will add 181 EOSs to its original EOSC of 117 EOSs bringing the total deployment to 298 EOSs. In the second and final phase, Telesat will add 1,373 EOSs, which will bring its total number of EOSs to 1,671.
}

%
\subsubsection{UK's OneWeb}
\label{sssec:OneWeb} 
{\color{black} \par In April 2016, the British satellite company WorldVu Satellies Limited, under the business name OneWeb, requested authorization from the FCC for Phase 1 of its LEO EOSC, which was to consist of 720 EOSs \cite{OneWeb:1} in 18 orbital planes at 1,200 km altitude and 87.9º inclination with 40 EOSs per plane. Due to its near-polar inclination, OneWeb’s EOSs could provide service to any location on Earth. In a later FCC filing in May 2020 \cite{OneWeb:2}, OneWeb proposed a modification to its EOSC that consisted of a small decrease in the number of EOSs deployed in Phase 1, from 720 to 716, and a huge increase in EOSs deployed in Phase 2, from 716 to 47,844 EOSs. 
	
\par The modified Phase 1 consisted of two shells, one polar with 588 EOSs at 87.9º inclination and one inclined with 128 EOSs at 55º inclination. The new Phase 2 comprised three shells, one polar and two inclined. The polar shell consisted of 1,764 EOSs at 87.9º inclination while the inclined shells comprised 23,040 EOSs each at different inclinations. However, like Phase 1, the altitude of the OneWeb EOSs remained the same at 1,200 km.

\par In March 2020, OneWeb filed for bankruptcy after its request for additional funding was rejected by investors. In November 2020, OneWeb emerged from bankruptcy protection after it was able to raise a \$1 billion investment from a consortium of the UK Government and India’s Bharti Enterprises. In a subsequent FCC filing in January 2021 \cite{OneWeb:3}, OneWeb declared that it was drastically reducing the size of its EOSC from 47,844 to 6,372 EOSs. This latest revised EOSC retains the same altitude and the same number of orbital planes but reduces the number of EOSs per plane in the two inclined shells from 720 to 72. The number of EOSs in the polar shell is unchanged, which reduces the total size of the OneWeb EOSC to 6,372 EOSs.

\par The four EOSCs described above (Starlink, Kuiper, Lightspeed, and OneWeb) are the four largest commercial EOSCs in low Earth orbit or very low Earth orbit. A summary of the different design parameters for these four EOSCs based on their latest FCC filings is given in Table \ref{tab:DesParamCons}. The number of shells column in this table indicates the number of sub-EOSCs within the main EOSC.}

\begin{table*}
	\centering
	\color{black}
	\caption{Design Parameters of Starlink, Kuiper, Lightspeed, and OneWeb.}
	\label{tab:DesParamCons}
	\begin{tabular}{|c|c|c|c|c|c|c|c|c|}
	\hline
	
    \multicolumn{1}{|c|}{\begin{tabular}[c]{@{}c@{}}\textbf{Final}\\\textbf{EOSC}\end{tabular}} & \multicolumn{1}{c|}{\begin{tabular}[c]{@{}c@{}}\textbf{Number}\\\textbf{of }\\\textbf{shells}\end{tabular}} & \multicolumn{1}{c|}{\begin{tabular}[c]{@{}c@{}}\textbf{Number}\\\textbf{of}\\\textbf{orbital}\\\textbf{planes}\end{tabular}} & 
	\multicolumn{1}{c|}{\begin{tabular}[c]{@{}c@{}}\textbf{EOSs}\\\textbf{per}\\\textbf{plane}\end{tabular}} & \multicolumn{1}{c|}{\begin{tabular}[c]{@{}c@{}}\textbf{Inclination}\\\textbf{(degrees)}\end{tabular}} & \multicolumn{1}{c|}{\begin{tabular}[c]{@{}c@{}}\textbf{Altitude}\\\textbf{(km)}\end{tabular}} & \multicolumn{1}{c|}{\begin{tabular}[c]{@{}c@{}}\textbf{EOSs}\\\textbf{per}\\\textbf{shell}\end{tabular}} & \multicolumn{1}{c|}{\begin{tabular}[c]{@{}c@{}}\textbf{Total }\\\textbf{EOSs}\end{tabular}} & \multicolumn{1}{c|}{\begin{tabular}[c]{@{}c@{}}\textbf{Final}\\\textbf{coverage}\end{tabular}}  \\ 	
	\hline
	\multirow{8}*{\bfseries Starlink Gen1} & 
	\multirow{8}*{8} &
	72 & 22 & 53 & 550 & 1,584 &
	\multirow{8}*{11,926} &
	\multirow{8}*{ Global}	\\
	\cline{3-7}
	& & 72	& 22	& 53.2	& 540	& 1,584 & & \\
	\cline{3-7}
	& & 36	& 20	& 70	& 570	& 720 & & \\
	\cline{3-7}
	& & 6	& 58	& 97.6	& 560	& 348 & & \\
	\cline{3-7}
	& & 4	& 43	& 97.6	& 560	& 172 & & \\
	\cline{3-7}
	& & -	& -	& 53	& 345.6	& 2,547 & & \\
	\cline{3-7}
	& & -	& -	 & 48	& 340.8	& 2,478 & & \\
	\cline{3-7}
	& & -	& -	& 42	& 335.9	& 2,493 & & \\
	\hline
	
	\multirow{8}*{\bfseries Starlink Gen2} & 
	\multirow{8}*{8} &
	1 & 7,178 & 30 & 328 & 7,178 &
	\multirow{8}*{30,000} &
	\multirow{8}*{Global}	\\
	\cline{3-7}
	& & 1	& 7,178	& 40	& 334	& 7,178 & & \\
	\cline{3-7}
	& & 1	& 7,178	& 53	& 345	& 7,178 & & \\
	\cline{3-7}
	& & 40	& 50	& 96.9	& 360	& 2,000 & & \\
	\cline{3-7}
	& & 1	& 1,998	& 75	& 373	& 1,998 & & \\
	\cline{3-7}
	& & 1	& 4,000	& 53	& 499	& 4,000 & & \\
	\cline{3-7}
	& & 12	& 12	 & 148	& 604	& 144 & & \\
	\cline{3-7}
	& & 18	& 18	& 115.7	& 614	& 324 & & \\
	\hline

	\multirow{3}*{\bfseries Kuiper} & 
	\multirow{3}*{3} &
	34 & 34 & 51.9 & 630 & 1,156 &
	\multirow{3}*{3,236} &
	\multirow{3}*{56ºN to 56ºS}	\\
	\cline{3-7}
	& & 36	& 36	& 42	& 610	& 1,296 & & \\
	\cline{3-7}
	& & 28	& 28	& 33	& 590	& 784 & & \\
	\hline

	\multirow{2}*{\bfseries Lightspeed} & 
	\multirow{2}*{2} &
	27 & 13 & 98.98 & 1,015 & 351 &
	\multirow{2}*{1,671} &
	\multirow{2}*{Global}	\\
	\cline{3-7}
	& & 40	& 33	& 50.88	& 1,325	& 1,320 & & \\
	\hline
	
	\multirow{3}*{\bfseries OneWeb} & 
	\multirow{3}*{3} &
	36 & 49 & 87.9 & 1,200 & 1,764 &
	\multirow{3}*{6,372} &
	\multirow{3}*{Global}	\\
	\cline{3-7}
	& & 32	& 72	& 40	& 1,200	& 2,304 & & \\
	\cline{3-7}
	& & 32	& 72	& 55	& 1,200	& 2,304 & & \\
	\hline		
	
\end{tabular}
\end{table*}

%
\subsubsection{USA Space Development Agency's Blackjack}
\label{sssec:Blackjack} 
{\color{black} \par To provide assured, resilient, and low-latency connectivity and communications for military data all over the globe to support a full range of warfighter platforms, the USA’s SDA with the support of DARPA is building its own transport layer in space that will consist of an EOSC comprising 300 to more than 500 EOSs with altitudes ranging from 750 km to 1,200 km \cite{Blackjack_SDA}. The Blackjack EOSC will provide the USA's Department of Defense with a global high-speed network in low Earth orbit. After the deployment of the full Blackjack EOSC, the objective is to ensure constant worldwide coverage, with 95\% of the Earth covered by at least two EOSs at any given time and 99\% covered with at least one EOS. 

\par DARPA is supporting the design and development of Blackjack. DARPA aims to exploit commercial sector advances in the design and manufacturing of LEO EOSs intended for broadband internet service. It wants to capitalize on these advances by redesigning commercial technology and manufacturing practices for building low-cost LEO EOSs for military uses \cite{Blackjack_DARPA}. The Tranche (or Phase) 0 of Blackjack will be deployed in 2022 and will consist of 20 EOSs with a limited networked capability. From Phase 1 onwards, the EOSs will have the ability to route data across a larger network. Deployed in 2024, Phase 1 of Blackjack will have 144 EOSs distributed over six orbital planes at 1,000 km altitude with near-polar inclination. With every new tranche, SDA will increase Blackjack’s size and capabilities.}
%

\subsubsection{AST’s SpaceMobile}
\label{sssec:Spacemobile} 

{\color{black} \par AST is a start-up EOS company that aims to provide cost-effective and high-speed wireless broadband services throughput the USA using off-the-shelf and unmodified UEs, such as an LTE mobile phone/smartphone. With their EOSC, they are looking to bridge the digital divide in rural and remote areas \cite{tuheen2022digital},\cite{liu2021digitaldivide}, by helping \glspl{mno} to fulfill their nationwide coverage obligations even in places that are without terrestrial infrastructure. This will be accomplished by working with MNOs to ensure that a UE without terrestrial base station coverage can still connect to an MNO’s network via EOSs in AST’s EOSC by using that MNO’s own spectrum resources \cite{ast:1}. 

\par AST expects its LEO EOSC, SpaceMobile, to begin operations in 2023. It will be able to provide services to mobile phones/smartphones using 2G, 3G, 4G, LTE, and 5G networks. SpaceMobile will consist of 243 EOSs operating in 16 orbital planes with 15 EOSs per plane except for the equatorial plane. In its initial phase, SpaceMobile will only have EOSs in the equatorial plane (i.e., at 0º inclination) and there will be 18 EOSs in this plane. The altitude of EOSs in SpaceMobile will range from 725 km to 740 km, and its 15 inclined orbital planes will have an inclination of either 40º or 55º \cite{ast:2}. Currently, AST is conducting tests with its small EOS, BlueWalker 1, and it is planning to launch another small EOS (BlueWalker 3) in 2022 for additional tests. Its gateway stations will be located both in USA and around the world.

}

%

\subsubsection{Sat-IoT Companies}
\label{sssec:IoT}

Satellite companies with large infrastructures, such as Iridium or Inmarsat, are able to provide real-time connectivity for mobile, maritime, cargo tracking, and different logistic applications in general. Although connectivity costs are relatively high, they are justified by the fact that communications are critical and require the delivery of considerable volumes of data with reduced latencies. On the other hand, there are IoT applications such as application to monitor water levels, air quality, humidity, temperature, etc., which require a few measurements per day over large areas with inexpensive sensors and low power consumption. These demands have been partially met thanks to terrestrial low power wide area networks (LPWANs) technologies such as Lora, Sigfox, and NB-IoT. However, the growing demand for such services on a global scale will call for a new type of network, specifically low power \textit{global} area Networks (LPGANs) which are based on constellations of small and inexpensive LEO satellites that are designed to meet the demands of billions of IoT devices distributed worldwide. A summary of several companies that are moving forward to cover this market is listed in Table \ref{tab:SatIoT}.

\begin{table*}[]
\centering
\color{black}
\caption{Sat-IoT Companies.}
\label{tab:SatIoT}
{%
\begin{tabular}{|c|c|c|c|}
\hline
\textbf{Company} & \textbf{Launched satellites} & \textbf{Planned satellites} & \textbf{Technology} \\ \hline
\textbf{Sateliot} \cite{sateliot}              & 1   & 100     & NB-IoT        \\ \hline
\textbf{OQ Technology} \cite{oqtechnology}         & 1   & 60     & NB-IoT        \\ \hline
\textbf{Lynk} \cite{lynk}                  & 5   & 10-5000 & -           \\ \hline
\textbf{Swarm Technologies} \cite{swarmtechnologies}    & 121 & 150     & VHF modem     \\ \hline
\textbf{Astrocast} \cite{astrocast}             & 12  & 100     & L-band module \\ \hline
\textbf{Fleet Space} \cite{fleetspace}           & 6   & 140     & LoRaWAN       \\ \hline
\textbf{Myriota} \cite{myriota}               & 2   & 50      & -           \\ \hline
\textbf{Lacuna Space} \cite{lacunaspace}          & 5   & 240     & LoRaWAN       \\ \hline
\textbf{Kepler Communications} \cite{kepler} & 15  & 140     & -           \\ \hline
\end{tabular}%
}
\end{table*}

%
\subsubsection{Very High Throughput GEO Satellites}
\label{sssec:VHTS} 
{\color{black}
A communications satellite that achieves a high capacity of tens or hundreds of Gbps is called a high throughput satellite (HTS). Although high throughput can be achieved by satellites in LEO or MEO orbits, this term has been associated with GEO satellites for a long time. Therefore, this section focuses on GEO HTS systems. GEO-based Internet services have been offered to the public since the 1990s. In 1996, Hughes Network Systems provided satellite Internet access via its DirectPC service in the United States. A GEO satellite was used for the downlink, whereas the return link was sent via a telephone modem. Later in 2001, this service was advanced to use both links via a GEO satellite. However, the users' experienced data rates were very low (lower than 2 Mbps download and 128 kbps upload) \cite{pratt2019satellite}. In the 2000s, HTS systems began to be used for Internet services in rural areas.  One of those early HTS satellites is Anik F2 \cite{AnikF2}, which was launched in 2004 by Telesat Canada to support voice, data, and broadcasting services in North America with a system capacity of 2 Gbps. In 2011, ViaSat Inc. launched the ViaSat 1 GEO satellite \cite{Viasat1} and successfully achieved a high capacity of 140 Gbps providing broadband services in North America. 

The high capacity achieved by HTS satellites is enabled by two major technologies: spot beams technology and Ka-band transponders. Traditional EOSs that provide conventional fixed satellite services (FSS) have a single broad beam for their transmission. This broad beam  uses the whole allocated frequency bands over the covered area, or footprint, which can be thousands of kilometres (i.e., a continent sometimes). However, advances in antenna design technology have enabled EOSs to transmit narrower beams that cover smaller areas and use multiple spot beams per EOS instead of a single broad one. With this technology, the allocated spectrum bands can be reused several times (depending on the number of spot beams) by utilizing frequency reuse techniques (e.g., the four-colour frequency reuse factor) to mitigate the interference between the spot beams. The other technique that enables HTS systems is the use of the Ka-band instead of the traditional C- and Ku-bands. Although Ku-bands are still used in several HTS EOSs (e.g., SES 15), the majority of the operating HTS GEO satellites use Ka-bands, which has several benefits. First, since the antenna's gain depends on its physical dimensions and operating frequency, the high frequency of the Ka-band provides a higher gain for the same antenna dimensions. In addition, the Ka-band allows for narrower beams, which means that a higher number of spot beams can be used, and it enables a higher frequency reuse accordingly. The benefits of using the Ka-band, inspired Eutelsat to name their HTS system (which provides broadband services across Europe and the Middle East) KA-SAT. Table \ref{tab:HTS} lists some examples of HTS GEO satellites, their launch time, number of beams, the frequency band, and the capacity.
\begin{table*}
	\centering
	\color{black}
	\caption{Examples of HTS GEO Satellites.}
	\label{tab:HTS}
	\begin {tabular}{|l|c|c|c|c|c|}
	\hline
	\bfseries Satellite system	& \bfseries Company & \bfseries Launch year	& \bfseries No. of beams &  \bfseries  Capacity &  \bfseries  Bands\\
	\hline
	\bfseries Anik F2	& Telesat & 	2004 &66  &2 Gbps & C-, Ku-, Ka-bands \\  
	\hline
	\bfseries WildBlue I	& Viasat  & 2007 &35  &7 Gbps & Ka-band \\
	\hline
	\bfseries KA-SAT	& Eutelsat  &2010 	&82  &90 Gbps & Ka-band \\
	\hline
	\bfseries ViaSat 1 & Viasat & 2011	& 72 &140 Gbps& Ka-band \\
	\hline
    \bfseries EchoStar 17	& Hughes & 2012	& 60 & 100 Gbps& Ka-band \\
    \hline
    \bfseries Viasat 2	& Viasat  &2017 & - & 260 Gbps& Ka-band \\
	\hline
    \bfseries Echostar 19	& Hughes &2017 	& 138 & 220 Gbps& Ka-band \\
    \hline
    \bfseries Intelsat 35e  & Intelsat  & 2017	&  - & 25-60 Gbps& C-, Ku-bands \\
	 \hline
\end{tabular}
\end{table*}
%

Next-generation HTS systems are expected to exceed $1$ Tbps in capacity. These are referred to as ultra or very high throughput satellite (VHTS) systems. The main bottleneck to increasing the capacity to such an extent is the feeder link. The use of radio-frequency in the feeder link limits the available bandwidth (uplink and downlink) to 1, 6, or 8 GHz for Ku-, Ka-, Q/V-bands, respectively \cite{calvo2019optical}. This means that a significant number of ground gateways are required to achieve Tbps capacity. For instance, more than 50 ground gateways are required to achieve a capacity of $500$ Gbps in the Ka-band \cite{mody2017operator}. Therefore, using free-space optical (FSO) communication in the feeder links has received great attention for realizing VHTS. In the optical bands, around 10 THz bandwidth is available. Accordingly, utilizing techniques such as wavelength division multiplexing (WDM), the Tbps capacity can be achieved. To this end, a demonstration of an FSO-based link between a ground optical station and a GEO satellite was carried out by the German Aerospace Center (DLR) resulting in a capacity of $1.72$ Tbps \cite{poliak2018demonstration}.
}
%
\subsection{National Initiatives and Projects}
\label{ssec:IandP} 

{\color{black} \par In this subsection, significant near-Earth, lunar, and martian projects led by the \acrfull{esa} and the \acrfull{nasa} are briefly reviewed. These include ESA’s \acrfull{edrs}, \acrfull{scylight}, \acrfull{hydron}, and Moonlight. We also briefly review NASA’s LunaNet and \acrfull{marco}. An overview of the activities undertaken by Canada’s \acrfull{osc} is also presented.
	
\par ESA’s EDRS geostationary EOSs use laser inter-space node (or inter-satellite) links to provide data relay services to LEO EOSs, its ScyLight program fosters the development of optical technologies for EOS communications, its HydRON project aims to build a “Fiber in the Sky” network by realizing a Tbps all-optical transport EOSN in space, and its Moonlight initiative will place a \acrfull{losc} of \glspl{los} around the moon to provide communication and navigation services for the moon. NASA’s LunaNet is envisioned as a services network to enable lunar operations while its MarCO mission demonstrated the concept of “carrying your own relay” by placing a CubeSat in Mars orbit for collecting data from its Mars lander to relay to Earth.
}

%
\subsubsection{ESA's EDRS}
\label{sssec:EDRS} 

{\color{black} \par The European Data Relay System uses geostationary EOSs to provide optical and microwave data relay services between LEO EOSs and ground terminals \cite{poncet2017edrs}. The system consists of two geostationary nodes, EDRS-A and EDRS-C (Fig. \ref{fig:EDRS}), over Europe at 9º East and 31º East and is a public-private partnership program between ESA and Airbus Defense and Space Germany \cite{rossmanith2017dlr}. While EDRS-C adds redundancy, there is a plan to add a third geostationary node, EDRS-D, over the Asia-Pacific region for the globalization of the EDRS \cite{hauschildt2017edrs}.
	
\par The first node, EDRS-A, became operational in November 2016, and it is a payload, which is hosted by the Eutelsat 9B EOS. Although the functions of the EDRS-A payload include the provision of optical and Ka band \glspl{isnl} as well as Ka band feeder links with ground terminals, the main focus is on a \acrfull{lct} for \acrfull{oisnl} (or \acrfull{lisnl}). The duration of an LISNL established with an LEO EOS to acquire and relay its data is in the order of minutes. The second node of the system, EDRS-C, was successfully deployed in 2019 \cite{calzolaio2020edrs}. The LCT for EDRS-A and EDRS-C was manufactured by Tesat and can provide data rates of up to 1.8 Gbps with the ability to switch between different LEO EOSs equipped with LCTs.

\begin{figure*}[htbp]
	\centerline{\includegraphics[scale=0.28]{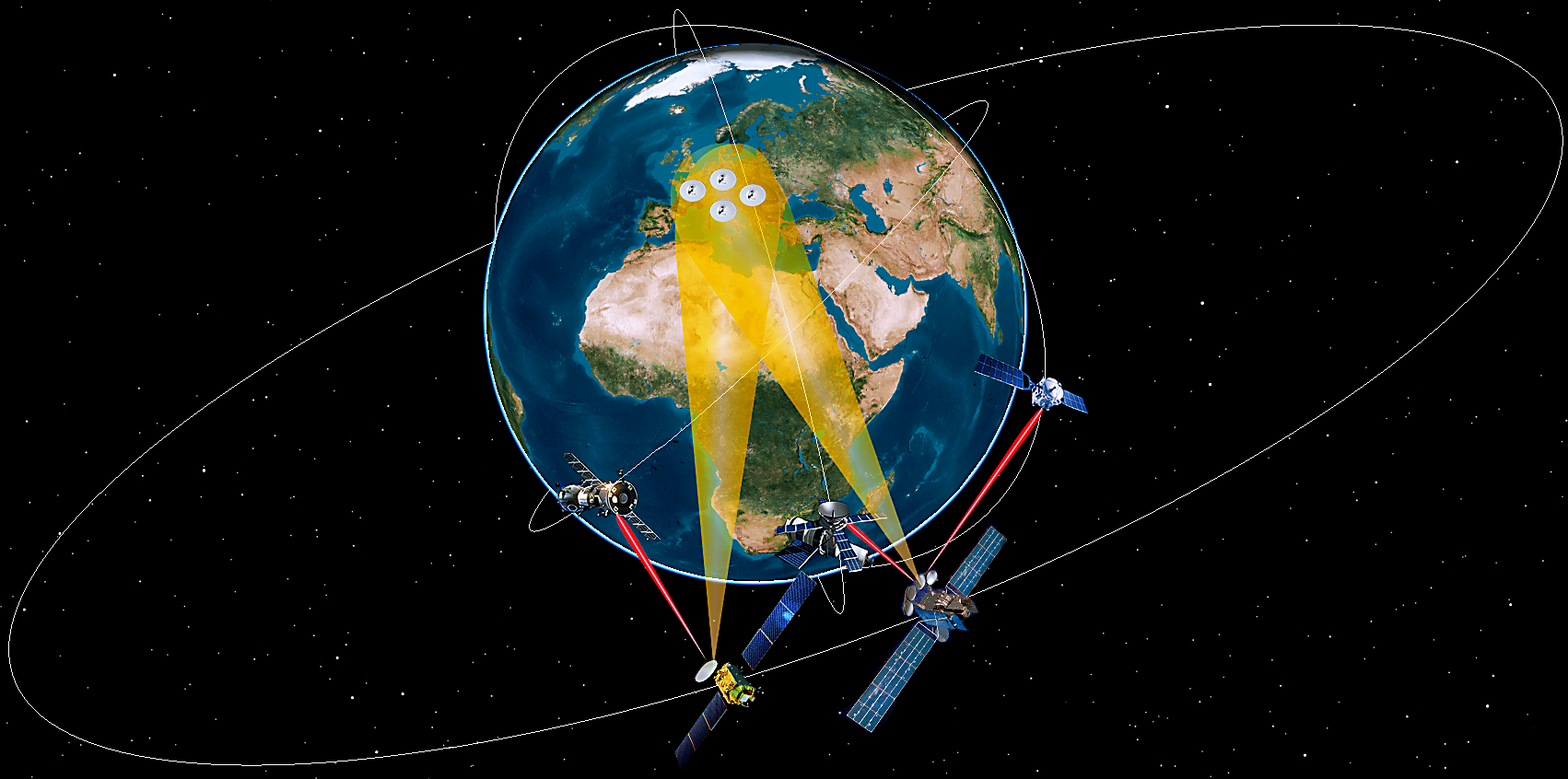}}
	\caption{EDRS-A and EDRS-C over Europe providing data relay services to LEO EOSs adapted from \cite{rossmanith2017dlr}.}
	\label{fig:EDRS}
\end{figure*}

} 
%
\subsubsection{ESA's ScyLight}
\label{sssec:ScyLight} 

{\color{black} \par In December 2016, ESA launched a new program called ScyLight (or SeCure and Laser communication Technology) and pronounced “SkyLight” to support the development of innovative optical technologies for EOS communications \cite{hauschildt2017scylight}. The objectives of the ScyLight program can be summarized as follows:
\begin{itemize}
	\item To address the development and usage of innovative optical technologies for EOS communications and assist industry in developing new market opportunities for these technologies;
	\item To demonstrate these technologies to the end user community via in-orbit demonstration; and
	\item To support industry in developing capabilities and competitiveness in this field.
\end{itemize}
	
\par To achieve these objectives, the ScyLight program concentrates European and Canadian research and development endeavors on optical communication technologies in the following three areas: optical communication terminal technology; intra-EOS optical payloads; and quantum cryptography technologies. The activities in these areas will support innovative developments in the following fields \cite{hauschildt2019scylightandhydron}:
\begin{itemize}
	\item Optical inter-space node (or inter-satellite) links;
	\item Optical feeder links;
	\item Optical user ground stations;
	\item Optical airborne-to-space/ground links;
	\item Extremely high bandwidth equipment;
	\item Optical fiber at spacecraft level; 
	\item Technologies resulting in equipment of low \acrfull{swap}; and
	\item In-orbit quantum key generation and distribution systems.
\end{itemize}

}
\subsubsection{ESA's HydRON}
\label{sssec:HydRON}

{\color{black} \par To support the development of optical communication technologies and to provide more opportunities for European and Canadian industry to test, demonstrate, and prove their technologies in orbit, ESA has initiated a new project called HydRON or High thRoughput Optical Network \cite{hauschildt2019scylightandhydron}. The aim of the HydRON program is to achieve a “Fiber in the Sky” network via a Tbps all-optical transport EOSN in space and its integration into the terrestrial high capacity fiber-based network infrastructure \cite{hauschildt2020hydron}. HydRON will employ all-optical payloads interconnected via optical inter-space node (or inter-satellite) links in the Tbps regime to realize a true “Fiber in the Sky” network.
	
\par An illustration of the HydRON system is shown in Fig. \ref{fig:HydRON} and is envisioned to have the following functionalities\cite{hauschildt2020hydron}:
\begin{itemize}
	\item Reliable optical feeder links;
	\item Bidirectional Tbps optical inter-space node links;
	\item Onboard high-speed transparent optical switching and high-speed regenerative electrical switching;
	\item Interface compatibility with different types of customer payloads;
	\item Flexible traffic distribution/collection to/from customer (RF/optical) payloads;
	\item Seamless integration of space nodes into terrestrial fiber-based networks; and
	\item Network optimization techniques based on AI. 
\end{itemize}
It is not within the scope of the HydRON program to implement the entire HydRON system. A subset of the key HydRON elements will be selected and implemented as individual HydRON demonstration missions as part of the \acrfull{hydron ds}. The HydRON DS will focus on validating the main HydRON system concepts and the end-to-end system functionalities.

\begin{figure*}[htbp]
	\centerline{\includegraphics[scale=0.34]{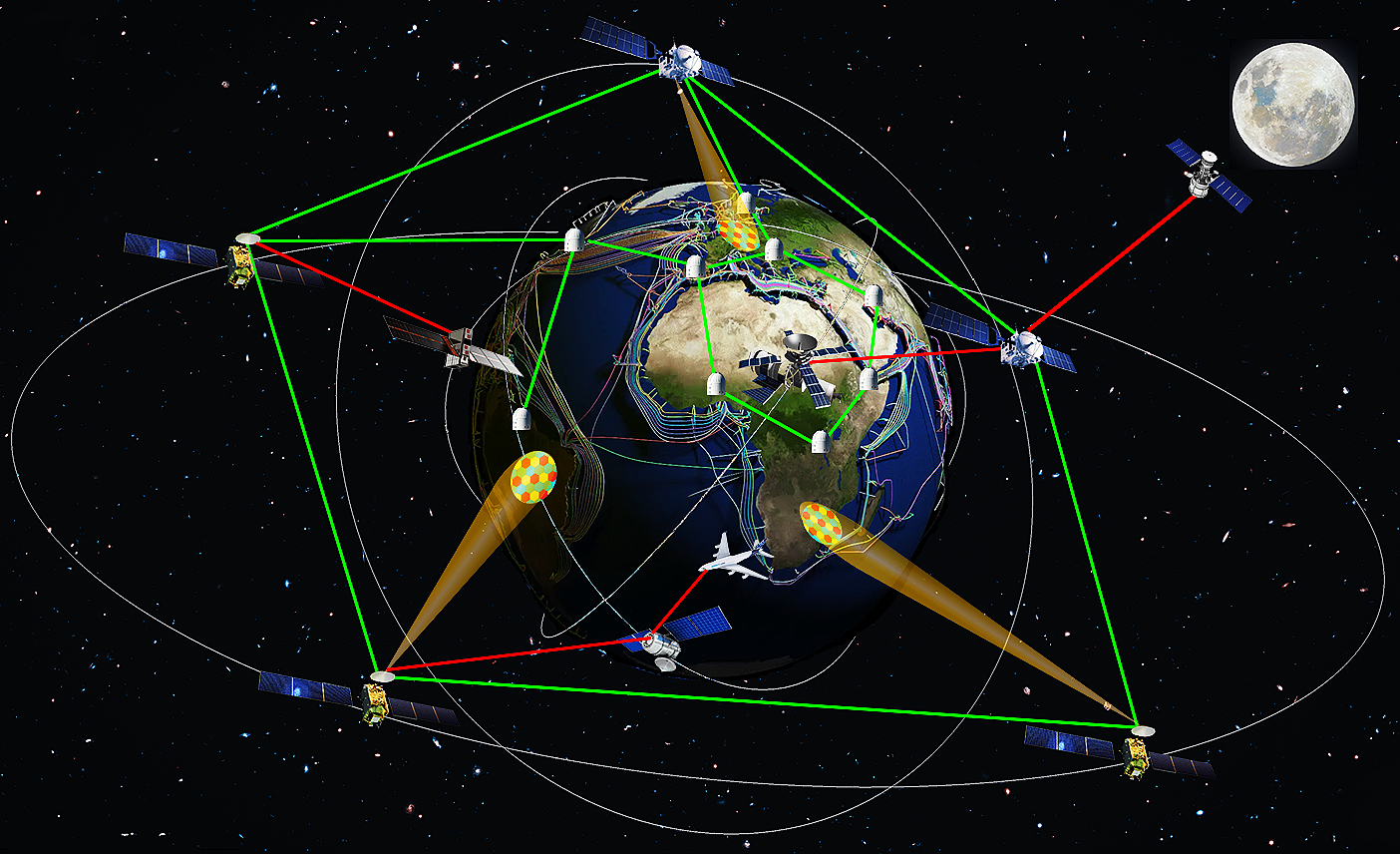}}
	\caption{HydRON – An all optical space EOSN integrated into a terrestrial fiber-based network adapted from \cite{hauschildt2020hydron}.}
	\label{fig:HydRON}
\end{figure*}

}
%
\subsubsection{ESA's Moonlight}
\label{sssec:Moonlight} 

{\color{black} \par Over the next decade, several programs are being planned by space agencies for the exploration of the moon. For example, NASA’s Artemis program is aiming to return humans to the lunar surface by 2024. In cooperation with other partners, including ESA, NASA is also planning to deploy a Gateway in lunar orbit. This Gateway will have living quarters and it will be home to astronauts from all around the world. ESA is working on a European Large Logistics Lander for different types of uncrewed missions. These ambitious plans need reliable navigation and communication to succeed. As part of its Moonlight initiative, ESA is analyzing the planned lunar missions to provide communication and navigation services for the moon by putting an LOSC of LOSs around the moon. This LOSC will allow lunar missions to communicate with Earth even when they land on the side of moon that has no direct visibility from Earth \cite{ESA_moonlight}.

\par A possible lunar navigation satellite system was studied by the authors in \cite{schonfeldt2020moonlight}. It is not easy to find stable orbits around the moon due to the irregular nature of its gravitational field. The following lunar orbits were considered in this study:
\begin{itemize}
	\item elliptical lunar frozen orbit (ELFO) to cover the South Pole area and another \acrshort{elfo} (termed as ELFO symmetric) for coverage of the North Pole area of the moon;
	\item \acrfull{nrho} for almost constant visibility from Earth and almost constant coverage of the lunar South Pole;
	\item distant retrograde orbit (DRO) for covering the moon’s equatorial regions; and
	\item \acrfull{lco} for good pole coverage. 
\end{itemize}
Different combinations of these orbits containing different number of LOSs were compared in terms of the quality of their global lunar coverage. It was shown that an LOSC combination of four LOSs in ELFO, four in ELFO symmetric, one in NRHO and three in \acrshort{dro} provided a better percentage of time a certain latitude was covered by three or more LOSs.

}
%
\subsubsection{NASA's LunaNet}
\label{sssec:LunaNet} 

{\color{black} \par LunaNet is a services network envisioned by NASA to enable lunar operations. Its architecture consists of building blocks called nodes. A node is a network access point for lunar orbital and surface users. The following are the three standard services provided by LunaNet \cite{israel2020lunanet}:
	\begin{itemize}
		\item Networking services;
		\item Position, navigation, and timing services; and
		\item Science utilization services. 
	\end{itemize}
LunaNet will have a flexible architecture with a variety of topology implementations and its infrastructure can have infinite instantiations. There can be multiple ways for a lunar surface user to communicate with Earth. 

\par LunaNet will provide its standard services to users during every stage of a lunar mission, including cruising to and from the moon, orbiting the moon, lunar descents, landings, and ascents as well as fixed and mobile surface operations \cite{NASA_lunanet}. It will set the precedent for MarsNet (i.e., a network of satellites around Mars and ground nodes on the surface of Mars) and an inter-planetary network. Imagine that the year is 2050. We have established a colony on Mars, and we have a base on the moon for re-fueling of missions to Mars and elsewhere. On the moon, astronauts need to stay connected to be able to communicate in real time. There is a need for a network that spans the entire solar system, and the first step to achieve this can be LunaNet.

}
%
\subsubsection{NASA's MarCO}
\label{sssec:MarCO} 

{\color{black} \par Entry, descent, and landing (EDL) is considered the riskiest of all phases of a space mission that aims to land on a planetary surface for scientific exploration. To closely monitor future Mars lander missions during their \acrshort{edl} phase and afterwards, the concept of CubeSat satellites around Mars to provide communication relay services to Earth was conceived by NASA in 2016. The mission was called Mars Cube One (or MarCO). It consisted of launching multiple identical CubeSats with specific communications capabilities on the InSight launch rocket. Two such CubeSats are envisioned in orbit above Mars in Fig. \ref{fig:MarCO} \cite{asmar2016marco}.  

\par Since the size of these SNs prevented component redundancy, SN redundancy was chosen. Initially, four such SNs were proposed but later it was decided to build and launch only two identical MarCO CubeSats, MarCO-A and MarCO-B. In May 2018, these two MarCO CubeSats were launched with the InSight mission. These were the first interplanetary CubeSats and served as a demonstration of the “carry-your-own-relay” concept for future missions. One of the CubeSats was sufficient to collect transmitted data from the InSight Mars lander and relay it back to the \acrfull{dsn} on Earth while the other was meant for redundancy. The MarCO mission also aimed at demonstrating the capability of a CubeSat sized, DSN compatible, deep space transponder \cite{klesh2018marco}.

\begin{figure*}[htbp]
	\centerline{\includegraphics[scale=0.37]{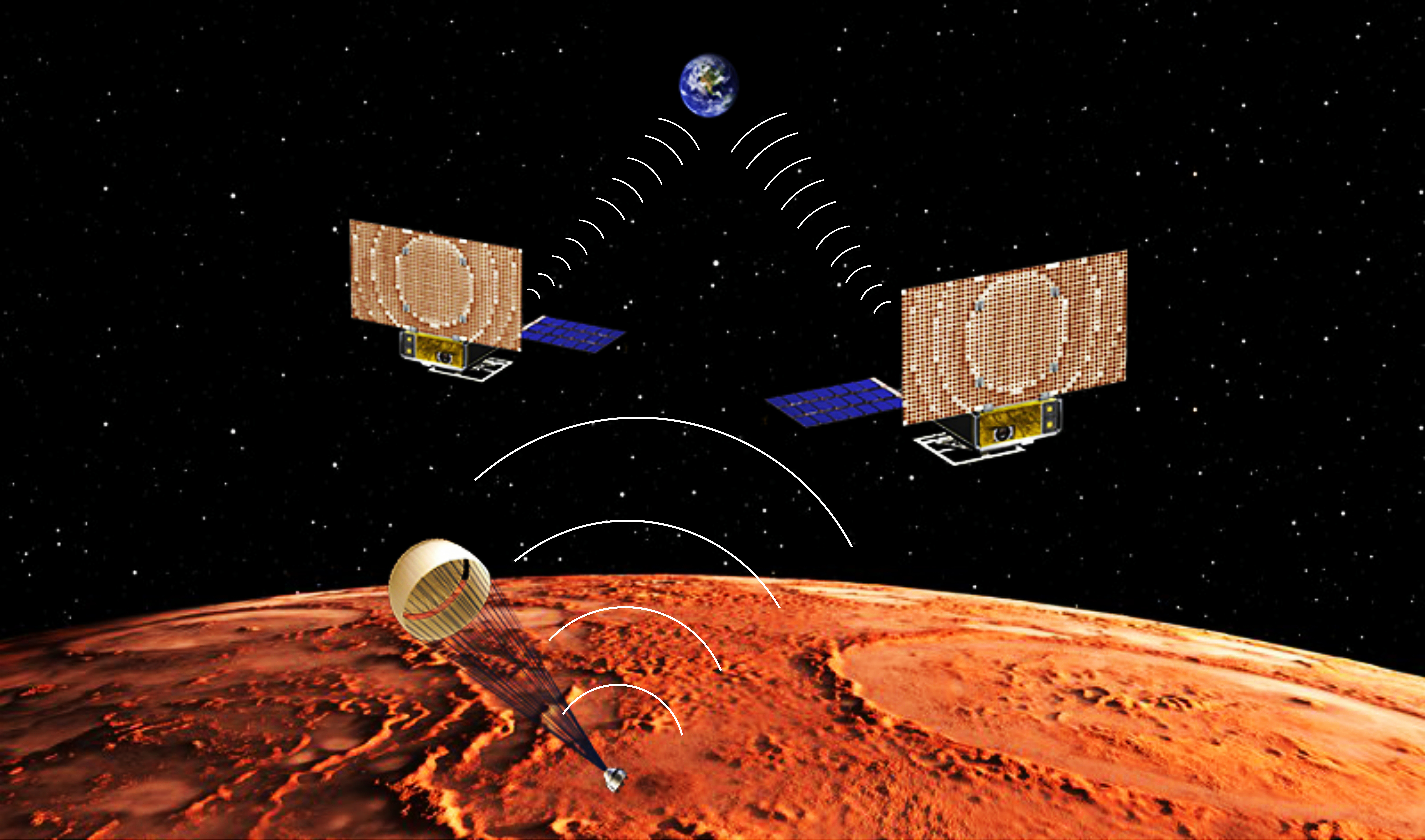}}
	\caption{MarCO-A and MarCO-B CubeSats envisioned in orbit above Mars adapted from \cite{asmar2016marco}.}
	\label{fig:MarCO}
\end{figure*}

}	
%
\subsubsection{Canada's OSC}
\label{sssec:OSC} 
{\color{black}
Due to its geography, around $20 \%$ of Canada's population lives in rural or remote areas, according to \cite{StatCanCatalog91003x}. In 2019, the Canadian Radio-television and Telecommunications Commission (CRTC) reported that only $45.6\%$ of rural households have access to broadband Internet, which the CRTC defines as 50 Mbps download and 10 Mbps upload speeds with unlimited data \cite{CRTCReport}. Therefore, the National Research Council of Canada (NRC) initiated a program called the High-Throughput and Secure Networks Challenge (HTSN) to improve broadband Internet connectivity in remote and rural areas in Canada. HTSN is aiming to achieve $1$ Gbps everywhere, which goes beyond the CRTC's objective of 50/10 Mbps. In 2019, HTSN formed the Optical SatCom Consortium (OSC) after discussions with the  Satellite Canada Innovation Network (SatCan) to pursue that target \cite{SatCanOSCWebsite}.

OSC is a member-based group composed of 14 founding corporations, institutions, organizations, and small- and medium-sized enterprises (SMEs) that NRC's HTSN leads. The main goal of OSC is to share ideas and research for developing technologies that enable network operators and service providers to offer affordable and secure broadband Internet services in remote and rural parts in Canada via optical satellite communications. Towards that end, several research and development projects have been implemented, and a road map is being established for Canada's future in optical EOS communications.
}
%
\section{Communications and Networking Challenges and Research Directions}
\label{sec:ResearchDirections}
To realize the envisioned SpaceNets, several challenges need to be addressed. In this section, we discuss the significant challenges, potential solutions proposed in the literature, and the research directions from communications and networking perspectives. Section \ref{ssec:MobilityManagement} discusses the mobility management of SpaceNets as well as the challenges and potential solutions of the two types of mobility management (i.e., handover management and location management). Section \ref{ssec:space-terrestrial-routing} provides an overview of routing mechanisms that emerged in the delay/disruption tolerant networking paradigm and that are intended to integrate terrestrial networks and SpaceNets. In Section \ref{ssec:LaserISLs}, we present a detailed discussion of the emergence of \glspl{lisnl}, the classification of LISNLs, the capabilities of currently available LCTs for establishing LISNLs, and different types of delays that exist in FSO satellite networks (FSOSNs) along with recent works that have attempted to address the issue of latency in FSOSNs. Section \ref{ssec:CrossLayerDesign} investigates an important aspect of SpaceNets, namely cross-layer design. In this regard, we highlight the motivation for this type of design, its major types, as well as representative works in the literature on this as it pertains to SpaceNets. Section \ref{sssec:PLS} presents a discussion of the importance of physical layer security in SpaceNets, which has recently emerged as an effective security approach.
%
\subsection{Mobility Management}
\label{ssec:MobilityManagement} 
{\color{black}
Due to the high mobility of SNs, SNTs need to switch between SNs frequently to remain connected to a SpaceNet. The process of switching between SNs is called a handover process. Besides, the logical location of the users should be updated, and the new data arrivals should be directed to the new location. Therefore, SN mobility management consists of two main components: handover management and location management. In this section, we discuss the major challenges associated with mobility management in future SpaceNets and the proposed solutions in the literature.
\subsubsection{Handover Management}
\label{sssec:HandoverManagement}
Handover management is required at both link-layer and network-layer levels. At the link-layer, the handover process is responsible for switching the communication link between the SNT and the SpaceNet by connecting the SNT to a different SN when the serving one is out of the SNT's visibility (communication range). On the other hand, a network-layer handover is required to switch higher-layer protocols (e.g., TCP and UDP) to the new IP address of the SNT when it is connected to a different home network during the SN link handover. 

\paragraph{Benefits of handover management in SpaceNets} There are several benefits to optimizing the handover process in SpaceNets. Efficient handover management improves the performance of a SpaceNet by reducing the signaling overhead on the SN-to-SNT and SN-to-controller links during the handover process. Besides, it mitigates the throughput losses due to frequent switching between the SNs. Due to the major challenge of long propagation delay associated with SN communications, the processing delays of space network management should be minimized. In this regard, efficient handover management reduces the time required to process a large number of handovers. Moreover, it reduces IP addressing issues, such as the data tunneling and forwarding required to direct the data packets to the new IP addresses of the user terminals. Therefore, the optimization of the handover process is crucial for an efficient and resilient SpaceNet.

\paragraph{Challenges of handover management in SpaceNets} Compared to terrestrial networks, handover management in SpaceNets faces several challenges. The major ones can be summarized as follows:
\begin{itemize}
\item \textbf{ Frequent handovers:} The high speed of the SNs relative to SNTs in most SpaceNet scenarios (especially LEO satellites serving ground and air UTs) results in frequent and unavoidable handover for all connected SNTs. Based on the analysis of 3GPP in \cite{3gpp38821}, the frequency of LEO satellite handovers can be similar to that of terrestrial users on high-speed trains. This directly impacts the QoS of all users, regardless of their speed, due to the possibility of losing the connection during a handover, the delay in processing handovers, and the additional signaling that consumes energy from the devices. In addition, SpaceNet performance is affected by frequent handovers due to the associated throughput losses, signaling, and processing.
\item \textbf{Latency of signaling}: Similar to the 3GPP specifications \cite{3GPPTR36-881}, the service interruption time due to a handover can be defined as the time difference between the times of end of transmission to the old serving node and the start of transmission to the target one. Based on this definition, and without considering other sources of processing delay, the interruption time can be 2 round-trip time (RTT) for the downlink and 1.5 RTT for the uplink \cite{3gpp38821} in EOSNs. Therefore, given the long propagation delay in space communications, this latency of handover signaling can significantly impact the network and user satisfaction.
\item \textbf{Lack of fixed anchors:} Due to the mobility of SNs (e.g., LEO satellites) and the long delay required to exchange the control signaling with ground stations, there are no fixed nodes to handle the handovers. This makes the process more complicated than terrestrial networks, which have fixed anchors capable of processing the handovers.
\item \textbf{Handover for a large number of SNTs:} In future SpaceNets, there will be a need to process a large number of handovers at the same time (i.e., group handover). This challenge is associated with high-load network segments (i.e., in the link between EOSs and ground and air SNTs). This is due to the mobility of the access points (e.g., LEO satellites), not the terminals only as in terrestrial networks. Therefore, a massive number of SNTs will need to be switched to another EOS simultaneously. This will require a large number of binding updates, new links, and signaling overhead. Accordingly, conventional fast mobility management protocols (e.g., fast handovers for Mobile IPv6 (FMIPv6) \cite{IETFRFC4068FMIPv6}), which are mainly designed for terrestrial networks, are not suitable for SpaceNets. 
\item \textbf{Target SN:} Selecting the handover target when multiple SNs are available is not straightforward, as this requires processing different SN parameters (e.g., channel conditions, visibility period, and load) to optimize the association between terminals and SNs.
\end{itemize}

\paragraph{Potential solutions for handover management in SpaceNets} Here, we discuss the state-of-the-art proposed solutions in the literature to tackle the aforementioned challenges of handover management in SpaceNets with a focus on LEO satellite networks. To reduce the handover rate, the authors in \cite{li2020forecast} proposed an architecture on the basis of several layers of terrestrial relays, HAPSs, LEOs, and GEOs for relay purposes. Then, the handover strategies among these systems were investigated, along with other aspects, such as spectrum and power allocation and user association. This is based on utility functions that depend on the throughput, dropping probability, minimum rate, and delay. However, this requires interoperability and coordination among these different systems. In \cite{wu2019satellite}, the authors studied a software-defined networking (SDN) architecture that used GEOs to connect the LEOs to the controller on the ground. In doing so, they represented the LEO satellites and ground terminals as a bipartite graph and proposed a game theory-based strategy to maximize the benefits of the terminals. Nevertheless, incorporating the GEOs in the loop entails a high handover processing delay due to the long propagation delay associated with GEO communication links. 

The authors in \cite{wu2016graph} proposed a graph-based handover strategy by considering the LEO satellite covering period as the node and the possible handover as the edge. Therefore, the handover strategy involved finding the best path in the directed graph. For this purpose, several handover criteria could be used to adjust the weights of the edges and to find the best path. In \cite{yang2016seamless}, the authors drew attention to the benefits of using SDN in EOSNs. In this regard, they proposed an SDN-based architecture for EOSNs. More specifically, they introduced a seamless handover scheme based on the proposed SDN architecture and evaluated its performance. The proposed SDN architecture composed of a centralized controller on the ground that controlled the resource allocation, handover, and data routing, which was used to separate the control and data planes. The controller was connected to a location server that stored tables for user locations (local address and associated LEO gateway). The LEOs were connected to the controller via GEOs using satellite network OpenFlow (SNOF) channels.

The authors in \cite{musumpuka2016performance} proposed an analytical queuing Markovian model for the handover and new call services in inter- and intra-LEO mobile satellite systems. They exploited the correlation between the service times in the adjacent spot beams resulting from the fact that the channel characteristics do not change dramatically from one spot beam to an adjacent one. The handover and new services blocking probabilities were derived using the moments of the probability generating functions of the correlated queue service times. In \cite{demir2022performance}, the authors evaluated the performance of LEO satellite handover on the basis of different metrics, such as measurements, elevation angle, distance, and timer-based techniques. Adopting the 3GPP model, the simulations show that the measurement-based handover outperforms other methods if optimized hysteresis/offset margins and time-to-trigger parameters are used.

On another front, MIMO techniques were employed to address the challenges associated with handover management in EOSNs. The authors in \cite{feng2020satellite} adopted a bipartite graph model for the ground gateway stations and the LEO satellites in their field of visibility, supposing that accurate information about the satellites was available at the gateways. Accordingly, the ground gateways could be connected to multiple LEO satellites in a MIMO manner when the maximum matching of this graph was determined. In \cite{abdelsadek2021ultra}, the authors utilized cell-free massive MIMO (CF-mMIMO) architecture to jointly optimize the power allocation and handover management processes in LEO satellite networks. Based on this, the handover rate was reduced while maximizing the network throughput. This approach achieved a substantial improvement in handover management, as the visibility of clusters of LEO satellites is longer than that of single satellites.  
}
{\color{black} \subsubsection{Location Management}
\label{sssec:LocationManagement}

According to the IETF IP-based mobility management protocols, location management has two components: (1) location updating, which is the process of identifying and updating the logical location of the MN in the network; and (2) data delivery (i.e., routing), which forwards the data packets directed to the MN to their new location. Due to the differences between terrestrial networks and SpaceNets in terms of topology, processing power, and communication links, the application of standard IP mobility management protocols, and more specifically their location management techniques, to SpaceNets has some drawbacks \cite{Han2016b}. IETF’s IP-based location management techniques were
designed to manage the logical location of MNs (terminals)
and deliver their data to wherever they move. However, in
SpaceNets, both terminals and BSs (satellites) are moving,
which creates new challenges that cannot be fully addressed using existing IETF’s location management techniques. In addition, IETF location management techniques are intended to work in centralized units that manage both control and data traffic (i.e., routing) \cite{Cordova2019}. As a result, IETF location
management techniques have poor scalability and may create
processing overload in core network devices. Moreover, even in terrestrial networks, these mobility management standards present several problems because of their low granularity mobility management and suboptimized routing. What makes things more challenging is the characteristics of future SpaceNets, such as the very frequent and rapid topology changes due to the high speed of LEO satellites, the very dense deployment of LEO satellites in the form of a network of megaconstellations, and the complete
integration between aerial, terrestrial, and even deep space
networks. In addition, future EOSNs will be utilized
in highly populated areas, where thousands or millions of
heterogeneous user devices can communicate directly with LEO satellites (without going through a gateway). Hence, future EOSNs will create unprecedented mobility scenarios that will require innovative solutions. For a detailed discussion of IETF IP-based standardized location management protocols and their limitations in future LEO satellite networks, and in EOSNs in general, the reader may refer to \cite{darwish2021location}.

To overcome the limitations of IETF IP-based location
management, three approaches have been adopted by works on EOS location management.

\begin{itemize}
    \item \textbf{IETF location management technique extensions for EOSNs:} This approach attempts to enhance or extend IETF IP-based location management techniques \cite{Zhang2019}, \cite{Dai2020}. The IETF IPv6 mobility management standards (e.g., MIPv6, PMIPv6, FMIPv6, HMIPv6) addressed the location management issue in terrestrial networks. Although some works have attempted to employ the location management techniques of IPv6 mobility management standards \cite{He2016}, \cite{Han2018}, \cite{Jaff2014}, such techniques have many limitations when applied to EOSNs. To enhance the performance of the IETF location management techniques, a number of extensions were proposed for EOSNs location management. These solutions are either distributed or centralized. The distributed IETF location management techniques’ extensions can be either anchor-based or anchorless. In anchor-based location management, the responsibilities of location management are permanently assigned to certain network entities (e.g., gateways) \cite{Shahriar2008}, \cite{Zhang2012}, \cite{Han2016}. In contrast, the anchorless location management approach shifts the location management role from one network entity (e.g., satellite) to another based on network topology changes \cite{Zhang2019} \cite{Dai2020}.

\item \textbf{Locator/identifier split in EOSNs:} Current EOSN architectures use IP addresses
as both identifiers (identify who is the endpoint) and
locators (to locate the endpoint in the routing system). Thus, the IP address has to be changed based on changes in the SN location in the network. In IP networks, mobility support depends heavily on the network topology that has static anchor nodes. Therefore, IP mobility solutions are considered impractical when applied to future EOSNs or SpaceNets, which are very dynamic and have pressing requirements for high scalability and tight time constraints \cite{Han2016b}.  IP-based mobility management consists of two main procedures: location management and handover management. However, in the location/identity split approach, mobility management is achieved through two correlated steps, namely
location (binding) update and location resolution \cite{Zhang2017}. With locator/identifier splitting, a node in the network can be identified using a unique identifier regardless of its location in the network. Thus, with locator/identifier separation, it is possible to keep ongoing communication continuous since the identifier does not change \cite{Wang2012}. When a node changes its location it has to perform a location update. To send a packet to a UE, the location resolution procedure needs to be executed first. In \cite{Han2016b}, GRIMM was proposed as a gateway-based
mobility management architecture for EOSNs based
on locator/identifier split. In \cite{Feng2016} and \cite{Feng2017b}, locator/identifier split network architectures for integrated satellite-terrestrial networks were proposed. Although location/identifier split can enhance mobility in EOSNs, employing the conventional binding (location) update schemes will create a large number of frequent binding updates for both UEs and satellites due to the high mobility of EOSN components. To mitigate the effect of frequent satellite handovers on the binding update rate, the authors of \cite{Zhang2017} and \cite{Zhang2016b} proposed the concept of virtual attachment point (\acrshort{vap}) to make a binding update
independent of a satellite’s motion, where the VAP stays in a fixed position relative to the ground. Thus, a virtual spherical network consisting of fixed VAPs is superimposed over the physical satellite topology in order to hide the mobility of satellites from the terrestrial endpoints. A VAP is created and maintained by the satellites that pass over the fixed network location of the VAP. For this, rapid mapping is necessary to resolve identifiers to network locations in a real-time manner in satellite networks. To overcome the geographical and delay constraints of ground station-based location resolution, \cite{Zhang2017b} presented a space-based distributed rapid mapping resolution system (RMRS) along with a dynamic replica placement algorithm. The goal of RMRS is
to achieve low location resolution latency, low update cost,
and high system availability (resilience to failures). There are a number of locator/ identifier split algorithms and architectures available in the literature, but they mainly focus on terrestrial networks. Further investigation is required to study the feasibility of applying such locator/ identifier split algorithms in the future megaconstellations of EOSNs.

\item \textbf{SDN-based location management in EOSNs:} The third approach involves using a software defined network (SDN) for the purpose of topology (location) management \cite{Xu2018b, Shuai2018}. Existing studies have proposed several architectures to implement SDN in EOSNs. A simple \acrfull{sdsn} architecture was proposed in \cite{Bao2014}. It contains three planes: a data plane (satellite infrastructure, terminal router), a control plane (a group of GEO satellites), a the management plane (\acrfull{nocc}). Similarly, the author of \cite{Li2017} proposed an SDSN where the controllers were located on GEO satellites and the switches were deployed in MEO and LEO satellites. The main factors that affected SDN performance were the number of deployed controllers and their positions, and how to assign switches to controllers. A dynamic SDN controller placement for EOSNs was considered in \cite{Papa2018}. The objective was to find the optimal controller placement and the number of satellites that would work as controllers, while minimizing the average flow setup time with respect to the traffic dynamics. The author considered an SDN architecture where the control plane layer consisted of several LEO satellites that varied on the basis of traffic demands in addition to seven satellite gateways placed on the ground to serve as entry points to the backbone network. The satellites that were part of the control plane served as both controllers and network switches. They managed, controlled, and updated the forwarding rules of the flow tables of the satellites of the data plane. On the other hand, the satellites of the data plane were only responsible for forwarding packets based on rules defined by the corresponding controllers. A three-layer hierarchical controller architecture for software-defined GEO and LEO satellite networks was proposed in \cite{Xu2018b}. The solution exploited the wide coverage ability of GEO satellites, the easy upgrading and maintenance of NOCC, and the stability of inter-satellite links in the same low Earth orbit. The control plane consisted of domain controllers, slave controllers, and a super controller. The GEO satellites were set as domain controllers because of their broadcast capabilities over a wide-coverage area and stable connection with the ground station. The domain controller monitored and managed the LEO satellites located in its coverage area. The LEO satellites forwarded and collected network status information, and were divided into different domains according to the GEO coverage. Several slave controllers were selected from LEO satellites. The GEO domain controllers communicated only with the slave controllers under their authority instead of with all LEO satellites in their domain. By using inter-satellite links, the slave controllers collected the status information of the LEO satellites under their own authority, and this information was then sent to the corresponding domain controllers. The NOCC was deployed as a super controller that could obtain knowledge of the overall network through the primary GEO satellites. Based on the aforementioned description, a logically centralized control plane with global knowledge was created through physically distributed LEO controllers. Implementing SDNs in future EOSNs and SpaceNets has a good potential due to the flexibility that SDN introduces. As future SNs will have software defined payloads,  reprogramming SNs to play different roles will enhance the performance of future EOSNs and SpaceNets and support optimum resource utilization. 

\end{itemize}

Although existing location management solutions have  potential for future EOSNs and SpaceNets, many challenges will be encountered as well. This is due to the complicated and new mobility and topology characteristics of future EOSNs and SpaceNets. Section \ref{ssec:OpenIssuesMM} highlights the key challenges and open issues for location management in future EOSNs and SpaceNets. 

}
%
{\color{black} \subsection{\bfseries Space-Terrestrial Integrated Routing}
\label{ssec:space-terrestrial-routing}

\subsubsection{Emergence of Delay/Disruption Tolerant Networking} 
\label{sssec:dtn_emergence}

Routing is the process of computing and using routes to send traffic from its origins to its destinations. In terrestrial networks, such as the Internet, connections between nodes are stable, with low delay and low error rates. This allows stable and continuous end-to-end multihop routes to be formed, and a source can negotiate communication parameters with the destination through sessions using the TCP/IP stack. On the other hand, in SpaceNets, nodes may face large delays and error rates, and frequent interruptions in communications due to orbital dynamics and resource or power constraints. This has led to the development of communication protocols and routing schemes that are particularly adapted to the space environment. 

Organizations such as 3GPP have proposed to extend and adapt the protocols developed for 3G/4G/5G networks to incorporate EOSNs into the terrestrial network infrastructure. However, because they seek to maintain legacy protocols, such as TCP and IP, they need to ensure the assumptions on which these protocols based their design. This implies the use of space networks in a very simple form (bentpipe), where space networks are not leveraged as networks themselves, and nodes simply act as repeaters. Another possibility is to carry out the costly deployment of a large number of nodes in the form of megaconstellations, so as to allow the formation of continuously connected end-to-end paths everywhere and at all times. 

There is, however, another paradigm called \acrfull{dtn} \cite{burleigh2003delay,fall2008dtn, mcmahon2009delay} that also proposes to integrate heterogeneous terrestrial and SpaceNets but relaxing the network infrastructure requirements, or by allowing for a gradual increase in the available infrastructure. In the \acrshort{dtn} architecture, generated traffic can be sent from an origin to a destination without the need for the next hop of a route to be available. That is, a \acrshort{dtn} node allows incoming traffic to wait in a persistent buffer until there is communication with the next neighboring node on the route. When such communication is available, the traffic is sent to that neighboring node, which repeats the same process until the traffic is eventually delivered to the final destination. This way of sending traffic is known as store-carry-forward and is implemented through an overlay protocol layer called \acrfull{bp}, which is being standardized by both the \acrshort{ietf} \cite{rfc9171} and \acrshort{ccsds} \cite{ccsds2015bp}. Among other things, \acrshort{bp} allows for different interruptions, delays, and bandwidths of the underlying communications. It is important to note that this architecture does not require traffic to wait at a node if the next hops of a route are available. In this way, a \acrshort{dtn} enables the generalized treatment of heterogeneous networks by considering end-to-end continuously connected (synchronous) networks as particular cases of disrupted (asynchronous) networks where the delay is close to 0 and there are no interruptions. A \acrshort{dtn} can thus help manage and accommodate traffic with different QoS requirements, by sending delay sensitive and delay tolerant traffic with the best possible resource utilization. 

\subsubsection{Routing in Delay/Disruption Tolerant Networks} 
\label{sssec:dtn_routing}
In DTNs, communication opportunities are called contacts and they have a limited time duration. Depending on how they occur, they can be classified as opportunistic, probabilistic, and scheduled. Different routing strategies have been proposed for each case. Opportunistic contacts occur randomly, for example when two nodes enter within each other's coverage range in an unplanned manner. Due to the high uncertainty in this type of communication, flooding or epidemic routing strategies that make multiple copies of the packets and send them for every possible communication have been devised \cite{wu2013performance}. This type of mechanism is inefficient and even detrimental in resource-limited and congestion-prone networks, so schemes that restrict the number of copies, such as Spray-and-Wait \cite{spyropoulos2005spray}, have also been developed and they obtain acceptable delivery rates with arbitrarily reduced resource usage in a parametric way. On the other hand, probabilistic contacts are those that exhibit a certain distribution and can be predicted with a certain level of confidence based on a history of previous encounters between nodes. The routing schemes proposed to leverage this type of communication seek to take advantage of the knowledge of these distributions to send packets in such a way that they have the highest chances of reaching their respective destinations \cite{lindgren2003probabilistic, burgess2006maxprop}. Surveys of opportunistic and probabilistic routing schemes for different use cases are discussed in \cite{sobin2016survey, benamar2014routing}. Finally, DSN and EOSN contacts are predictable and can be scheduled with high accuracy thanks to the existence of orbital propagators and the control of node orientation. Routing schemes such as those developed in \cite{merugu2004routing} can leverage such planning to make much more efficient decisions than routing schemes in other types of networks. The \acrfull{cgr} scheme described in \cite{burleigh1contact, Fraire2021Routing}, and \cite{ccsds2019sabr}, is the most developed solution for scheduled networks, and it has been the subject of numerous contributions from the research community \cite{araniti2015contact, segui2011enhancing, bezirgiannidis2014contact, bezirgiannidis2016analysis, edward2011improving, caini2021schedule, de2019efficient, fraire2018route, raverta2021routing, dhara2019cgr, burleigh2016toward, madoery2018congestion}. These contributions include source routing extensions \cite{edward2011improving}, the adaptation of Dijkstra \cite{segui2011enhancing}, the prevention of routing loops and consideration of multiple destinations \cite{birrane2012analysis, caini2021schedule}, overbooking management techniques \cite{bezirgiannidis2014contact, bezirgiannidis2016analysis}, congestion mitigation by proper volume annotations in routes \cite{madoery2018congestion},
opportunistic enhancement so that unplanned contacts can be included in the routing decisions \cite{burleigh2016toward}, route table management strategies and the incorporation of Yen’s algorithm \cite{fraire2018route}, a spanning-tree formulation
to compute routes to several destinations \cite{de2019efficient}, and a partial queue information sharing \cite{dhara2019cgr}.

\subsubsection{Hierarchical Routing} 
\label{sssec:hierarchical_routing}
Because DTNs function in principle as a network of networks, their routing can be conceptualized hierarchically \cite{musolesi2008framework, qi2016unified, harras2006inter, bhotmange2015region, wen2009rena, liu2007scalable, madoery2018managing} such that each network with similar characteristics can use an internal routing (intra-regional routing) tailored to function appropriately in that type of network, while routing between networks (inter-regional routing) can be done with a different scheme, such as those described above. The Internet works similarly with the border gateway protocol (BGP) that connects different autonomous systems, and where each autonomous system can implement a particular routing protocol, such as the open shortest path first (OSPF) or the routing information protocol (RIP).
A hierarchical routing scheme has the following benefits. On the one hand, it overcomes the problem of having to implement a unique and homogeneous routing solution that works efficiently in very heterogeneous types of networks. On the other hand, by making it possible to reuse routing protocols that have already been studied, optimized, and that currently operate on the ground and in SpaceNets, it facilitates rapid adoption and decreases the likelihood of faults or errors. Finally, this type of solution allows a trade-off between performance and scalability, which is vital for the seamless operation of networks with hundreds of thousands of nodes, which will be the case for the next-generation space-terrestrial integrated networks.
}
%
\subsection{Laser Inter-SN Links}
\label{ssec:LaserISLs}

{\color{black} \par In this subsection, we discuss the rationale for the development of laser links between space nodes (e.g., EOSs) within VLEO and LEO EOSCs and the classification of LISNLs based on the location of SNs within a constellation and the duration of LISNLs between SNs. We also discuss the capabilities of currently available LCTs for establishing LISNLs, especially in terms of link capacity (or data rate), and different types of latencies (or delays) that exist in \glspl{fsosn}. In so doing, we discuss recent works that have endeavored to address this last issue.
}

\subsubsection{Why LISNLs?}
\label{why:Why}

{\color{black} \par RF is the most widely used wireless communications technology, its fifth generation is being deployed for mobile communications, and it is also being used to interconnect EOSs in the Iridium NEXT EOSC. However, FSO is rapidly becoming a more appealing substitute for RF in inter-space node (or inter-satellite) communications. The EOSs that will be deployed in upcoming VLEO or LEO EOSCs, like Telesat’s Lightspeed and SpaceX’s Starlink, will be equipped with LCTs to enable them to establish LISNLs and form FSOSNs in space \cite{erwin2021telesat},\cite{jewett2021starlink}. Space FSO includes ground-to-EOS and EOS-to-ground FSO links as well as EOS-to-EOS FSO links. A brief history of the developments in space FSO is given in Table \ref{tab:histSpaceFSO}.

\begin{table*}
	\centering
	\color{black}
	\caption{Brief History of the Developments in Space FSO.}
	\label{tab:histSpaceFSO}
	\begin{tabular}{|c|p{5.5cm}|c|p{5.5cm}|} 
	\hline
	\textbf{ Year } & \textbf{ Event }                                                                                            & \textbf{ Type }      & \textbf{ Description }                                       \\ 
	\hline
	1967            & Space optical uplink transmission \cite{fried1967laser}                            & Theoretical 
	study & Ground-to-EOS                                          \\ 
	\hline
	1972            & Space uplink transmission using a continuous-wave argon laser \cite{minott1972laser} & Demonstration        & Ground-to-geodetic
	Earth orbiting satellite-II (GEOS-II)   \\ 
	\hline
	1985            & Space laser beam transmission \cite{aruga1985laser}                                & Demonstration        & Ground-to-geostationary
	Japanese meteorological EOS  \\ 
	\hline
	1992            & Deep space uplink optical communication at 6 million km \cite{wilson1993gopex}     & Demonstration        & Optical ground
	station-to-Galileo spacecraft               \\ 
	\hline
	1994            & Space bi-directional laser link using adaptive optics \cite{wilson1997cemerll}        & Demonstration        & Earth-to-Moon                                                \\ 
	\hline
	1996            & First optical link between a ground station and an EOS \cite{toyoda1996laser} & Demonstration        & Optical ground
	station-to-ETS-VI EOS                 \\ 
	\hline
	2001            & First LISNL at 50 Mbps for optical data-relay service \cite{nielsen2002oisl}       & Demonstration        & SPOT-4
	EOS-to-ARTEMIS EOS                      \\ 
	\hline
	2005            & First bi-directional LISNL \cite{katsuyoshi2012oicets}                                  & Demonstration        & OICETS
	EOS-to-ARTEMIS EOS                      \\ 
	\hline
	2008            & First high-speed LISNL at 5.625 Gbps between LEO EOSs \cite{gregory2011lisl} & Demonstration        & Terra SAR-X
	EOS-to-NFIRE EOS                   \\ 
	\hline
	2016–2021       & 1.8 Gbps LISNLs between EOSs for data-relay \cite{calzolaio2020edrs}           & Data-relay
	service & Sentinel LEO EOSs-to-EDRS
	GEO EOSs             \\
	\hline
	\end{tabular}
\end{table*}

\par Compared to RFISNLs, LISNLs operate at a higher frequency (or a smaller wavelength), which means smaller antenna sizes, less weight, and less volume. Additional benefits of LISNLs include the following: higher bandwidth, which means higher capacity (or data rate); narrower beam divergence (or beam spread), which means narrower beams with no interference and more security; and much higher directivity, which means lower transmit power requirements \cite{chaudhry2020fso}. A comparison of these two types of ISNLs is presented in Fig. \ref{fig:xticsISLs}. Due to their smaller SWaP requirements, LCTs use fewer onboard space node resources, are easy to integrate into SN platforms, and their smaller form factor also helps in minimizing SN launching and deployment costs.

\begin{figure*}[htbp]
	\centerline{\includegraphics[scale=0.73]{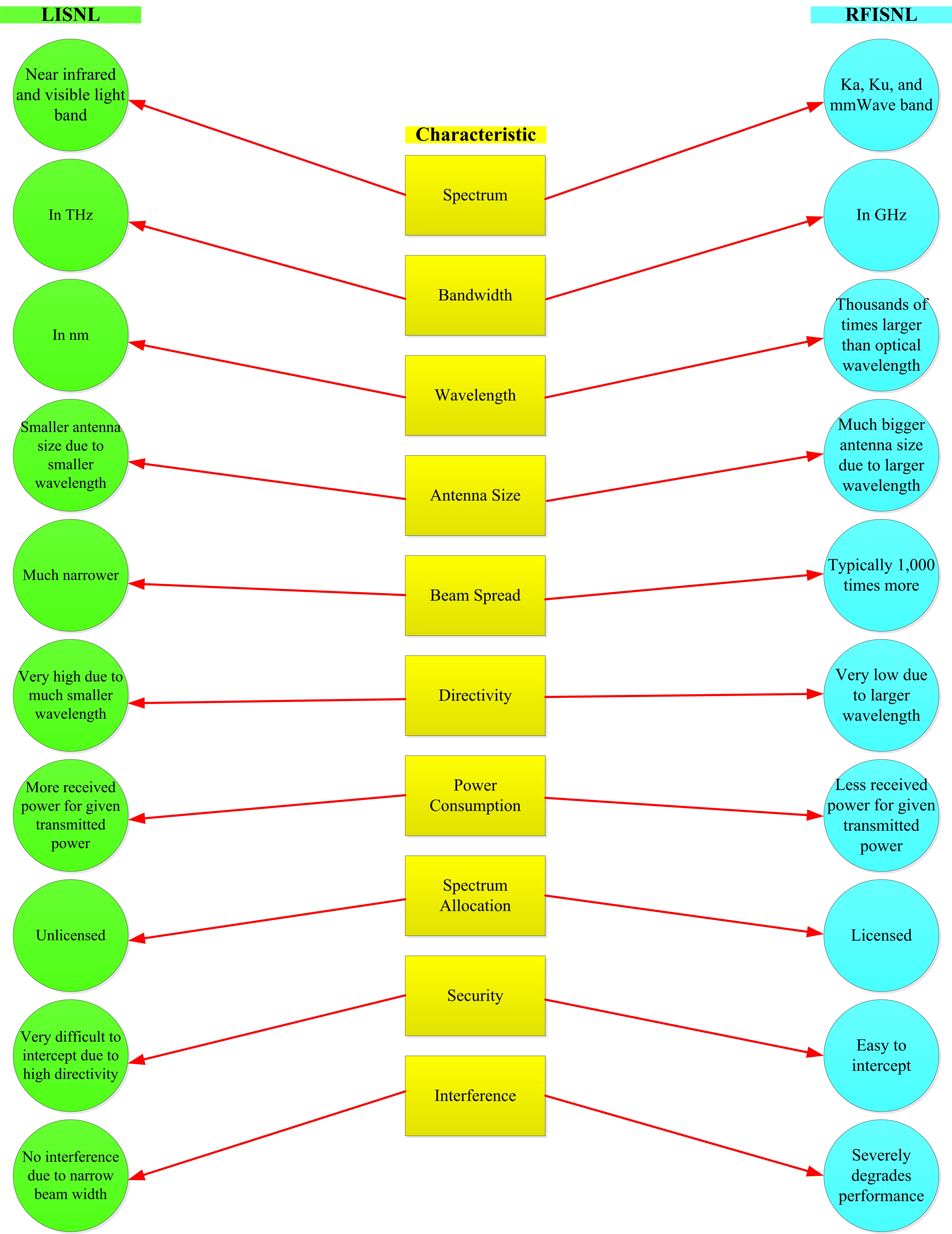}}
	\caption{Characteristics of LISNLs vs. RFISNLs.}
	\label{fig:xticsISLs}
\end{figure*}

}

\subsubsection{Classification of LISNLs}
\label{sssec:Classification}

{\color{black} \par As illustrated in Fig. \ref{fig:classificationLISLs}, LISNLs can be broadly classified on the basis of the location of SNs and the duration of LISNLs \cite{chaudhry2020fso}, \cite{chaudhry2021lisl}. Depending on the location of the space nodes (e.g. satellites), LISNLs can be classified into five types:
\begin{itemize}
	\item Intra-OP LISNLs – which can be between SNs in the same \acrfull{op} of a constellation;
	\item Inter-OP LISNLs – which can be established between SNs located in different OPs of a constellation;
	\item Inter-Orbit LISNLs – which can be created between SNs in different orbits;
	\item Inter-Planet LISNLs – which can be formed between SNs that are located near two different planets; and
	\item Deep-Space LISNLs – which can be set up between an SN located near Earth or Mars and a deep-space spacecraft, like Galileo.
\end{itemize}

\par Inter-OP LISNLs can be further divided into three types:
\begin{itemize}
	\item Adjacent OP LISNLs – which are between SNs in adjacent OPs;
	\item Nearby OP LISNLs – which are between SNs in nearby OPs; and
	\item Crossing OP LISNLs – which are between SNs in crossing OPs.
\end{itemize}

\par Inter-Orbit LISNLs can also be of two types:
\begin{itemize}
	\item Inter-Shell LISNLs – which can be initiated between SNs in two different shells (or sub-EOSCs) of a megaconstellation; and
	\item Inter-Constellation LISNLs – which can be formed between SNs in two different constellations.
\end{itemize}

\par LISNLs can also be classified as permanent or temporary, depending on their duration. SNs in the same OP of a constellation move at the same speed and in the same direction, and intra-OP LISNLs are permanent, which means that once established, they exist continuously. They are easy to establish and maintain due to the same velocities of the SNs. The velocity of SNs in different OPs is different, and adjacent OP LISNLs and nearby OP LISNLs are harder to establish and maintain. These LISNLs are usually permanent in nature; however, LISNLs between some SNs in adjacent and nearby OPs can also exist temporarily near Polar regions or at high latitudes. 

\par SNs in crossing OPs move at high relative velocities, and it is hard to establish crossing OP LISNLs. These LISNLs can only exist temporarily between SNs in crossing OPs when they come within range of each other. SNs in different orbits have different altitudes, speeds, and relative velocities, and inter-orbit LISNLs are also temporary in nature. The LISNLs between SNs located near different planets or between a SN near a planet and a deep-space spacecraft can only be created when a clear line-of-sight exists between the two SNs or between the SN and the deep-space spacecraft, and inter-planet LISNLs as well as deep-pace LISNLs can also be considered temporary in nature.
		
\begin{figure*}[htbp]
	\centerline{\includegraphics[scale=0.67]{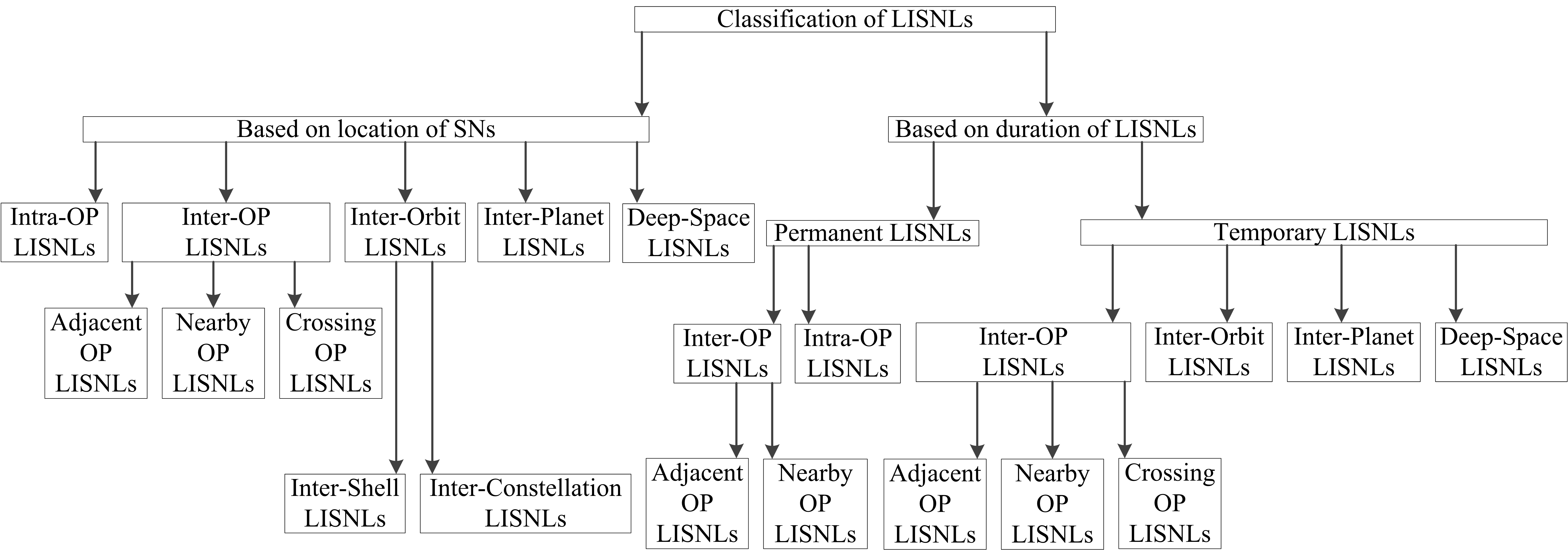}}
	\caption{Classification of LISNLs.}
	\label{fig:classificationLISLs}
\end{figure*}

}

\subsubsection{Laser Communication Terminals}
\label{sssec:lcts} 

{\color{black} \par The first successful LISNL between two LEO EOSs, TerraSAR-X and NFIRE, was demonstrated in 2008 using LCTs developed by Tesat. A bidirectional transmission was achieved at 5.625 Gbps between these EOSs. The LCTs in this demonstration used a laser operating at 1,064 nm in wavelength, a telescope with an aperture diameter of 125 mm, and an LISNL range of 5,100 km \cite{gregory2017tesat}. Subsequently, Tesat developed LCTs for data-relay service between Earth observation LEO EOSs, such as Sentinel-1A/-1B/-2A/-2B and GEO EOSs EDRS-A/-C for quasi real-time delivery of Earth observation data to customers. The EDRS is currently using these LCTs to establish LEO-to-GEO LISNLs at a data rate of 1.8 Gbps over an LISNL range of 45,000 km \cite{zech2017tesat}.

\par LCTs for establishing LISNLs between EOSs in upcoming VLEO and LEO EOSCs, like SpaceX’s Starlink and Telesat’s Lightspeed, will be critical in the formation of an FSOSN in space. Tesat, Mynaric, and General Atomics are the three main companies that are actively working to develop such LCTs, and the capabilities of their LCTs are summarized in Table \ref{tab:LCTs}.

\par {\bfseries CONDOR:} Mynaric’s LCT for VLEO and LEO EOSCs will provide a data rate of 10 Gbps over an LISNL range of 4,500 km \cite{muncheberg2019mynaric}. Recently, Mynaric named this LCT CONDOR, and declared that it could provide a reduced data rate of 5 Gbps at an LISNL range of 7,780 km \cite{carrizo2020mynaric} and that it would employ \acrfull{imdd}. 

\par {\bfseries ConLCT1550:} Tesat is developing an LCT specifically for EOSs in VLEO and LEO EOSCs. Tesat refers to this LCT as ConLCT1550 \cite{motzigemba2019tesat}. Using \acrfull{ook} modulation, this LCT will enable the creation of LISNLs at a 10 Gbps data rate with an LISNL range of 6,000 km.

\par {\bfseries GA-LCT:} General Atomic’s GA-LCT uses OOK but it can support other modulation schemes as well. It has a modular design in terms of amplifier stages, which enables it to support LISNLs at a data rate of 5 Gbps over LEO-to-LEO LISNL ranges varying from hundreds to thousands of kilometers \cite{leboffe2020ga}, \cite{freeman2021ga}.

\par LCTs for establishing LISNLs between EOSs in upcoming VLEO and LEO EOSCs are in their infancy and offer capacities of up to 10 Gbps. LCTs providing data rates in Tbps will need to be developed to realize a true global communications network in space.

\begin{table*}
	\centering
	\color{black}
	\caption{LCTs for Establishing LISNLs in VLEO and LEO EOSCs.}
	\label{tab:LCTs}
	\begin{tabular}{|p{1.6cm}|p{1.7cm}|p{1.2cm}|p{1.4cm}|p{1.4cm}|p{1.4cm}|p{1.5cm}|p{1cm}|p{1.3cm}|} 
		\hline
		\textbf{ Company } & \textbf{ LCT } & \textbf{LISNL} \textbf{capacity} \textbf{~(Gbps)} & \textbf{LISNL} \textbf{range} \textbf{~(km)} &  \textbf{Operating} \textbf{wavelength} \textbf{~(nm)}& \textbf{Modulation} & \textbf{Aperture} \textbf{diameter} \textbf{~(mm)} &  \textbf{Mass} \textbf{~(kg)}  &  \textbf{Power} \textbf{consumption} \textbf{~(W)} \\ 
		\hline
		Mynaric            & CONDOR         & 10                                                                                            & 4,500                                                                                    & 1,550                                                                                               & IMDD                  & 80                                                                                              & 18                                                                      & 60                                                                                              \\ 
		\hline
		Tesat              & ConLCT1550     & 10                                                                                            & 6,000                                                                                    & 1,550                                                                                               & OOK                   & unknown                                                                                         & 15                                                                      & 80                                                                                              \\ 
		\hline
		General Atomics    & GA-LCT         & 5                                                                                             & 100s–1000s                                                                               & 1,550                                                                                               & OOK                   & 72                                                                                              & unknown                                                                 & unknown                                                                                         \\
		\hline
	\end{tabular}
\end{table*}
	
}

\subsubsection{Latency in FSOSNs}
\label{sssec:Latency}

{ 
\color{black} \par In the following, we discuss various delay types that are part of end-to-end latency in FSOSNs, and we provide an overview of recent works that investigate latency in FSOSNs.
	
\paragraph{Types of delay in end-to-end latency in FSOSNs} The end-to-end latency is the delay from the source to the destination in the network and consists of the following main components: transmission delay, processing delay, queueing delay, and propagation delay \cite{kurose2009delays}. The transmission delay is the time that is required to transmit all bits of a packet onto the communication link, and it is dependent upon the packet size and the data rate of the link. The queueing delay is the time that a packet spends in the buffer at a node (e.g., a SN in the FSOSN that is on the path between the source and destination ground stations) while it waits for transmission onto the communication link, and it is influenced by the number of packets that are waiting for transmission in the buffer.

\par The processing delay is the time that is used by an SN to process a packet, such as the time used to read the packet header to make appropriate routing and switching decisions, before sending the packet to the appropriate next hop. This delay is influenced by the number of hops or SNs between the source and destination ground stations and becomes significant when the data communication must pass through several intermediate hops. In FSOSNs, the propagation delay is the time taken by the transmission of the optical signal along the medium, which is the vacuum in space for LISNLs between SNs or a combination of vacuum in space and the planet's atmosphere for optical uplink/downlink. This delay depends on the end-to-end distance, which is the distance between the source and destination ground stations over the FSOSN, and it becomes significant for long-distance data communications. 
	
\par For congestion-free FSOSNs with LISNLs having Gbps data rates, the queueing and transmission delays can be considered as negligible, and the end-to-end latency consists of processing and propagation delays. In SNs with high-speed onboard switching and routing capability, the processing delay can be assumed to be in milliseconds, such as 1 ms \cite{hermenier2009delay}. 
	
\par Due to PAT during the creation of an LISNL between a pair of SNs equipped with LCTs, the LISNL setup delay is another major component of the end-to-end latency in FSOSNs. The LISNL setup delay is incurred when a new LISNL needs to be established between a pair of SNs. It depends upon the time incurred during the initial setup of the LISNLs between the SNs on the path and whenever there is a change in the path and one or more new SNs are introduced in the path requiring the establisshment of new LISNLs. LISNL setup delay currently ranges from a few seconds to tens of seconds (e.g., Mynaric’s CONDOR laser communication terminal requires approximately 30 seconds to set up LISNLs between pairs of SNs in the FSOSN for the first time but once the position and altitude of the SNs are exchanged, the LISNL setup delay is reduced to two seconds \cite{carrizo2020mynaric}). A summary of the different types of delay that are part of the end-to-end latency in FSOSNs is provided in Table \ref{tab:typesDelay}.

\begin{table*}
	\centering
	\color{black}
	\caption{Summary of Different Types of Delay in End-to-End Latency in FSOSNs.}
	\label{tab:typesDelay}
	\begin{tabular}{|l|p{5cm}|c|} 
		\hline
		\textbf{Type of delay } & \textbf{ Dependency }                                                                                                                                       & \textbf{ Significance }                                   \\ 
		\hline
		Transmission delay       & Packet size and link data rate                                                                                                                              & Negligible in FSOSNs with LISNLs
		having Gbps data rates  \\ 
		\hline
		Processing delay         & Number of hops                                                                                                                                              & Can be assumed as 1 ms per hop                            \\ 
		\hline
		Queueing delay           & Number of packets waiting in the
		buffer                                                                                                                   & Negligible in congestion-free
		FSOSNs                    \\ 
		\hline
		Propagation delay        & End-to-end distance                                                                                                                                          & Significant for long-distance
		data communications       \\ 
		\hline
		LISNL setup delay         & Time incurred: (a) during initial setup of LISNLs;  (b) when new LISNLs need to be created  whenever the path changes & Ranges from 2 to 30 seconds                       \\
		\hline
	\end{tabular}
\end{table*}

\paragraph{Investigations on latency in FSOSNs} The impact of LISNLs on the latency in FSOSNs has attracted the attention of the research community. Optical fibers are typically made of glass and have a refractive index of approximately 1.5. Therefore, the speed of light in optical fiber is approximately 2/3 of the speed of light in vacuum, which translates to the speed of light in vacuum being approximately 50\% higher than in optical fiber. This fact is of critical importance in optical wireless satellite networks, and it gives them an advantage over optical fiber terrestrial networks in terms of latency when the data communications takes place over long distances. In \cite{chaudhry2020fso}, the authors examined the suitability of an optical wireless satellite network, consisting of LISNLs between EOSs in Starlink’s Phase I LEO EOSC, for low-latency communications over long distances in comparison with the optical fiber terrestrial network. Multiple optical fiber relay stations and multiple EOSs were considered on the optical fiber terrestrial network and the optical wireless satellite network, respectively. It was shown that the optical wireless satellite network operating at an altitude of 550 km outperformed the optical fiber terrestrial network in terms of latency when the terrestrial hop distance exceeded 3,000 km.

\par Repetitive patterns in the network topology, called motifs, were proposed to improve the latency of an EOSN, where an EOS’s connectivity over LISNLs was limited to four neighbors \cite{bhattacherjee2019topology}. The same connectivity pattern was assumed for all EOSs within a motif, and different motifs offered different connectivity patterns. The performance of different motifs was observed to be widely different, and to choose the motif with the best performance out of all possible motifs required exhaustively evaluating all possible motifs. Using a hypothetical constellation of 1,600 EOSs at 550 km altitude, a median round trip-time improvement of 70\% with the EOSN was shown in comparison with internet latency. However, this comparison was unduly favorable to the EOSN as delays due to sub-optimal routing, congestion, queuing, and forward error correction were not counted in the EOSN, while such delays were taken into consideration in measuring internet latency.

\par A preliminary evaluation of an EOSN using LISNLs to provide low-latency communications was conducted in  \cite{handley2018latency}. Starlink’s original Phase I sub-EOSC of 1,600 EOSs at 1,100 km altitude was used, LISNLs between EOSs were considered, and the latency of the resulting FSOSN was examined. The connectivity of an EOS was limited to five nearby neighbors, which included two nearest neighbors of an EOS in the same OP (i.e., one in the front and one at the rear). Instead of connecting to nearest neighbors in left and right adjacent OPs, connectivity to adjacent OP neighbors in east-west directions was considered as it could benefit most of the population in developed countries, especially the United States. A nearby EOS in a crossing OP was considered as the fifth neighbor to improve routing options by providing inter-mesh links (i.e., connectivity between EOSs in crossing OPs) even though such LISNLs are temporary in nature and require tracking and connecting to a new crossing EOS as the old one moves away.

\par It was concluded that LISNLs between EOSs in a FSOSN can provide lower latency than ground-based relays \cite{handley2019isls}. The 1,584-EOS sub-EOSC for Phase I of Starlink at 550 km altitude was considered and the use of ground-based relays was investigated as an alternative for LISNLs to provide low-latency wide area networking. It was suggested that in case a LISNL does not have sufficient capacity to meet the offered load, multipath routing could be considered to spread the load and a ground-based relay could be used to supplement the LISNL. Another similar study showed that the use of LISNLs between EOSs in a FSOSN significantly reduced the temporal variations in latency of the EOSN in comparison to the bent-pipe scenario (i.e., the connectivity scenario where EOSs in the EOSC were inter-connected through ground stations only and LISNLs between EOSs did not exist \cite{hauri2020isls}). 

\par A simulator was developed for studying the network behavior of EOSNs, including latency \cite{kassing2020hypatia}. This simulator, named Hypatia, provided a packet-level simulation environment incorporating the orbital dynamics of an EOSN based on a LEO EOSC. It had a visualization module to render views of EOS trajectories, ground station-perspective on overhead EOSs, etc. Using this simulator, the behavior of individual end-to-end connections across EOSNs based on three upcoming LEO EOSCs was analyzed on the basis of their changing latencies. However, the connectivity pattern of an EOS was limited to the four nearest neighbors in the same and adjacent OPs, and crossing OP (or temporary) LISNLs were not supported. The simulator would need to be modified if it were to be used to analyze the performance of an FSOSN with permanent as well as temporary LISNLs between EOSs.

\par A comparison of an optical wireless satellite network (or FSO satellite network) and the optical fiber terrestrial network in terms of latency in different scenarios for long-distance inter-continental data communications was investigated in \cite{chaudhry2021latency}. Starlink’s Phase I sub-EOSC of 1,584 EOSs was considered and LISNLs were assumed between EOSs to realize an FSOSN. It was observed that the ground station-to-EOS, EOS-to-EOS, and EOS-to-ground station links and/or their latency changes at every time slot (or second) due to the high-speed movement of EOSs along their OPs. As a result, the shortest path (and/or its latency) over the FSOSN between the source and destination ground stations in two different cities also changes at every time slot. It was mentioned in this study that this comparison was favorable to the optical fiber terrestrial network since the shortest distance between two cities in different continents over this network was considered along the surface of the Earth. In reality, long-haul submarine optical fiber cables do not follow the shortest path to connect two points on Earth’s surface. It was observed that the FSOSN performed better than the optical fiber terrestrial network in terms of latency in all scenarios and it was noted that FSOSNs could be an ideal solution to inter-connect financial stock markets around the world for low-latency inter-continental \acrfull{hft} as a one millisecond advantage in HFT can be worth \$100 million a year in revenues for a major brokerage firm.

\par A crossover function was proposed in \cite{chaudhry2022crossover}, which enabled to find the crossover distance (i.e., a distance between two points on Earth beyond which switching from an optical fiber terrestrial network to an optical wireless satellite network is beneficial for data communications between these points in terms of latency). It was shown that the crossover distance varied with the refractive index of the optical fiber in an optical fiber terrestrial network as well as with the altitude of satellites and the end-to-end propagation distance in an optical wireless satellite network. Different optical fiber terrestrial networks with different refractive indices and different optical wireless satellite networks with different satellite altitudes were studied in different scenarios for long-distance inter-continental data communications for a comparative analysis of these networks in terms of latency.

\par The effect of LISNL range on network latency in FSOSNs was investigated in \cite{chaudhry2022lislrange}. Using the satellite constellation for Phase I of Starlink, six different LISNL ranges for satellites in this constellation were examined in three different scenarios for long-distance inter-continental data communications. In \cite{chaudhry2022templisls}, the authors studied the impact on network latency in next-generation FSOSNs (having only permanent LISNLs) versus next-next-generation FSOSNs (having permanent LISNLs as well as temporary LISNLs) using different long-distance inter-continental data communications scenarios and the satellite constellation for Phase I of Starlink. 

\par A discussion of open issues related to LISNLs, including an accurate understanding of latency in FSOSNs realized from LISNLs, the number of LCTs per SN, LISNL setup delay, and capacity, cost, and standardization of LCTs, is provided later in Section \ref{sec:OpenIssues}.

}

%
\subsection{Cross-Layer Design}
\label{ssec:CrossLayerDesign}
{\color{black}
Most networking protocol suites adopt a layered approach (e.g., using the open systems interconnection (OSI) model) to characterize and standardize different communications and networking functions. In this approach, each layer is optimized independently without taking other layers into consideration. However, this approach can be challenging in SpaceNets due to the following reasons:
\begin{itemize}
\item Unlike terrestrial networks that have some segments with wired links (e.g.,  fiber),  almost all the communication segments of SpaceNets are wireless links (i.e., using RF or FSO). Therefore, the design of the physical and data link layers has a significant impact on the networking aspects. 
\item Due to the mobility of SNs (especially NGSOs) relative to SNTs, ground stations, and other SNs, there are different handovers associated with the network nodes and terminals. Therefore, handling these various handovers can timeout the higher layers due to delay.
\item Low throughput due to bad channel conditions at any communication segment can cause packet congestion at the higher layers.
\end{itemize}
It is therefore crucial to consider the upper and lower layers in a cross-layer manner while designing the network management and signaling procedures for SpaceNets. In this regard, cross-layer design can provide several benefits for SpaceNets as investigated in \cite{giambene2006cross}. In this section, we highlight two major approaches for cross-layer design and discuss representative works from the literature that adopt this cross-layer perspective in SpaceNets.

\subsubsection{Indirect Cross-Layer Design}
In this approach, the parameters of the different layers are taken into consideration while designing the network management procedures. The main advantage of this method is that it does not require major modification in the communication protocols. However, these techniques require adopting complicated mathematical models to incorporate different system parameters from the different layers. 

A good example of this approach is the design of the radio resource management (RRM) techniques in a cross-layer manner \cite{abdelsadek2019optimal}. Among the different satellite network management functionalities, RRM plays a vital role in maximizing the utilization of radio resources. In this regard, RRM operations, such as spectrum allocation, power allocation, beamforming, and frequency reuse, are designed to maximize the network utility function while ensuring user satisfaction and considering the network and SNT capabilities.  In addition,  they can be designed in a cross-layer manner to optimize the upper layers of the network. 

Based on this concept, the authors in \cite{abdelsadek2021ultra} proposed a CF-mMIMO architecture for ultra-dense LEO satellite networks. In this architecture, the ground user terminals were connected to a cluster of LEO satellites instead of traditional single-satellite connectivity. This cluster of satellites was composed of a set of satellite access points (SAPs) that were connected to a super satellite node (SSN), which acted as a central processing unit (CPU) for the cluster. Accordingly, the cluster of EOSs exchanged signals with SNTs cooperatively without considering cell boundaries. In addition, the authors proposed a joint optimization framework for the power allocation and handover management processes for CF-mMIMO-based LEO satellite networks in a cross-layer manner. In the formulated multi-objective optimization problem, the throughput was maximized, and the handover rate was minimized while considering the QoS parameters of the user terminals. Consequently, lower layer processes (i.e., power allocation and beamforming) were utilized to improve the upper layers (i.e., handover process and its associated binding updates and packets forwarding and tunneling). 

Another example of indirect cross-layer design was demonstrated in the joint optimization of routing and wavelength assignment in optical LEO satellite networks in \cite{wen2017ant}. The authors formulated an optimization problem to find a light path for every source-to-destination connection request minimizing a cost function. The cost function was expressed in terms of the transmission delay and Doppler wavelength shift. The constraints of the optimization problem were the maximum transmission delay, the maximum Doppler wavelength shift, and wavelength-continuity. This cross-layer problem was then solved using an ant colony-based technique. The simulation results showed that the proposed cross-layer design achieves superior performance in communication success probability.
\subsubsection{Direct Cross-Layer Design}
In this approach to cross-layer design, some information is exchanged among the different layers to represent the status and parameters of the different layers such that a global coordinator is realized  \cite{giambene2006cross}. The main advantage of this strategy is that it enables global optimization for the different system procedures. However, this requires major modifications in the protocols and service level agreements among different  operators (e.g., networking and service operators).

As an example, the authors in \cite{jiang2021novel} proposed allowing the parameters of every layer to be shared with other layers to be taken into consideration in the decision process of each layer. In this regard, the process was a loop of three major phases. The first phase was to collect the state information of all layers, such as the modulation scheme, transmit power, BER, packet loss, and topology. Then, in the second phase, effective information is extracted from these states (e.g., link delay, link reliability, and frame structure). Finally, the control process was implemented on the basis of this effective information.
}
\subsection{Physical Layer Security}
\label{sssec:PLS}
{\color{black}  \subsubsection{Why Physical Layer Security is Important}
On the one hand, as SpaceNets have a wide footprint that can cover a massive number of terrestrial users, SpaceNets are fundamentally ideal for offering broadcast or multicast services. On the other hand, this inherent characteristic makes SpaceNets more vulnerable to security threats from various IoT devices and equipment. As the wireless signal propagates in free-space, not only authorized users are able to access information, but illegitimate users as well who may be within the coverage area of the SN beam, obtain wireless power and recover secure information via power leakage in wireless signals. As a result, privacy and security issues in such communication networks have received much attention. The confidentiality of SpaceNets from eavesdroppers, there are often two approaches: 1) the upper-layer encryption and 2) physical layer security (PLS). The former uses encryption methods and a secret key to encrypt the information, making it difficult to decipher the original message. However, the computational cost of encryption or decryption is typically so high that it becomes a significant burden for both sides of communication, while secret key management and distribution are significantly more complicated due to the need for more complex protocols and structures. Cryptographic techniques at higher layers are not always reliable or robust enough once eavesdroppers have strong computation abilities. The principle behind PLS as a compelling complementary solution is that it uses randomness and the time-varying features of wireless links (e.g., noise and fading) to defend against unauthorized receivers in wireless communications without requiring a shared private key. This is as long as the main channel quality is better than the wiretap channel. PLS has the advantage of requiring essentially little upper-layer protocol adjustment, as it does not require sophisticated cryptographic calculations or additional computer resources from communication entities. As a result, the eavesdroppers will be unable to receive or interpret the transmitted information correctly, and thus the security level will be considerably improved. It should be noted that advances in signal processing and beamforming technologies, with the use of MIMO techniques, have been shown to have a significant impact on security enhancement. These obvious advantages paved the way for the use of PLS in the development of secure wireless communication networks. 

In addition, due to poor link stability brought on by frequent changes of LEO satellites, the security of space data transmission is significantly compromised. In particular, internal attack (that is, malicious attacks from inside an LEO satellite network) will bring great security threats to data transmission in space \cite{li2022secure}. Unfortunately, existing solutions, such as cryptography technology, which aims at external attacks of satellite network cannot solve the internal attack of satellite network. In addition, spoofing attacks where an active attacker tries to impersonate the legitimate transmitter to infiltrate and falsify
the network is a critical issue for space networks. Therefore, before starting the communication, any transmitter should establish its credibility with the receiver. The process of demonstrating this claim is known as authentication, and it is often accomplished at a layer higher than the physical one using cryptographic key-based methods. Given the high complexity requirements of the LEO satellite networks, the applicability of these methods would necessitate a key management architecture for satellite or spacecraft networks, which is impracticable. Therefore, physical layer authentication (PLA), will allow wireless devices to have an unclonable identity.

\subsubsection{Overview of Physical Layer Security Studies in SpaceNets}
A large literature on secure transmission in the context of SpaceNets has developed using the PLS paradigm, where extensive designs and solutions have been proposed. More specifically, the first research on PLS focused on land mobile satellite communication networks (LMS). This was followed by PLS studies on hybrid terrestrial SN-relay networks. More recently, research on PLS has focused on integrated SN-terrestrial networks following recent improvements in spatial information networks. The authors in \cite{petraki2010secrecy} were the first to investigate the PLS over SN channels with frequencies greater than 10 GHz while considering rain attenuation. Analytical expressions for the outage probability and probability of nonzero secrecy capacity were obtained while considering the frequency of operation, climatic conditions, and separation angles. This was extended to the scenario of two legitimate receivers and two eavesdroppers. In \cite{guo2016secure}, the authors studied the performance of an SN-to-ground wiretap network while assuming the shadowed Rician channel fading, which is considered to be the closest channel to reality. Later, they extended their work to the case of multi-cooperating eavesdroppers \cite{guo2017secure}. System performance has been shown to degrade when a channel experiences severe fading, and it has been shown to improve as the number of main receiver antennas increases. The authors in \cite{tao2018secrecy} investigated the secrecy performance for integrated SN-to-ground relay networks in the presence of various malicious eavesdroppers. They developed a problem to optimize the secrecy rate of the proposed system model, taking into account the perfect instantaneous CSI of SN-relay channel and statistical CSI of relay-receiver and relay-eavesdroppers channels. The authors in \cite{9473708} studied the secrecy performance of a HPAS-aided FSO-RF SatCom model. They showed that the use of the HAPS as a relay improved the security performance. To the best of our knowledge, the authors in \cite{9511459} are the first to investigate the secrecy performance of two practical scenarios by considering SN-to-HAPS and HAPS-to-UAV communications. It was shown that the SN-to-HAPS scenario is more secure than the HAPS-to-UAV transmission. Using a similar approach, the authors in \cite {9779543} introduced eavesdropping in the space. They proposed a novel system in which a sophisticated SN tries to capture the uplink and downlink communication between an SN and a HAPS node. The SN-to-HAPS downlink communication was shown to be slightly more secure than HAPS-to-SN uplink communication. Notably, in \cite{9473708,9511459,9779543}, atmospheric conditions were shown to have a direct impact on the security performance. }

\subsubsection{Physical Layer Security from a Cross-Layer Perspective}
{\color{black}
PLS has been investigated from a cross-layer design perspective in \cite{ding2019security,sharma2020secure,cui2020secure,liu2020terrestrial} and the references therein. In \cite{ding2019security}, the authors studied the PLS of an integrated satellite-terrestrial network with spectrum sharing in the presence of an eavesdropper, where a satellite scheduling scheme was designed to guarantee the wireless transmission of the EOS against eavesdropping attacks. The authors also analyzed the security-reliability trade-off of the multi-user scheduling scheme, and they obtained the probability of an interception occurring. In \cite{sharma2020secure}, the authors 
considered the communication of an EOSN with a UAV for hybrid satellite-terrestrial network in the presence of an aerial eavesdropper. In so doing, they adopted a stochastic mixed mobility model to mobilize the UAVs in a cylindrical cell with a ground UE. The authors also considered different deployments of eavesdroppers and obtained closed-form non-zero secrecy capacity and secrecy outage probability expressions. In addition to these studies, \cite{cui2020secure} considered a cooperative jamming idea for the multi-beam EOSN with secrecy outage probability constraints. To optimize the total power consumption of the satellites, they proposed a joint beamforming and power allocation algorithm with alternating optimization. Furthermore, \cite{liu2020terrestrial} proposed a novel secure communication method for satellite-terrestrial system that obviated the need for spread spectrum technology. The proposed communication method ensured the security of a confidential user by utilizing non-confidential users whose communications did not need to be encrypted. In addition, the proposed method was shown to improve the security of a confidential user at a very low cost without affecting the communication performance of non-confidential users.
}

Although the aforementioned studies provide insightful contributions to improving security in SpaceNets from a physical layer perspective, they are limited to space-to-ground downlink and ground-to-space uplink communications. Considering the PLS approach for space segments, there is a huge gap in the literature at the time of writing and this is due to unique features of this channel. Therefore, to extend PLS techniques to the space and improve the security based on the physical characteristics, new features should be extracted including mobility, position, Doppler shift, and misalignment distribution. The authors in \cite{9771727} are the first to propose a PLA scheme for
the LEO satellites by using Doppler frequency shifts as
the digital fingerprints to identify the legitimate LEO satellites.

%
%
\section{Open Issues}
\label{sec:OpenIssues}
{\color{black}
In this section, we present the open issues in the literature on communications and networking of SpaceNets. The organization follows that of the significant aspects identified and discussed in Section \ref{sec:ResearchDirections}.
}
\subsection{Mobility Management}
\label{ssec:OpenIssuesMM}
{\color{black}
The major open issues of mobility management in SpaceNets can be summarized as follows:
\subsubsection{How to Exploit the Opportunities of New EOS Constellations}
While some recent studies (e.g., \cite{abdelsadek2021ultra}) have drawn attention to opportunities presented by new EOSs, characteristics of these new satellites can be exploited further. For example, one main difference between old and new LEO satellite constellations is the ultra-dense deployment of new constellations. Tens of thousands of LEO satellites are planned to be launched by 2030 (e.g., SpaceX is planning to add $30,000$ satellites to its Starlink Phase 1 constellation). This can be contrasted with old LEO constellations, such as Iridium which had $66$ satellites. In addition, technologies such as FSO are being utilized to establish high-speed, low-latency ISNLs. That is, future SpaceNets will be interconnected with advanced links that will enable the exchange of large amounts of data at high speeds with low latency and high reliability. Moreover, there is much interest in adding the flexibility of software-based operation and configuration to SpaceNets. These are just a few of the opportunities that can be exploited for developing more efficient handover management techniques in future SpaceNets.
\subsubsection{Feedback and Measurement Validity} Link adaptation and handover generally depend on feedback received from a SNT regarding the power of the received signal. However, handover decisions based on the measurements of the received signal and sending feedback reports to the network could be inefficient. This is because of the long propagation delay associated with satellite and space communications that makes this information outdated. Therefore, decisions based on these outdated factors would be inaccurate, and handover could be missed accordingly. Satellite ephemeris or SNT location could be used to tackle this issue. However, this would entail extra power consumption and require the availability of SN's data and SNT's location. This issue requires a more thorough investigation to balance these factors.
\subsubsection{Different Types of SNTs and Stations}  The SpaceNet is required to communicate with different gateway stations and numerous types of SNTs, such as ground and air terminals (e.g., VSATs, IoT devices, and handheld devices) and space terminals (e.g., telescopes, planetary rovers, and space probes). These SNTs and stations are different in their communications capabilities (e.g., power, antenna, QoS demands). Therefore, for efficient handover management, suitable techniques for SNTs and station categories need to be identified.

}
{\color{black}
\subsubsection{SpaceNet Topology Management Using IP-Based Solutions} For route optimization, the nodes that execute topology management functions (in both anchor based and anchorless methods) must be carefully chosen. Clearly, neither the deployment of topology management functionality in space nor the placement of topology management functionality on the ground can provide optimal routing performance in all forwarding cases. Location management in future SpaceNets will face the challenge of high propagation delays and limited communication resources when executing binding updates or data transmission while managing the topology from Earth. Assigning topology management to LEO satellites, on the other hand, may face other issues, such as restricted onboard storage and processing, as well as the high speed of satellites. Thus, it is recommended to further develop the hybrid and dynamic topology management functions placement. Furthermore, IP-based location management systems need complicated signaling, such as tunnel dynamic creation and release, which places additional strain on satellite OBP units. Furthermore, existing IP-based location management improvements in SpaceNets confront the issue of high location management overhead because of the unique network design, in which satellite-mounted BSs move at fast speeds, resulting in frequent  handover of users in large numbers.

\subsubsection{SpaceNet Topology Management Using a Location/Identifier Split Approach} 
    It's critical to have scalable and fast location update or resolution techniques that work with reduced signaling cost and complexity. The use of ground-based location resolvers may cause delays in location update and resolution, and the Earth's geographical nature may not allow the even distribution of controllers. The placement of a region's location resolvers on satellites necessitates the transfer of location-ID mapping records from current satellite to an approaching one on a regular basis, which may consumes the ISLs resources. Placing location resolvers on a high altitude platform station (HAPS) system might be a good option in this case, because HAPS systems are located between satellites and ground stations or users and are considered quasi-stationary with respect to Earth. Any location/identifier split system, however, should consider the backward compatibility of existing IP locator and identifier systems. 
    Some location/identifier split systems use global updates, but these will not be feasible when there are thousands of satellites and millions of user devices connected to EOSNs. Thus, location/identifier split solutions must take into account the characteristics of future satellite networks, which will be made up of several megaconstellations that will provide Internet services and connectivity not only in remote  and  rural areas, but also in densely populated urban areas.

\subsubsection{SpaceNet Topology Management Using Software Defined Networks} Millions or billions of user devices will be communicating with LEO satellites in future EOSNs. A large number of flow records is created at each switch mounted on an LEO satellite. However, such records will expire as soon as the satellite moves and starts serving  a different set of users. Creating new flow records with every satellite handover causes delays and consumes a satellite's resources. In this regard, the idea of transferring flow tables from a departing satellite to an approaching one is worth exploring. It's a difficult and time-consuming task to store, manage, and search through these flow table entries. A variety of factors, including traffic loads, user mobility, user distribution, communication cost, and the dynamic nature of EOSNs, should be taken into consideration in order to reap the benefits of dynamic SDN controller placement in future EOSNs. In future EOSNs and SpaceNets, AI and ML should be used with SDNs to automate  network management procedures. Therefore, artificial intelligence should be implemented in SDN-based SpaceNets location management. In order to adapt to the dynamic nature of EOSNs and SpaceNets, AI and ML learning techniques can be valuable in choosing the SDN controllers and where they should be placed. In terrestrial networks, the flow tables are updated using the information collected through topology discovery protocols. However, predicting satellite movement can save a significant amount of topology update overhead in EOSNs and SpaceNets. User devices, on the other hand, will have many candidate SNs as handover targets due to the availability of thousands of SNs in future EOSNs and SpaceNets. Updating users' associated flows based on mobility prediction will be difficult in this case.

}

\subsection{Space-Terrestrial Integrated Routing}
\label{ssec:OpenIssuesRouting}

{\color{black}
\begin{table*}[!t]
	\centering
	\color{black}
	\setlength{\belowcaptionskip}{9pt}
	\caption{Comparison of centralized and distributed routing.}
	\label{tab:centralized_distributed}
	\begin{tabular}{cc}
		\hline
		\multicolumn{1}{|c|}{\textbf{Centralized routing}}                                                                                                                                                     & \multicolumn{1}{c|}{\textbf{Distributed routing}}                                                                                                                                                             \\ \hline
		\multicolumn{1}{|c|}{\begin{tabular}[c]{@{}c@{}}A central node (i.e., mission control on Earth) \\ calculates and distributes route tables in advance.\end{tabular}} & \multicolumn{1}{c|}{\begin{tabular}[c]{@{}c@{}}Each node is capable of calculating route tables \\in orbit and on demand based on a previously distributed \\(or learned/discovered) contact plan.\end{tabular}}                      \\ \hline
		\multicolumn{1}{|c|}{\begin{tabular}[c]{@{}c@{}}Calculates several routes, \\ some of which may not be used \\(high provisioning and memory overhead).\end{tabular}} & \multicolumn{1}{c|}{\begin{tabular}[c]{@{}c@{}}Only calculates routes that are necessary to forward traffic \\(scales better).\end{tabular}}                                                                   \\ \hline
		\multicolumn{1}{|c|}{\begin{tabular}[c]{@{}c@{}}Reaction to unexpected events will require \\ alternatives routes distributed in the route table \\ (high provisioning and memory overhead).\end{tabular}} & \multicolumn{1}{c|}{\begin{tabular}[c]{@{}c@{}}Better adaptation and reaction to unexpected events \\ (unexpected traffic sources, contact failures, etc).\end{tabular}}                                     \\ \hline
		\multicolumn{1}{|c|}{\begin{tabular}[c]{@{}c@{}}Allows for more controlled calculation and \\ optimization of the whole network.\end{tabular}} & \multicolumn{1}{c|}{\begin{tabular}[c]{@{}c@{}}Difficult to optimize and control what each node \\ calculates in the end, thus how the traffic will flow.\end{tabular}}                                                                                          \\ \hline
		\multicolumn{1}{|c|}{\begin{tabular}[c]{@{}c@{}}Free in-orbit nodes resource-constrained\\ processors from having to calculate complex routes.\end{tabular}} & \multicolumn{1}{c|}{\begin{tabular}[c]{@{}c@{}}In-orbit nodes need to use their local processor \\ to calculate route tables \\ (high in-orbit processing overhead).\end{tabular}}                                                                           \\ \hline
	\end{tabular}
\end{table*}
}

{\color{black} \par Current space-terrestrial integrated routing open issues are described below:

    \subsubsection{Novel hierarchical routing schemes} In general, the proposed inter-regional routing schemes consider that contacts between networks can be \textit{only} one of four possible types: opportunistic, probabilistic, scheduled, or permanent. Recently, some works \cite{kang2020improved, zekkori2019hybrid, krug2018hybrid, mao2019optimized} have started to consider the possibility that different types of contacts could coexist (scheduled and probabilistic, scheduled and permanent) between networks, and to design routing algorithms that work efficiently in such cases. However, there is still ample room for improvement, as the proposed solutions consider very specific use cases and leave out many others. This is the main reason why no general routing solution has yet been standardized within the \acrfull{bp} in \acrshort{ietf}, although there are many proposed solutions that work well in different cases.  
    
   \subsubsection{Multi-objective routing} Generally, the proposed routing solutions consider a fixed metric to evaluate and select the best routes. However, in heterogeneous SpaceNets there may be very different needs or limitations. On the one hand, some networks, such as DSNs, have unavoidable limitations in terms of delay, or sensor networks have limitations in terms of energy consumption. On the other hand, it is possible to have traffic with different required qualities of service. There may be telemetry traffic that is critical to the development of a space mission, but there may also be IoT traffic related to certain measurements that do not require high reliability or low latencies. Given such different needs, it is essential that routing algorithms that seek to integrate space and terrestrial networks are capable of considering multi-objective metrics that are aware of the required quality of service and the available network infrastructure capabilities and limitations.
    
   \subsubsection{Data-driven routing} Advances in artificial intelligence and machine learning techniques are proving to be a powerful enabler to solve the routing problem in heterogeneous networks that need to adapt to changing traffic and quality of service needs. How best to use these tools to improve the various key performance indicators remains an open problem. In particular, questions arise as to how to generate the data to feed the different models, how to validate such models, how to obtain explainable models and how to achieve models capable of performing acceptably in the face of changes in traffic patterns.
    
   \subsubsection{Centralized vs distributed} As discussed in \cite{madoery2018trade} and summarized in Table \ref{tab:centralized_distributed}, there is a dichotomy between centralized and decentralized routing. In general, traditional EOSNs, composed of a reduced number of SNs, have advocated centralized schemes in which a control center is in charge of calculating the routes and sending them to the SNs for them to forward the traffic. The argument is that a strict control over the flow of data is needed and SNs may not have a large computing capacity, so it is better to perform the computationally intensive part on the ground where there is greater capacity and cheaper energy. On the other hand, a distributed scheme provides a degree of autonomy and intelligence to the SNs, which allows greater robustness against disconnections or unexpected events and avoids possible bottlenecks or delays due to communication overhead with a central entity. This results in better scalability at the expense of having less controlled data flows.
    
   \subsubsection{Scalability} As networks grow in size, it may be necessary to sacrifice performance in order to reduce the computational load that the SNs must handle to be able to route traffic in a reduced time and with a given energy consumption \cite{madoery2018managing}. Investigating different mechanisms to deal with this trade-off is one of the major open problems in routing integrating terrestrial and space networks.

}

\subsection{Laser Inter-SN Links }
\label{ssec:OpenIssuesLaserISLs}
{\color{black} \par In the following, we highlight some open issues related to LISNLs. In particular, we discuss latency in FSOSNs, the setup delay for LISNLs, the number of LCTs expected per SN, and the capacity, standardization, and cost of LCTs.

\subsubsection{Latency in FSOSNs} A reduction of one millisecond of latency in HFT can produce \$100 million in revenue per year for a single brokerage firm. The higher speed of light in FSOSNs gives them an advantage over optical fiber terrestrial networks in terms of latency, and developing a better understanding of latency in FSOSNs has been the focus of research in this area. However, a realistic modeling of different types of delays, such as processing delay, propagation delay, and LISNL setup delay, which are major sources of end-to-end latency in FSOSNs, is still an open issue that needs to be addressed for a more accurate understanding of latency in FSOSNs.
    
\subsubsection{LISNL Setup Delay} LISNL setup delay is incurred due to pointing and acquisition during the formation of an LISNL between a pair of SNs equipped with laser communication terminals. Current LISNL setup times, which range from a few seconds to tens of seconds, are prohibitive, and in NG-FSOSNs that are expected to become fully operational by mid to late 2020s, an SN will be limited to establishing only permanent LISNLs with neighboring SNs that are always within its LISNL range. In NNG-FSOSNs that are likely to come into existence in the early to mid 2030s, LISNL setup times in milliseconds will be needed to enable an SN to instantaneously set up an LISNL with any neighboring SN that is currently within its LISNL range. However, reducing the LISNL setup delay from a few seconds to a few milliseconds will require extreme advancements in SN LCT’s PAT technology.

\subsubsection{Number of LCTs per SN} Four LCTs are expected per Starlink EOS according to SpaceX’s latest FCC filings. This will restrict the connectivity of a Starlink EOS to four permanent LISNLs including to two neighbors in the same OP and to two neighbors in two different OPs. Next-next-generation SNs will need to be equipped with several LCTs to establish dynamic LISNLs to provide robust connectivity within the EOSN, which will ensure low latency paths within the network, and to establish inter-orbit LISNLs for communications between different FSOSNs.

\subsubsection{Capacity of LCTs} Tesat’s ConLCT1550 will offer a capacity of 10 Gbps, while Mynaric’s CONDOR LCT is promising to provide LISNLs with data rates of 10 Gbps, and General Atomics GA-LCT will support data rates of up to 5 Gbps. LCTs for creating LISNLs are still in their infancy and are expected to offer capacities of up to 10 Gbps. LCTs offering capacities in Tbps will need to be developed to support initiatives like HydRON as well as to establish a true global communications network in space.

\subsubsection{Standardization of LCTs} Different LCTs for LEO and VLEO applications with different capabilities are being developed by different companies. These LCTs are not standardized and it is unclear whether LCTs from different vendors will be interoperable. It is also unclear whether LEO and VLEO EOSs in different EOSCs equipped with LCTs from different vendors will be able to seamlessly communicate with each other. A standardization of LCTs for FSOSNs is needed to ensure interoperability between different LCTs for seamless communications between SNs in different constellations.
    
\subsubsection{Cost of LCTs} At present, some LCTs for intra-OP and inter-OP LISNLs are ready for deployment in EOSs for LEO and VLEO EOSCs, but they are not easily available commercially. For instance, ten Starlink EOSs launched on January 24, 2021 are reported to be equipped with LCTs. However, the vendor who supplied LCTs to SpaceX for these EOSs is not known. Also, information on the number of LCTs per EOS and the capabilities of these LCTs is not currently available from SpaceX. Over the next five to ten years, LCTs are expected to become more widely available commercially, but the cost will remain a concern. Low cost and standardized LCTs supporting dynamic or on-demand intra-OP, inter-OP, and inter-orbit LISNLs will be needed to support NNG-FSOSNs.

}

\subsection{Cross-Layer Design}
\label{ssec:OpenIssuesCrossLayerDesign}
{\color{black} 

The major open issues of cross-layer design in SpaceNets can be summarized as follows:
\subsubsection{Sharing Information Across Layers} As discussed previously, direct cross-layer design can provide global benefits by optimizing the different layers, but it requires the availability of the information of each layer. It is challenging to design the interactions and sharing of the information among the different layers. This needs further investigation.
\subsubsection{Combined Direct and Indirect Design} Combining direct and indirect cross-layer design could produce an optimized system with high levels of adaptability and scalability. However, this requires developing sophisticated mathematical models that consider different parameters, status, and operating conditions. 
\subsubsection{Standardization} For wide deployment of cross-layer design and optimization,  it is vital to have standardized interactions among the different layers.  However, standardization efforts focus on each layer separately and may be developed by different organizations.

}
%
\subsection{Physical Layer Security}
\label{ssec:OpenIssuesPHYSecurity}
{\color{black} In this subsection, we highlight the open issues and challenges related to PLS:
\subsubsection {Smart and Active Eavesdropping} Since different types of users with advanced capacities and abilities to extract the transmitted information are expected to be integrated into future SpaceNets, new threat scenarios must be considered. Therefore, more serious security concerns must be addressed by including more powerful eavesdropping models in SpaceNets.
\subsubsection{Securing Massive MIMO} Massive MIMO is one of the leading physical layer technologies that is expected to be used in future networks, so it is important to examine its vulnerabilities in the face of possible malicious attacks, including jamming and eavesdropping. In addition, massive MIMO technology has been shown to be vulnerable to distributed jamming techniques. Providing various beamforming techniques as countermeasures to distributed jammers could be an interesting direction of research.
\subsubsection{Integrating AI/ML with PLS Techniques} In the previous section, we shed the light on the importance of AI/ML techniques in SpaceNets. In fact, integrating AI/ML techniques with PLS can be an effective solution for future SpaceNets by learning the different behaviors of unauthorized users and detecting active eavesdroppers. Thus, PLS can be further enhanced by incorporating AI/ML technology.
\subsubsection{Cross-Layer Security} Improving the performance of physical layer-based key generation by considering the new technologies of future SpaceNets is another challenge that needs to be investigated.
\subsubsection{Intelligent Reflective Surfaces (IRSs)} IRS is a revolutionary transmission method that changes the amplitude, phase, and frequency of incident signals to provide different transmission paths. In the literature, there are a few works that consider IRS-based satellites. In \cite{Tekbiyik_IOT}, the authors propose an architecture
involving the use of IRS units to mitigate the path loss associated with long transmission distances while providing significant gains in signal transmission. Therefore, IRS can enhance the overall PLS performance of SpaceNets. 
\subsubsection{Securing the Envisioned Use Cases for Future SpaceNets} New scenarios are envisioned in future SpaceNets, including ISNL communications, interplanetary communications, the use of flying platforms, cooperative communications, and mobile submarines, as discussed in previous sections. Therefore, new PLS techniques must be developed considering these use cases.
\subsubsection{PLS for Millimeter Wave Communications} Existing PLS studies focus on RF communications, and recently the PLS for FSO communication has been considered in the literature. However, investigating PLS for millimeter wave communications in future SpaceNets is still an open issue.

}
%
%
\section{Conclusions}
\label{sec:Conclusions}
{\color{black}
This paper has presented a visionary reference architecture for future SpaceNets capturing different kinds of space nodes, space network terminals, and their interactions via various types of sub-networks. This vision takes into account the most recent advances in different domains related to SpaceNets. We showed that the architectures and technologies investigated could revolutionize current Earth orbit satellite networks and space-related activities and bring about the envisioned SpaceNet.  In addition, we showed that the activities and projects of different standardization bodies and organizations are aligned with the envisioned SpaceNet. However, to realize such a vision, several communications and networking challenges need to be addressed. These challenges, potential solutions in the literature, and open issues have been highlighted.
}
\section*{Acknowledgment}
This work has been supported in part by the National Research Council  Canada’s (NRC) High Throughput Secure Networks program (CSTIP Grant  \#CH-HTSN-625) within the Optical Satellite Communications Consortium  Canada (OSC) framework, in part by MDA Space, and in part by Mitacs  Canada.

\ifCLASSOPTIONcaptionsoff
  \newpage
\fi



%
\balance
\bibliographystyle{ieeetr}
\bibliography{ref}

%
\begin{IEEEbiography}[{\includegraphics[width=1in,height=1.25in,clip]{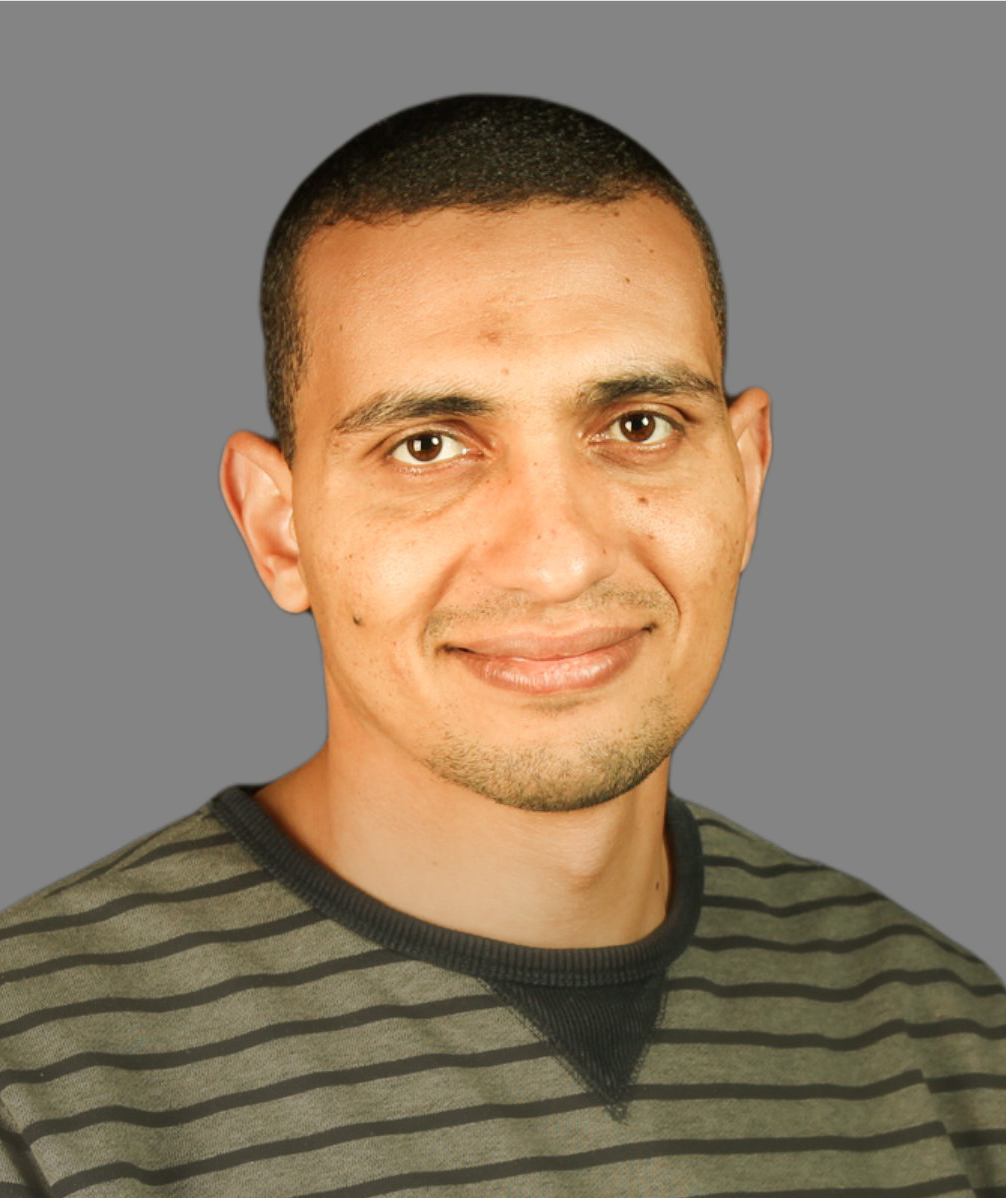}}]{Mohammed Y. Abdelsadek}
(Senior Member, IEEE) received the B.Sc. (with honors) and M.Sc. degrees from Assiut University, Assiut, Egypt, in 2011 and 2014, respectively, and his Ph.D. degree from the Memorial University of Newfoundland, St. John's, NL, Canada, in 2020, all in electrical and computer engineering. He is currently a Postdoctoral Fellow at Carleton University, Ottawa, ON, Canada. He is also an Assistant Professor (on leave) in the Department of Electrical Engineering, Assiut University, Assiut, Egypt. Earlier, he was a Sessional Instructor at the Memorial University of Newfoundland. His current research interests include satellite communication networks, ultra-reliable low-latency communications, radio resource management, and AI/ML for wireless networks. He has served as a Reviewer and TPC member for several IEEE journals and conferences.
\end{IEEEbiography}
\begin{IEEEbiography}[{\includegraphics[width=1in,height=1.25in,clip]{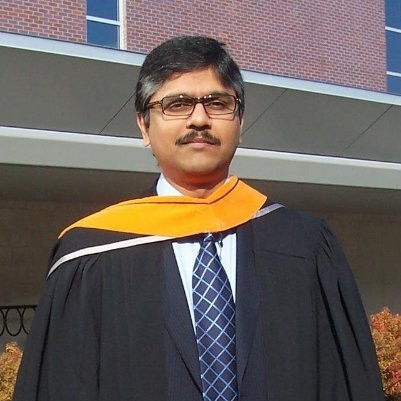}}]{Aizaz U. Chaudhry}
(Senior Member, IEEE)  received the B.Sc. degree
in electrical engineering from the University of Engineering and Technology Lahore, Lahore, Pakistan, in 1999, and the M.A.Sc. and Ph.D. degrees in electrical and computer engineering from Carleton University, Ottawa, Canada, in 2010 and 2015, respectively.
He is currently a Senior Research Associate with the Department
of Systems and Computer Engineering, Carleton University. Previously, he worked as an NSERC Postdoctoral Research Fellow with the Communications Research Centre Canada, Ottawa. His research work has been published in refereed venues and has received several citations. He has authored and coauthored more than 35 publications. His research interests include the application of machine learning and optimization in wireless networks.
Dr. Chaudhry is a Licensed Professional Engineer in the Province of
Ontario and a Member of IEEE ComSoc’s Technical Committee on Satellite and Space Communications. He serves as a technical reviewer for conferences and journals on a regular basis. He has also served as a TPC Member for conferences, such as IEEE ICC 2021 Workshop 6GSatComNet, IEEE ICC 2022 Workshop 6GSatComNet, IEEE VTC 2022-Spring, IEEE VTC 2022-Fall, Twenty-First International Conference on Networks (ICN 2022), 11th Advanced Satellite Multimedia Systems Conference (ASMS 2022), 17th Signal Processing for Space Communications Workshop (SPSC 2022), First International Symposium on Satellite Communication Systems and Services (SCSS 2022), and 15th International Conference on Communication Systems and Networks (COMSNETS 2023). 
\end{IEEEbiography}

\begin{IEEEbiography}[{\includegraphics[width=1in,height=1.25in,clip]{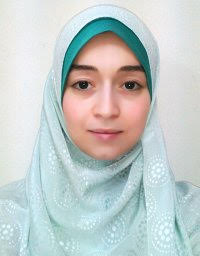}}]{Tasneem Darwish}
(Senior Member, IEEE) is currently an assistant professor at St. Francis Xavier University, Canada. She received the MSc. degree with merit in Electronics and Electrical Engineering from the University of Glasgow, UK, and her Ph.D. degree in Computer Science from Universiti Teknologi Malaysia (UTM), Malaysia. From  2019 to June 2022, Tasneem was a postdoctoral fellow at the Department of Systems and Computer Engineering, Carleton University, Canada. She was working on a collaborative project with MDA Space to investigate mobility management in future LEO satellite networks. From  2017 to 2019, Tasneem was a postdoctoral fellow at UTM. From 2019 to 2020, she was a research associate at Carleton University, Canada. Her current research interests include mobility management in future LEO satellite networks, edge/fog computing and data offloading in HAPS, vehicular ad hoc networks, and intelligent transportation systems. Tasneem is a senior IEEE member and an active reviewer for several IEEE journals.
\end{IEEEbiography}

\begin{IEEEbiography}[{\includegraphics[width=1in,height=1.25in,clip]{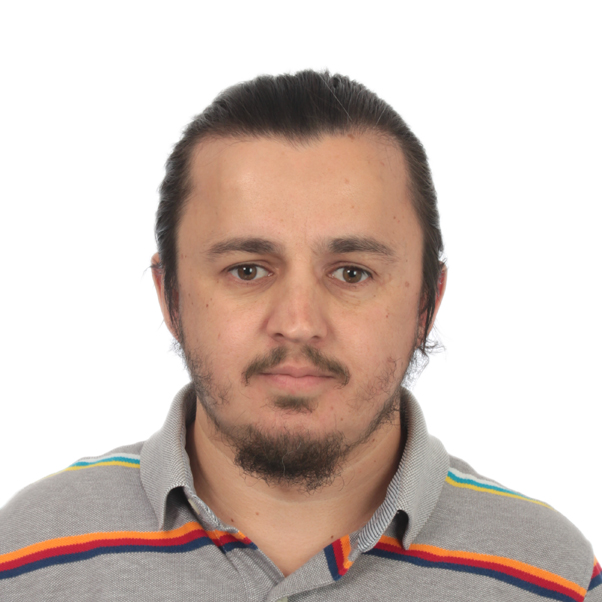}}]{Eylem Erdogan}
(Senior Member, IEEE) received B.Sc. and M.Sc. degrees from Işık University, Istanbul, Turkey and the Ph.D. degree from Kadir Has University, Istanbul, Turkey in 2014 all in electronics engineering. He is currently an Associate Professor in the Department of Electrical and Electronics Engineering, Istanbul Medeniyet University. He was a Post-Doctoral Fellow in Electrical Engineering department, Lakehead University, Thunder Bay, ON, Canada from March 2015 to September 2016 and a visiting professor in Carleton University, Ottawa, Canada during summer 2019. He has authored or coauthored more than 40+ papers in peer reviewed SCI/SCI-E journals and international conferences. Dr. Erdogan has served as a Technical Program Committee member for various IEEE conferences including IEEE Global Communications Conference (GLOBECOM), IEEE Vehicular Technology Conference (VTC-Fall), IEEE International Symposium on Personal, Indoor and Mobile Radio Communications (PIMRC) and etc. His research interests are in the broad areas of wireless communications, including signal processing for wireless communications, the performance analysis of cooperative relaying in cognitive radio networks, unmanned aerial vehicle communications and networks, free space optical communications and optical satellite networks.
\end{IEEEbiography}

\begin{IEEEbiography}[{\includegraphics[width=1in,height=1.25in,clip]{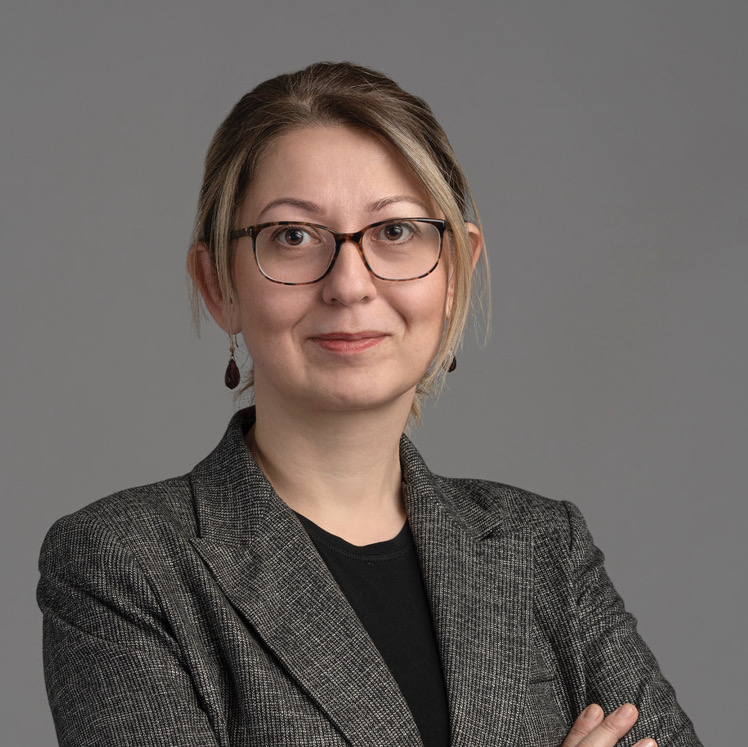}}]{Gunes Karabulut Kurt}
(Senior Member, IEEE) received the B.S. degree with high honors in electronics and electrical engineering from Bogazici University, Istanbul, Turkey, in 2000 and the M.A.Sc. and the Ph.D. degrees in electrical engineering from the University of Ottawa, ON, Canada, in 2002 and 2006, respectively. From 2000 to 2005, she was a Research Assistant with the CASP Group, University of Ottawa. Between 2005 and 2006, she was with TenXc Wireless, Canada. From 2006 to 2008, Dr. Karabulut Kurt was with Edgewater Computer Systems Inc., Canada. From 2008 to 2010, she was with Turkcell Research and Development Applied Research and Technology, Istanbul. Between 2010 and 2021, she was with Istanbul Technical University. She is currently an Associate Professor of Electrical Engineering at Polytechnique Montréal, Montréal, QC, Canada. She is a Marie Curie Fellow and has received the Turkish Academy of Sciences Outstanding Young Scientist (TÜBA-GEBIP) Award in 2019. In addition, she is an adjunct research professor at Carleton University. She is currently serving as an associate technical editor of the \textit{IEEE Communications Magazine}, an associate editor of \textit{IEEE Communication Letters}, an associate editor of \textit{IEEE Wireless Communications Letters}, and an area editor of \textit{IEEE Transactions on Machine Learning in Communications and Networking}. She is a member of the IEEE WCNC Steering Board. She is serving as the secretary of IEEE Satellite and Space Communications Technical Committee and also the chair of the IEEE special interest group entitled "Satellite Mega-constellations: Communications and Networking". She is a Distinguished Lecturer of Vehicular Technology Society Class of 2022.
\end{IEEEbiography}

\begin{IEEEbiography}[{\includegraphics[width=1in,height=1.25in,clip]{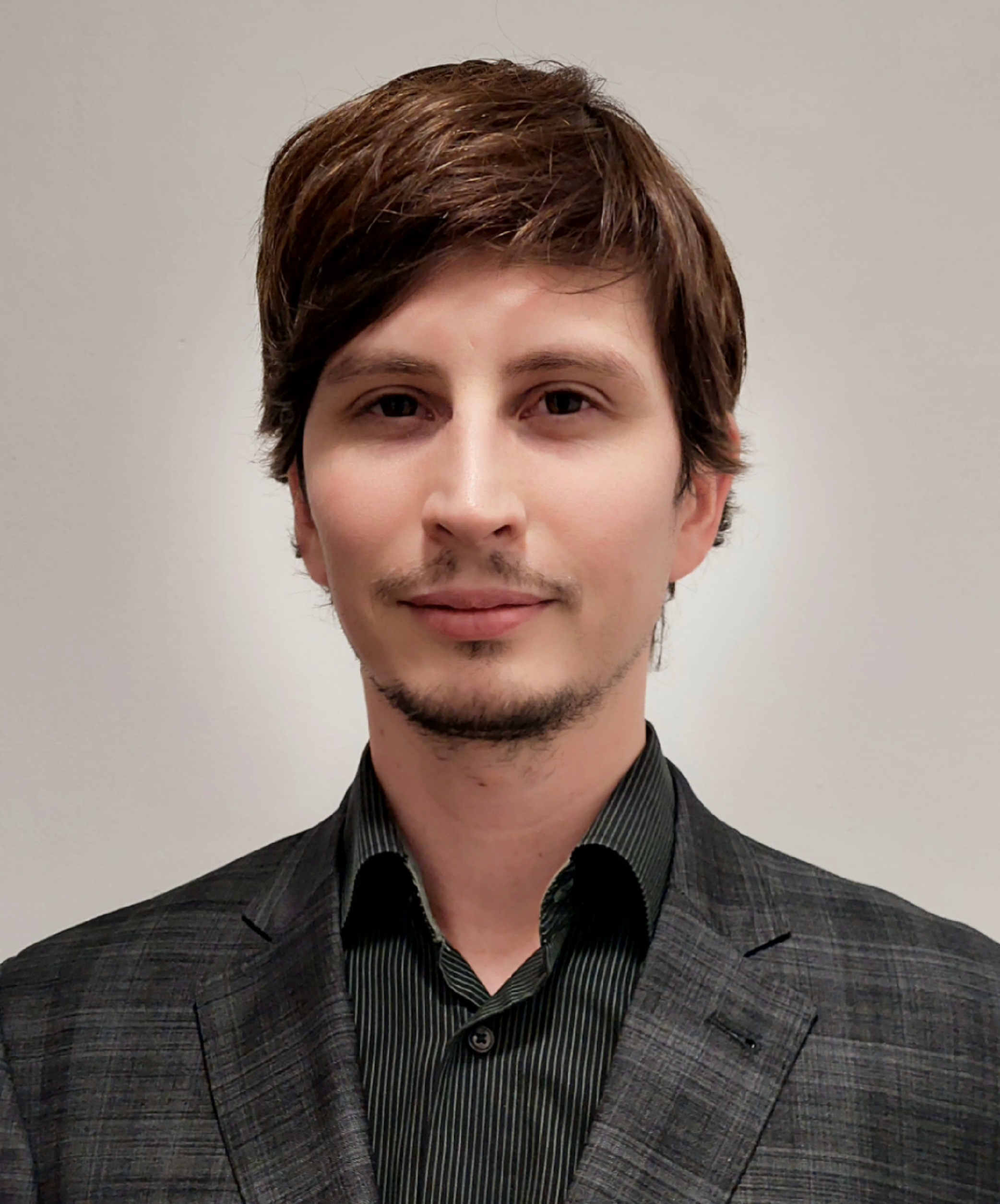}}]{Pablo G. Madoery}
(Member, IEEE) received the B.S. degree in telecommunication engineering from the National Defense University, Argentina, in 2015, and the Ph.D. degree in engineering sciences from the National University of Cordoba, Argentina, in 2019. He is a postdoctoral fellow at Carleton University on a MITACS fellowship, and an assistant professor of Computer Science at the National University of Cordoba, Argentina. He has co-authored 10 articles in international journals and 20 in leading conferences in the area of communication protocols and routing algorithms for delay-tolerant satellite networks. Currently, he is researching AI/ML-assisted routing and transport solutions applied to satellite megaconstellations.
\end{IEEEbiography}

\begin{IEEEbiography}[{\includegraphics[width=1in,height=1.25in,clip]{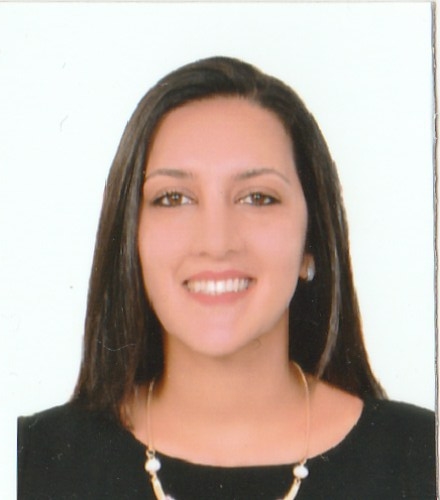}}]{Olfa Ben Yahia}
(Member, IEEE) received the Engineering degree with honors in Telecommunications from the Ecole Nationale d'Electronique et des Télécommunications de Sfax, Sfax, Tunisia, in 2016 and the Ph.D. degree in Telecommunications Engineering from Istanbul Technical University, Istanbul, Turkey in 2022. She is currently a Postdoctoral Fellow with the Department of Electrical Engineering, Polytechnique Montréal, Montreal, QC, Canada. 
Her research interests include optical wireless communications, performance analysis of physical layer security, satellite communication, aerial platforms, and cooperative communications. Currently, she is researching routing and load balancing solutions for high-throughput satellites. She is an active reviewer for several IEEE journals and conferences.
\end{IEEEbiography}

\begin{IEEEbiography}[{\includegraphics[width=1in,height=1.25in,clip]{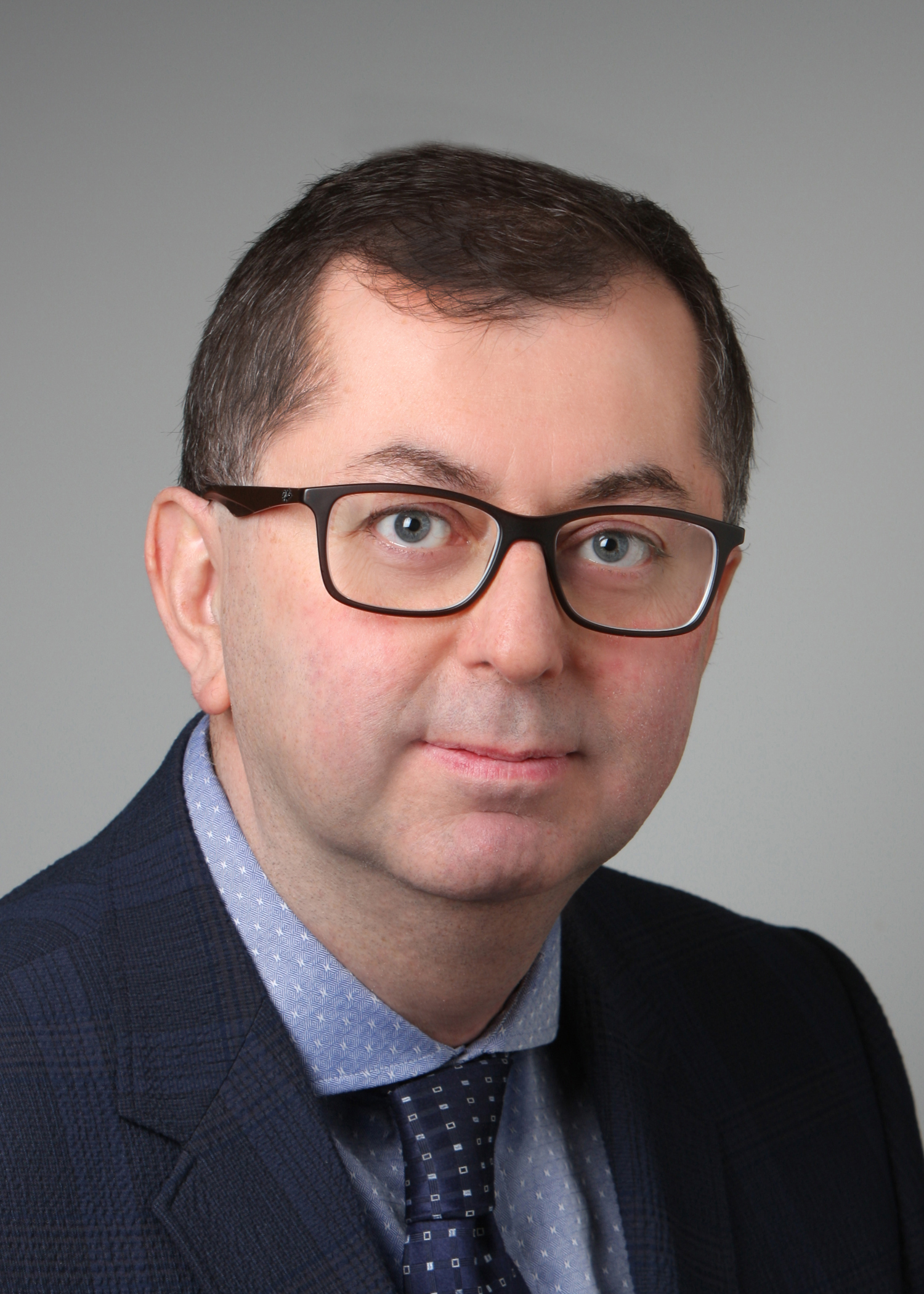}}]{Halim Yanikomeroglu}
(Fellow, IEEE) received the BSc degree in electrical and electronics engineering from the Middle East Technical University, Ankara, Turkey, in 1990, and the MASc degree in electrical engineering (now ECE) and the PhD degree in electrical and computer engineering from the University of Toronto, Canada, in 1992 and 1998, respectively. Since 1998 he has been with the Department of Systems and Computer Engineering at Carleton University, Ottawa, Canada, where he is now a Full Professor. His research interests cover many aspects of wireless communications and networks. Dr. Yanikomeroglu has coauthored 560+ published peer-reviewed research papers including 250+ in 29 different IEEE journals. He has given 110+ invited seminars, keynotes, panel talks, and tutorials in the last five years. Dr. Yanikomeroglu’s collaborative research with industry resulted in 39 granted patents. Dr. Yanikomeroglu is a Fellow of the IEEE, the Engineering Institute of Canada (EIC), and the Canadian Academy of Engineering (CAE). He is a Distinguished Speaker for the IEEE Communications Society and the IEEE Vehicular Technology Society, and an Expert Panelist of the Council of Canadian Academies (CCA|CAC). 

Dr. Yanikomeroglu is currently serving as the Chair of the Steering Committee of IEEE’s flagship wireless event, Wireless Communications and Networking Conference (WCNC). He is also a member of the IEEE ComSoc Conference Council and IEEE PIMRC Steering Committee. He served as the General Chair and Technical Program Chair of several IEEE conferences. He has also served in the editorial boards of various IEEE periodicals. 

Dr. Yanikomeroglu received several awards for his research, teaching, and service, including the IEEE ComSoc Fred W. Ellersick Prize (2021), IEEE VTS Stuart Meyer Memorial Award (2020), and IEEE ComSoc Wireless Communications TC Recognition Award (2018). He received best paper awards at IEEE Competition on Non-Terrestrial Networks for B5G and 6G in 2022 (grand prize), IEEE ICC 2021, IEEE WISEE 2021 and 2022.

\end{IEEEbiography}





\end{document}